\documentclass[11pt]{article}
\usepackage{latexsym}
\usepackage[dvips]{epsfig}
\usepackage{color}
\usepackage{amsmath}
\usepackage{amssymb}
\usepackage{amsthm}
\usepackage{amsfonts}
\usepackage{mathrsfs}
\usepackage{graphicx}
\usepackage{curves}
\usepackage{subfigure}
\usepackage{framed}
\usepackage{fancybox}
\usepackage{graphics}
\usepackage{graphicx}
\usepackage{epic,eepic}
\usepackage{algorithm}
\usepackage{algorithmic}
\usepackage{longtable}
\usepackage{multirow}
\usepackage{pdflscape}
\usepackage{hhline}
\usepackage{tikz}
\usepackage{threeparttable}
\usepackage{lineno}
\newtheorem{thm}{Theorem}
\newtheorem{lem}{Lemma}
\newtheorem{defn}{Definition}
\newtheorem{cor}{Corollary}
\newtheorem{prop}{Proposition}

\newtheorem{exam}{Example}

\newtheorem{obs}{Observation}
\newtheorem{conjec}{Conjecture}
\def\proof{\pn {Proof.} }
\def\solution{\pn {Solution.} }
\def\endproof{\hfill $\Box$ \vskip .5cm}
\def\endsolution{\hfill $\Box$ \vskip .5cm}

\def\pn{\par\smallskip\noindent}

\newcommand{\R}{\mathbb{R}}
\newcommand{\Q}{\mathbb{Q}}
\newcommand{\Z}{\mathbb{Z}}

\newcommand*\circled[1]{\tikz[baseline=(char.base)]{
  \node[shape=circle,draw,inner sep=1.8pt] (char) {#1};}}

\definecolor{Gray}{gray}{0.8}
\definecolor{LightCyan}{rgb}{0.88,1,1}
\definecolor{LightOrange}{rgb}{1.00,0.72,0.44}
\definecolor{Blue}{rgb}{0.00,0.00,1.00}
\definecolor{Red}{rgb}{1.00,0.00,0.00}
\definecolor{Green}{rgb}{0.00,1.00,0.00}
\definecolor{DarkRed}{rgb}{0.80,0.00,0.00}
\definecolor{DarkGreen}{rgb}{0.00,0.80,0.00}

\title{Tackling A Class of Hard Subset-Sum Problems:
Integration of
Lattice Attacks with Disaggregation Techniques}
\author{
Bojun Lu
\thanks{School of Data Science, The Chinese University of Hong Kong (Shenzhen), P.R.C.
Email: bojunlu@cuhk.edu.cn.}
\and Duan Li
\thanks{School of Data Science,
City University of Hong Kong,
Hong Kong.
Email: dli226@cityu.edu.hk.}
\and Rujun Jiang
\thanks{School of Data Science, Fudan University, Shanghai, P.R.C. Email: rjjiang@fudan.edu.cn.}
}
\date{August 2019}

\begin{document}
\maketitle
\renewcommand{\thefootnote}{\fnsymbol{footnote}}
\maketitle

\begin{abstract}

\noindent
Subset-sum problems belong to the NP class and play an important role in both complexity theory and knapsack-based cryptosystems, which have been proved in the literature to become hardest when the so-called \emph{density} approaches one. Lattice attacks, which are acknowledged in the literature as the most effective methods, fail occasionally even when the number of unknown variables is of medium size. In this paper we propose a modular disaggregation technique and a simplified lattice formulation based on which two lattice attack algorithms are further designed. We introduce the new concept ``jump points" in our disaggregation technique, and derive inequality conditions to identify superior jump points which can more easily cut-off non-desirable short integer solutions. Empirical tests have been conducted to show that integrating the disaggregation technique with lattice attacks can effectively raise success ratios to 100\% for randomly generated problems with density one and of dimensions up to 100. Finally, statistical regressions are conducted to test significant features, thus revealing reasonable factors behind the empirical success of our algorithms and techniques proposed in this paper.
\vspace{1cm}


\end{abstract}

\section{Introduction}

\subsection{Background}
Subset-sum problems defined as follows,
\begin{align}\label{eq:prob_subset_sum}
  \boldsymbol{a} \boldsymbol{x} := a_1x_1 + a_2x_2 + \cdots + a_n x_n = b
\end{align}
with $\boldsymbol{a} = (a_1, a_2, \ldots, a_n)\in\Z_+^n$, $b\in\Z_+$ and $\boldsymbol{x}\in{\mathcal{X}}=\{0,1\}^n$ are important problems in complexity theory and knapsack-based cryptosystems design (see \cite{lagarias1985solving}, \cite{coster1992improved}, \cite{woeginger2003exact}, \cite{impagliazzo1996efficient}, and \cite{hayes2002computing}).
Meanwhile subset-sum problems are also a special class of knapsack problems which are important in combinatorial optimization field and always of great interest to researchers (see \cite{KellererPP:2004}, \cite{Pisinger:2005}, \cite{Pisinger:2017}, and \cite{Feng:2017}).
Without loss of generality, we assume that
\begin{align}\label{eq:assump}
\max\{a_1,a_2,\ldots,a_n\}< b\leq \frac{\sum_{i=1}^{n}a_i}{2}.
\end{align}
Otherwise,
the complementary problem of \eqref{eq:prob_subset_sum} defined as follows,
\begin{align}\label{eq:complem_subset_sum}
  \boldsymbol{a} \boldsymbol{y} := a_1y_1 + a_2y_2 + \cdots + a_ny_n = \tilde{b} :=\sum_{i=1}^{n}a_i - b,
\end{align}
with $y_i = 1-x_i\in\{0,1\}$, $i=1,2,\ldots,n$, satisfies assumption~\eqref{eq:assump}.

\qquad
Identifying the feasibility of any subset-sum problem is NP-complete in general,
as the partition problem with $b=\frac{\sum_{i=1}^{n}a_i}{2}$
is NP-complete in its feasibility form (see \cite{Garey:1979}).
Meanwhile, to identify a solution of a feasible subset-sum problem is NP-hard.

\qquad A class of hard subset-sum problems can be utilized to design public-key cryptosystems
(see \cite{Merkle:1978}, \cite{ChorRivest:1984}, \cite{OTU:2000},
and \cite{micciancio2011lattice})
for transmitting 0-1 information.
Lattice attacks, which are the most critical cryptanalysis against knapsack cryptosystems,
are proposed
(see \cite{Brickell:1983}, \cite{lagarias1985solving}, \cite{coster1992improved},
and \cite{Schnorr-Euchner:1994}) to break
knapsack-based cryptosystems with relatively low density,
where \emph{density} is defined as follows,
\begin{align}\label{eq:density}
density = \frac{n}{\max_{1\leq i\leq n}(\log_{2}a_i)}.
\end{align}
The literature has revealed that subset-sum problems with their density close to one constitute the hardest subclass of subset-sum problems
(see \cite{lagarias1985solving}, \cite{Coster:1991}, and \cite{SchnorrS:2012}).
A subset-sum problem with density lower than one or higher than one is vulnerable to lattice attacks.
Beside of the \emph{density} feature defined in \eqref{eq:density},
some other structure features of subset-sum problems have also been proposed in the literature to describe the difficulty level
of the problem.
Nguyen and Stern (see \cite{NguyenStern:2005}) defined \emph{pseudo-density} to generalize the definition of
\emph{density} defined in \eqref{eq:density}, they claimed that if the
value of the pseudo-density is less than a critical value,
then the subset-sum problem is vulnerable to lattice attacks, even if the value of density
is within the critical range proposed in the literature.
Kunihiro (see \cite{Kunihiro:2008}) introduced the problem structure feature \emph{density $D$}
which unifies the notion of \emph{density} and \emph{pseudo-density}
and he also derived conditions under which
subset-sum problems are vulnerable to lattice attacks.
Jen et al. (see \cite{JLLY:2012} and \cite{JLLY2:2012}) also conducted their research study on the reliance
of the density feature defined for knapsack cryptosystems.

\qquad
The \emph{disaggregation problem} was first proposed by Glover and Woolsey (see \cite{GloverWoolsey:1972}) in 1972,
and can be described
in general as follows: How to decompose the following single Diophantine equation,
\begin{align}\label{eq:1_main_prob_Disaggregation}
\emph{\textbf{a}} \emph{\textbf{x}} := a_1x_1+a_2x_2+\cdots+a_nx_n = b,\quad \emph{\textbf{x}} \in \mathcal{X} \subseteq \Z^n
\end{align}
into an \emph{equivalent} system of two Diophantine equations,
\begin{align}\label{eq:2_disaggregated}
\left\{\begin{array}{l}
\boldsymbol\alpha \emph{\textbf{x}} :=\alpha_1x_1+ \alpha_2 x_2 + \cdots + \alpha_n x_n = b_1\\
\boldsymbol\beta \emph{\textbf{x}} := \beta_1x_1 + \beta_2 x_2 + \cdots + \beta_n x_n = b_2
\end{array}\right.,\quad \emph{\textbf{x}} \in \mathcal{X} \subseteq \Z^n
\end{align}
with $\emph{\textbf{a}} \in \Z^n_+$, $\boldsymbol\alpha$, $\boldsymbol\beta\in \Z^n$, $b\in\Z_+$, $b_1$, $b_2\in\Z$,
and $\mathcal{X}$ is a bounded set,
under the constriction that the feasible solution sets of \eqref{eq:1_main_prob_Disaggregation}
and \eqref{eq:2_disaggregated} are identical to each other.
However, after decades, research studies on the techniques dealing with the disaggregation problem are limited.
Mardanov and Mamedov studied the disaggregation problem with unknowns being binary
in their paper \cite{Mardanov:2000} and \cite{Mardanov:2004}, and
later extended their results with unknowns taking values
over a more general but still bounded set in \cite{Mamedov:2006}.

\qquad
An extension of subset-sum problems~\eqref{eq:prob_subset_sum} is the
so-called
linear Diophantine equations (LDEs), which can be presented as follows,
\begin{align}\label{eq:1_prob_main}
A \emph{\textbf{x}} = \emph{\textbf{b}},
\end{align}
where $A\in \Z^{m\times n}$ is a matrix of full row rank,
$\emph{\textbf{b}}\in\Z^{m}$, and $\emph{\textbf{x}}\in \mathcal{X} \cap {\Z^{n}}$ with $\mathcal{X} = \{x~|~0\leq x_i\leq u_i,~i=1,2,\ldots,n\}$.
In 2010, Aardal and Wolsey (see \cite{AardalW:2010}) studied and extended formulations
for system of LDEs in \eqref{eq:1_prob_main} based on lattice theory,
and also obtained a solution scheme for disaggregation problem as a by-product.
Beside of the fact of limited studies on disaggregation,
we are also inspired by
the cell enumeration method proposed by Li et al. (see \cite{Li:2011}).
Their method solves the linear Diophantine equations with the complexity $O\left(\left(n \max\{u_1, \ldots, u_n\}\right)^{n-m}\right)$,
which depends on the magnitude of $n-m$.
Hence, increasing the magnitude of $m$ and thus reducing the magnitude of $n-m$ directly
improve the complexity bound.




\qquad
As hard subset-sum problems are used in cryptosystem protocol designs
(see \cite{Sharma:2011}, \cite{Odlyzko:1990}, \cite{Kate:2011}, and
\cite{buhler1998lattice}),
attacking algorithms are proposed to
tackle these hard problems, among which lattice attack algorithms are important 
(see \cite{lagarias1985solving} \cite{coster1992improved}, \cite{Aardal:2000}, and \cite{SchnorrS:2012}).
Lagarias and Odlyzko (see \cite{lagarias1985solving}) proposed the lattice transformation
of a subset-sum problem and then used LLL basis reduction (see \cite{LLL:1982},
\cite{LovaszS:1992}, and \cite{divason2018formalization}) to derive short vectors in the lattice, which can help identify feasible solutions.
Coster et al. (see \cite{coster1992improved}) improved the work proposed in \cite{lagarias1985solving}
by shifting the lattice, and then can solve even sparser and harder subset-sum problems.
Aardal et al. (see \cite{Aardal:2000}) proposed a lattice transformation for the system of linear Diophantine equations with upper and lower bounds on the unknowns, thus to first identify
an integer solution, and then to use branch-and-bound methods to further search feasible solutions.

\subsection{Our motivation and contributions}
The following considerations motivate our study in this paper.
On one hand, lattice attack algorithms proposed in the literature on subset-sum problems
with density approaching one fail occasionally,
when the number of unknown variables is only of medium size.
On the other hand, disaggregation problem has been proposed for decades but with very limited research studies
on it, meanwhile increasing the number of equations can uncover more information and thus benefits the computation.
Most importantly, so far there are limited research work on combing lattice attacks with cutting methods,
especially with disaggregation techniques.

\qquad
In this paper, we propose an improved and simplified lattice formation of subset-sum problems and also propose modular disaggregation technique, then novelly integrate lattice formulation and
disaggregation technique.
The modular disaggregation technique
aims to reveal more information of the given system, and to cut-off the non-binary integer
solutions with small Euclidean length initially returned by lattice attack algorithms.
Numerical tests support that this mechanism work efficiently, and
can increase the probability of returning valid binary solutions.

\qquad
In this paper, we also introduce and define the concept of ``jump points of subset-sum problems",
which specially play an important role in the modular disaggregation technique.
Conditions are derived to identify superior jump points which can more easily help cut-off
non-desirable non-binary integer solutions with small Euclidean length.

\qquad
The algorithm $\textbf{Reduce}(\emph{\textbf{x}}_\emph{\textbf{b}}, D)$ proposed in this paper, can return the same integer solution
as that returned by algorithm \textbf{AHL-Alg}, but with a simpler lattice formulation and with just one big auxiliary integer number $N$.
Moreover,
after adopting a one-half modification to $\textbf{Reduce}(\emph{\textbf{x}}_\emph{\textbf{b}}, D)$, we further propose algorithm $\textbf{Reduce}_{1/2}(\emph{\textbf{x}}_\emph{\textbf{b}}, D)$ in this paper,
which can return valid binary solutions with higher probability compared with algorithm $\textbf{Reduce}(\emph{\textbf{x}}_\emph{\textbf{b}}, D)$.

\qquad
Worth mentioning that, Theorem~\ref{thm:matrix B_reduced} proved in this paper is
a more general result compared with the study in
Havas et al.'s paper (see \cite{HMM:1998}) for a single equation.

\subsection{Organization}

In Section~2, we first have a quick review on the classic LLL algorithm and three
important lattice attack algorithms in the literature.
In Section~3, we introduce our simplification and improvement on one such lattice attack algorithm,
and also propose the modular disaggregation technique.
Results of numerical experiments are reported in Section~4.
More analysis on the possible mechanisms behind the efficiency of integrating modular disaggregation technique
with lattice attack, are summarized in Section~5.
Finally, Section~6 contains conclusion and further research.

\section{Technical Preliminaries}

\subsection{Review of the Lattice Basis Reduction Algorithm}\label{review of LLL}

\begin{figure}[H]
\begin{center}
    \begin{tikzpicture}[scale=0.6]
    \draw [Gray, fill=Gray] (7.2,1.3)--(8.2,1.3)--(8.4,2.6)--(7.4,2.6);
    \draw [Gray, fill=Gray] (7.2,1.3)--(4.6,3.9)--(2.8,5.2)--(5.4,2.6);
    \draw [fill=orange](1,0) circle(2pt);
    \draw [fill=orange](2,0) circle(2pt);
    \draw [fill=orange](3,0) circle(2pt);
    \draw [fill=orange](4,0) circle(2pt);
    \draw [fill=orange](5,0) circle(2pt);
    \draw [fill=orange](6,0) circle(2pt);
    \draw [fill=orange](7,0) circle(2pt);
    \draw [fill=orange](8,0) circle(2pt);
    \draw [fill=orange](9,0) circle(2pt);
    \draw [fill=orange](1.2,1.3) circle(2pt);
    \draw [fill=orange](2.2,1.3) circle(2pt);
    \draw [fill=orange](3.2,1.3) circle(2pt);
    \draw [fill=orange](4.2,1.3) circle(2pt);
    \draw [fill=orange](5.2,1.3) circle(2pt);
    \draw [fill=orange](6.2,1.3) circle(2pt);
    \draw [fill=LightCyan](7.2,1.3) circle(2pt); 
    \draw [fill=orange](8.2,1.3) circle(2pt);
    \draw [fill=orange](9.2,1.3) circle(2pt);
    \draw [fill=orange](1.4,2.6) circle(2pt);
    \draw [fill=orange](2.4,2.6) circle(2pt);
    \draw [fill=orange](3.4,2.6) circle(2pt);
    \draw [fill=orange](4.4,2.6) circle(2pt);
    \draw [fill=orange](5.4,2.6) circle(2pt);
    \draw [fill=orange](6.4,2.6) circle(2pt);
    \draw [fill=orange](7.4,2.6) circle(2pt);
    \draw [fill=orange](8.4,2.6) circle(2pt);
    \draw [fill=orange](9.4,2.6) circle(2pt);
    \draw [fill=orange](1.6,3.9) circle(2pt);
    \draw [fill=orange](2.6,3.9) circle(2pt);
    \draw [fill=orange](3.6,3.9) circle(2pt);
    \draw [fill=orange](4.6,3.9) circle(2pt);
    \draw [fill=orange](5.6,3.9) circle(2pt);
    \draw [fill=orange](6.6,3.9) circle(2pt);
    \draw [fill=orange](7.6,3.9) circle(2pt);
    \draw [fill=orange](8.6,3.9) circle(2pt);
    \draw [fill=orange](9.6,3.9) circle(2pt);
    \draw [fill=orange](1.8,5.2) circle(2pt);
    \draw [fill=orange](2.8,5.2) circle(2pt);
    \draw [fill=orange](3.8,5.2) circle(2pt);
    \draw [fill=orange](4.8,5.2) circle(2pt);
    \draw [fill=orange](5.8,5.2) circle(2pt);
    \draw [fill=orange](6.8,5.2) circle(2pt);
    \draw [fill=orange](7.8,5.2) circle(2pt);
    \draw [fill=orange](8.8,5.2) circle(2pt);
    \draw [fill=orange](9.8,5.2) circle(2pt);
    \draw [->, thick] (7.2,1.3)--(8.2,1.3);
    \draw [->, thick] (7.2,1.3)--(7.4,2.6);
    \draw [->, thick] (7.2,1.3)--(4.6,3.9);
    \draw [->, thick] (7.2,1.3)--(5.4,2.6);
    \end{tikzpicture}
\caption{Illustration of an LLL-reduced basis.}
\label{fig:LLL-reduced basis}
\end{center}
\end{figure}
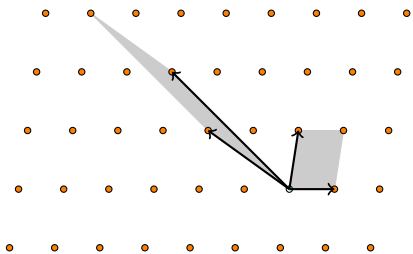

In this paper
we abbreviate the LLL basis reduction algorithm (see \cite{LLL:1982},
\cite{LovaszS:1992}, \cite{nguyen2009lll}, \cite{may2009using},
\cite{schneider2009probabilistic}, \cite{buhler1998lattice}, and \cite{divason2018formalization})
to the \emph{LLL algorithm}
and
name the basis obtained by applying the LLL basis reduction algorithm as the \emph{LLL-reduced basis}.
The LLL algorithm is a polynomial time and complexity algorithm which is a milestone algorithm
for integer programming problems with fixed dimension.
Algebraically speaking, to obtain the LLL-reduced basis,
a series of unimodular column operations need to be conducted on an ordered
pre-given basis.
Geometrically speaking, vectors consisting an LLL-reduced basis are relatively
\emph{short} and \emph{nearly orthogonal} to one another.
In this paper, we use the Euclidean norm of
a vector (see \cite{filaseta1999factorization},
\cite{martinet2013perfect}) to measure the length of a solution vector. For instance,
a well known result has been reviewed in Lemma~\ref{lem:6_LLL_3}, which is about
the upper bound of the length of a vector in the LLL-reduced basis.
Figure~\ref{fig:LLL-reduced basis} is an illustration of an arbitrary pre-given basis and the LLL-reduced basis for the same \emph{lattice}.

\begin{defn}[lattice, see \cite{Murray:2011}]
Let $\textbf{b}_1,\textbf{b}_2,\ldots,\textbf{b}_n \in \R^{\tilde{n}}$ be linearly independent column vectors with $n\leq \tilde{n}$, the set $\mathcal{L}$ defined as follows,
$$
\mathcal{L} := \Z \textbf{b}_1 + \Z \textbf{b}_2 + \cdots + \Z \textbf{b}_n := \left\{\sum_{i=1}^{n}z_i\textbf{b}_i~|~z_i\in\Z,~i=1,2,\ldots,n\right\},
$$
is called a lattice of dimension $n$.
Moreover, $\{\textbf{b}_1, \textbf{b}_2,\ldots,\textbf{b}_n\}$ is called a basis for
the lattice $\mathcal{L}$.
\end{defn}

\begin{thm}[see \cite{Murray:2011}]\label{thm:lattice_equivalence}
Given a lattice $\mathcal{L}$,
column vectors of matrix $B$ and column vectors of matrix $\tilde{B}$ are two equivalent bases for $\mathcal{L}$,
if and only if there exists a unimodular matrix $U$,
such that $B = \tilde{B}U$.
\end{thm}

\vspace{3mm}

\qquad
In the LLL algorithm, the Gram-Schmidt orthogonalization (GSO) process is a crucial
component,
which is reviewed in Algorithm~\ref{alg:GSO}.
A review of the LLL algorithm is presented in Algorithm~\ref{alg:LLL}.
A more detailed description of the theory, techniques, and applications of the
LLL algorithm, can be reached in Bremner's book \cite{Murray:2011},
and the book \cite{Nguyen:2010} edited by Nguyen and Vall$\rm \acute{e}$e.

\qquad
The major steps of the LLL algorithm can be described as follows.
\vspace{-3mm}
\begin{itemize}
  \item
  Firstly, the GSO process is conducted on the input ordered basis
    $\{\emph{\textbf{b}}_1, \emph{\textbf{b}}_2, \ldots, \emph{\textbf{b}}_n\}$
    and an orthogonal basis
    $\{\emph{\textbf{b}}_1^*, \emph{\textbf{b}}_2^*, \ldots, \emph{\textbf{b}}_n^*\}$ is obtained as follows,
    \begin{align*}\begin{array}{l}
    \emph{\textbf{b}}_1^* = \emph{\textbf{b}}_1,\\
    \emph{\textbf{b}}_2^* = \emph{\textbf{b}}_2 - \mu_{1,2}\emph{\textbf{b}}_1^*,\quad \mu_{1,2} = \frac{\emph{\textbf{b}}_2\cdot \emph{\textbf{b}}_1^*}{\emph{\textbf{b}}_1^*\cdot \emph{\textbf{b}}_1^*},\\
    \cdots\\
    \emph{\textbf{b}}_i^* = \emph{\textbf{b}}_i - \mu_{i-1,i}\emph{\textbf{b}}_{i-1}^* - \mu_{i-2,i}\emph{\textbf{b}}_{i-2}^*
    - \cdots - \mu_{1,i}\emph{\textbf{b}}_{1}^*, \quad \mu_{j,i} = \frac{\emph{\textbf{b}}_i \cdot \emph{\textbf{b}}_j^*}{\emph{\textbf{b}}_j^*\cdot \emph{\textbf{b}}_j^*},
    \quad 1 \leq j<i,\\
    \cdots\\
    \emph{\textbf{b}}_n^* = \emph{\textbf{b}}_n - \mu_{n-1,n}\emph{\textbf{b}}_{n-1}^*-\mu_{n-2,n}
    \emph{\textbf{b}}_{n-2}^* - \cdots - \mu_{1,n}\emph{\textbf{b}}_{1}^*.
    \end{array}
    \end{align*}
  \item
   Secondly, two crucial operations which are so-called as `Reduce' and `Exchange'
   will be applied to the input basis vectors
   $\{\emph{\textbf{b}}_1, \emph{\textbf{b}}_2, \ldots, \emph{\textbf{b}}_n\}$
   with $\frac{1}{4} < \alpha <1$ being a parameter with pre-given value.
   Empirically, the closer of $\alpha$ to 1, the higher quality of the LLL-reduced basis.
    \begin{itemize}
        \item
        (Reduce) If $|\mu_{j,i}|>\frac{1}{2}$,
        then $\emph{\textbf{b}}_i \leftarrow \emph{\textbf{b}}_i-\lceil \mu_{j,i} \rfloor \emph{\textbf{b}}_j$.
        \item
        (Exchange) If $||\emph{\textbf{b}}_i^*+\mu_{i-1,i}\emph{\textbf{b}}_{i-1}^*||^2 < \alpha||\emph{\textbf{b}}_{i-1}^*||^2$, then exchange $\emph{\textbf{b}}_i$ and $\emph{\textbf{b}}_{i-1}$.
    \end{itemize}
\end{itemize}

\qquad
As the output, the LLL algorithm returns the LLL-reduced (depends on the choice of $\alpha$) basis
which satisfies the following conditions,
\vspace{-3mm}
\begin{itemize}
\item $|\mu_{j,i}| \leq \frac{1}{2}$, $1\leq j<i\leq n$,
\item $||\emph{\textbf{b}}_i^*+\mu_{i-1,i}\emph{\textbf{b}}_{i-1}^*||^2 \geq \alpha||\emph{\textbf{b}}_{i-1}^*||^2$, $1<i\leq n$.
\end{itemize}

\begin{lem}[see \cite{LLL:1982}]\label{lem:6_LLL_3}
If $\{\textbf{b}_1, \textbf{b}_2, \ldots, \textbf{b}_n\}$ is the LLL-reduced basis with parameter $\alpha$ for the lattice $\mathcal{L}\in \R^{\tilde{n}}$ with $n\leq \tilde{n}$.
Let $\textbf{y}_1, \textbf{y}_2, \ldots, \textbf{y}_t \in \mathcal{L}$
be any $t$ linearly independent vectors.
Then for any $j$ with $1\leq j\leq t$, the following inequality holds,
$$
||\textbf{b}_j||^2\leq \beta^{n-1}\max\{||\textbf{y}_1||^2, ||\textbf{y}_2||^2, \ldots,||\textbf{y}_t||^2\}
$$
with $\beta = \frac{4}{4\alpha - 1}$.
\end{lem}


\begin{algorithm}[h!]
\caption{$(M, B^*)$ = \textbf{GSO}($B$)}
\label{alg:GSO}
{\small
\textbf{input} Matrix $B$, where $\emph{\textbf{b}}_i$, $i=1,2,\ldots,n$, denotes the $i$th column of $B$.\\
\textbf{output} Matrix $M$ and matrix $B^*$, such that $B = {B^*}M$ with columns of $B^*$
being orthogonal with one another.
\begin{algorithmic}[1]
\STATE $M$ $\leftarrow$ $\emph{\textbf{0}}^{n\times{n}}$
\FOR {$i = 1:n$}
    \STATE $\mu_{i,i}$ $\leftarrow$ 1
\ENDFOR
\STATE $\emph{\textbf{b}}_1^*$ $\leftarrow$ $\emph{\textbf{b}}_1$
\FOR {$i = 2:n$}
    \STATE $\emph{\textbf{b}}_i^*$ $\leftarrow$ $\emph{\textbf{b}}_i$
    \FOR {$j=1:i-1$}
        \STATE $\mu_{j,i}$ $\leftarrow$ $\frac{\emph{\textbf{b}}_i \cdot \emph{\textbf{b}}_j^*}{\emph{\textbf{b}}_j^* \cdot \emph{\textbf{b}}_j^*}$
        \STATE $\emph{\textbf{b}}_i^*$ $\leftarrow$ $\emph{\textbf{b}}_i^* - \mu_{j,i}\emph{\textbf{b}}_j^*$
    \ENDFOR
\ENDFOR
\RETURN $M$ and $B^*$
\end{algorithmic}
}
{\footnotesize Note:
$\mu_{i,j}$ denotes the $(i,j)$th entry of matrix $M$. $\emph{\textbf{b}}_i^*$ denotes the $i$th column of matrix $B^*$.
}
\end{algorithm}

\begin{algorithm}[h!]
\caption{$\tilde{B}$ = \textbf{LLL}($B$, $\alpha$) (or \textbf{LLL}($B$) as an abbreviation)}
\label{alg:LLL}
{\small
\textbf{input} Matrix $B$, where $\emph{\textbf{b}}_i$, $i=1,2,\ldots,n$, denotes the $i$th column of $B$.
Scalar $\alpha$ with $\frac{1}{4}<\alpha<1$.\\
\textbf{output} Matrix $\tilde{B}$, columns of which form the LLL-reduced basis.\\
\vspace{-4mm}
\begin{algorithmic}[1]
\STATE $\tilde{B}$ $\leftarrow$ $B$
\STATE $(M, \tilde{B}^*)$ = $\text{GSO}(\tilde{B})$
\STATE $k \leftarrow 2$
\WHILE {$k\leq n$}
    \STATE Reduce $\tilde{\emph{\textbf{b}}}_{k}$ by $\tilde{\emph{\textbf{b}}}_{k-1}$
            and update entries of $M$
    \IF {$||\tilde{\emph{\textbf{b}}}_{k}^*+\mu_{k-1,k}\tilde{\emph{\textbf{b}}}_{k-1}^*||^2
            < \alpha||\tilde{\emph{\textbf{b}}}_{k-1}^*||^2$}
        \STATE Exchange $\tilde{\emph{\textbf{b}}}_{k}$ and $\tilde{\emph{\textbf{b}}}_{k-1}$
                and update entries of $M$
        \IF {$k>2$}
            \STATE $k$ $\leftarrow$ $k-1$
        \ENDIF
    \ELSE
        \FOR {$h = k-2:1$}
            \STATE Reduce $\tilde{\emph{\textbf{b}}}_{k}$ by $\tilde{\emph{\textbf{b}}}_{h}$
        \ENDFOR
        \STATE $k$ $\leftarrow$ $k+1$
    \ENDIF
\ENDWHILE
\RETURN $\tilde{B}$
\end{algorithmic}
}
{\footnotesize Note:
$\mu_{i,j}$ denotes the $(i,j)$th entry of $M$.
$\tilde{\emph{\textbf{b}}}_i$ denotes the $i$th column of matrix $\tilde{B}$,
and $\tilde{\emph{\textbf{b}}}_i^*$ denotes the $i$th column of matrix $\tilde{B}^*$.
In order to save efforts in updating entries of $M$ in step~5 and step~7,
conclusions in Lemma~\ref{lem:3_GSO_reduction} and Lemma~\ref{lem:2_GSO_exchange} can
help.
}
\end{algorithm}

\qquad
We next review two important lemmas for the GSO process which play important roles
in simplifying computational efforts in the LLL algorithm.
Specifically, Lemma~\ref{lem:3_GSO_reduction} tells the property of reducing one basis vector by
another, and Lemma~\ref{lem:2_GSO_exchange} tells the property of exchanging one basis vector with another.

\begin{lem}[see \cite{Murray:2011}, or Lemma~2.2.1 of \cite{Bojun:2014} with proof]\label{lem:3_GSO_reduction}
Let $\textbf{b}_1, \textbf{b}_2, \ldots, \textbf{b}_n$ be a basis of the lattice $\mathcal{L}\in\R^{\tilde{n}}$ with $n\leq\tilde{n}$.
Let $\hat{\textbf{b}}_1, \hat{\textbf{b}}_2, \ldots, \hat{\textbf{b}}_n$ be another basis of the lattice $\mathcal{L}$, and
$$
\hat{\textbf{b}}_k = \textbf{b}_k - \gamma \textbf{b}_l,~~
\hat{\textbf{b}}_i = \textbf{b}_i~(i\neq k,~1\leq i\leq n),
$$
where $\gamma\in\Z$ and $1\leq l<k\leq n$ with $\gamma$, $k$ and $l$ being fixed.
Let $\textbf{b}_i^*$, $\mu_{j,i}$, $1\leq j<i\leq n$ and $\hat{\textbf{b}}_i^*$, $\hat{\mu}_{j,i}$, $1\leq j<i\leq n$ be the GSO output of these two bases respectively.
Then the following properties hold,
\begin{itemize}
\item [(a)] $\hat{\textbf{b}}_i^* = \textbf{b}_i^*$, for all $i$ with $1\leq i\leq n$.
\item [(b)] $\hat{\mu}_{i,j} = \mu_{i,j}$, for all $i$, $j$ with $i\neq k$, $1\leq j<i\leq n$.
When $i=k$, the following is true,
    \begin{align*}
    \hat{\mu}_{j,k} = \left\{\begin{array}{ll}
    \mu_{j,k} - \gamma \mu_{j,l}, &  1\leq j<l,\\
    \mu_{l,k} - \gamma, & j = l,\\
    \mu_{j,k}, &  l < j <k.
    \end{array}\right.
    \end{align*}
\end{itemize}
\end{lem}

\begin{lem}[see \cite{Murray:2011}, or Lemma~2.2.2 of \cite{Bojun:2014} with proof]\label{lem:2_GSO_exchange}
Let $\textbf{b}_1, \textbf{b}_2, \ldots, \textbf{b}_n$ be a basis of the lattice $\mathcal{L}\in\R^{\tilde{n}}$ with $n\leq \tilde{n}$.
Let $\hat{\textbf{b}}_1, \hat{\textbf{b}}_2, \ldots, \hat{\textbf{b}}_n$ be another basis of the lattice $\mathcal{L}$, and
$$
\hat{\textbf{b}}_{k-1} = \textbf{b}_{k},~~\hat{\textbf{b}}_{k} = \textbf{b}_{k-1},~~\hat{\textbf{b}}_i = \textbf{b}_i~(1\leq i\neq k-1, k \leq n).
$$
Let $\textbf{b}_i^*$, $\mu_{j,i}$, $1\leq j<i\leq n$ and $\hat{b}_i^*$, $\hat{\mu}_{j,i}$, $1\leq j<i\leq n$ be the GSO output of the two bases respectively.
Then the following properties hold,
\begin{itemize}
\item [(a)]$\hat{\textbf{b}}_{i}^*=\textbf{b}_i^*$ for all $i$ with $1\leq i\leq n$, $i\neq k-1,~ k$.
\item [(b)] $\hat{\textbf{b}}_{k-1}^*=\textbf{b}_k^* + \mu_{k-1,k}\textbf{b}_{k-1}^*$, and
$\hat{\textbf{b}}_{k}^*=\frac{||\textbf{b}_k^*||^2}{||\hat{\textbf{b}}_{k-1}^*||^2}\textbf{b}_{k-1}^*
-  \mu_{k,k-1}\frac{||\textbf{b}_{k-1}^*||^2}{||\hat{\textbf{b}}_{k-1}^*||^2}\textbf{b}_k^*$.
\item [(c)] $\hat{\mu}_{i,j}=\mu_{i,j}$, for all $i$, $j$ with $1\leq i\leq n$ and $i\neq k-1, i\neq k$, and $1\leq j<i$ and $j\neq k-1, j\neq k$.
\item [(d)] For all $i$ with $k+1 \leq i\leq n$, the followings hold,
$$\hat{\mu}_{i,k-1}
= \frac{\mu_{i,k}||\textbf{b}_k^*||^2 + \mu_{i,k-1}\mu_{k,k-1}||\textbf{b}_{k-1}^*||^2}
{||\hat{\textbf{b}}_{k-1}^*||^2},$$
$$\hat{\mu}_{i,k}
= \mu_{i,k-1} - \mu_{i,k}\mu_{k,k-1}
.$$
\item [(e)] $\hat{\mu}_{k-1,j} = \mu_{k,j}$, for all $j$ with $1\leq j\leq k-2$.
\item [(f)] $\hat{\mu}_{k,j} = \mu_{k-1,j}$, for all $j$ with $1\leq j\leq k-2$,
and $\hat{\mu}_{k,k-1} = \mu_{k,k-1}\frac{||\textbf{b}_{k-1}^*||^2}{||\hat{\textbf{b}}_{k-1}^*||^2}$.
\end{itemize}
\end{lem}

\subsection{Lattice Formulations for Linear Equation Problems}

In this section we first briefly illustrate how subset-sum problems and its extension systems of LDEs can be transformed into lattice formulations in the literature.
Specifically, we would briefly summarize here that how the three important
lattice attack algorithms
(see \cite{lagarias1985solving} \cite{coster1992improved}, and \cite{Aardal:2000})
transform the equation problems into lattice formulations.
Lattice formulations proposed in this paper are detailed described and explained in Section~\ref{subsec:improvements_of_lattice}. For convenience, we abbreviate the algorithms proposed in
\cite{lagarias1985solving} \cite{coster1992improved}, and \cite{Aardal:2000})
as \textbf{LO-Alg},
\textbf{CJLOSS-Alg}, and \textbf{AHL-Alg},
respectively.

\qquad
In \textbf{LO-Alg} (see \cite{lagarias1985solving}),
the following matrix of dimension $(n+1) \times (n+1)$,
\begin{align}\label{eq:matrix_B_LO}
B_{LO} = \left(\begin{array}{cc}
I^{n\times n} & \emph{\textbf{0}}^{n\times 1}\\
-\emph{\textbf{a}}^{1\times n} & b
\end{array}
\right),
\end{align}
is proposed for problem~\eqref{eq:prob_subset_sum}.
In \textbf{CJLOSS-Alg} (see \cite{coster1992improved}),
the following matrix of dimension $(n+1) \times (n+1)$,
\begin{align}\label{eq:matrix_B_CJ}
B_{CJLOSS} = \left(
\begin{array}{cc}
I^{n\times n} & \frac{1}{2}\times \emph{\textbf{1}}^{n\times 1}\\
\emph{\textbf{a}}^{1\times n}N & bN
\end{array}
\right)
\end{align}
is proposed for problem~\eqref{eq:prob_subset_sum}.
In \textbf{AHL-Alg} (see \cite{Aardal:2000}), the following
matrix of dimension
$(n+m+1) \times (n+1)$,
\begin{align}\label{eq:matrix_B_AHL}
B_{AHL} = \left(
\begin{array}{cc}
I^{n\times n} & \emph{\textbf{0}}^{n\times 1}\\
\emph{\textbf{0}}^{1\times n} & N_1\\
A^{m\times n}N_2 & -\emph{\textbf{b}}^{m\times 1}N_2
\end{array}
\right)
\end{align}
is proposed for problem~\eqref{eq:1_prob_main}.

\qquad
In order to use lattice basis reduction theory, columns of matrix $B_{LO}$, columns of matrix $B_{CJLOSS}$, and columns of matrix $B_{AHL}$ are regarded as bases of three different lattices, respectively.
Then the LLL reduced bases would be derived, respectively.
Specifically, after transforming the subset-sum problem or its extension system of LDEs into lattice problems,
the three algorithms \textbf{LO-Alg}, \textbf{CJLOSS-Alg} and \textbf{AHL-Alg} tried to identify the feasible integer solution in the following ways, respectively.
\vspace{-3mm}
\begin{itemize}
\item
{Let $\tilde{B}_{LO}$ denote the matrix whose columns consist the LLL-reduced basis of
  the lattice generated by columns of $B_{LO}$.
\textbf{LO-Alg} checked whether \emph{any} $j$th column of $\tilde{B}_{LO}$
with $j\in\{1,2,\ldots,n+1\}$
is of the form that
$\tilde{b}_{i,j}\in\{0, \lambda\}$ for \emph{all} $i \in \{1,2,\ldots, n\}$
for some fixed scalar $\lambda$, and $\tilde{b}_{n+1, j}=0$.
The desired and identified column vector $(b_{1,j}, b_{2,j}, \ldots, b_{n,j})^T$ is divided
by the scalar $\lambda$, which becomes a binary column vector, then we check whether the binary column vector
is a feasible binary solution to Problem~\eqref{eq:prob_subset_sum}.
If no such targeted column vector appears,
the procedure can be applied to the Complementary Problem~\eqref{eq:complem_subset_sum}
as well, with $b$ replaced by $\tilde{b}:=\sum_{i=1}^{n}a_i - b$.
An analysis derived for \textbf{LO-Alg} is presented in \cite{Frieze:1986}.
}
\item
{Let $N>\frac{1}{2}\sqrt{n}$, and let
$\tilde{B}_{CJLOSS}$ denote
the matrix whose columns consist the LLL-reduced basis of
the lattice generated by columns of $B_{CJLOSS}$.
\textbf{CJLOSS-Alg} checked whether \emph{any} $j$th column of $\tilde{B}_{CJLOSS}$
with $j\in\{1,2,\ldots,n+1\}$
is of the form that
$\tilde{b}_{i,j}\in\{-\frac{1}{2}, \frac{1}{2}\}$ for \emph{all} $i \in\{1,2,\ldots, n\}$,
and $\tilde{b}_{n+1, j}=0$.
If no such column vector appears,
The desired and identified column vector $(b_{1,j}, b_{2,j}, \ldots, b_{n,j})^T$ is
added by $\frac{1}{2}$, which becomes a binary column vector, then we check whether the binary column vector
is a feasible binary solution to Problem~\eqref{eq:prob_subset_sum}.
If no such targeted column vector appears,
the procedure can be applied to the Complementary Problem~\eqref{eq:complem_subset_sum}
as well, with $b$ replaced by $\tilde{b}:=\sum_{i=1}^{n}a_i - b$.
}
\item {Let $N_1>N_{01}$
and $N_2>2^{n+m}N^2_1 + N_{02}$, where $N_{01}$ and $N_{02}$ are big enough finite positive integers and the existence of them can be guaranteed theoretically.
Let $\tilde{B}_{AHL}$ denote the matrix whose columns consist the LLL-reduced basis of
the lattice generated by columns of $B_{AHL}$.
\textbf{AHL-Alg} checked whether the $(n-m+1)$th column of $\tilde{B}_{AHL}$
is of the form that $|\tilde{b}_{n+1, n-m+1}| = N_1$ and $\tilde{b}_{i, n-m+1} = 0$
for \emph{all} $i\in\{n+2, n+3, \ldots,n+m+1\}$.
If so, then $(\tilde{b}_{1, n-m+1}, \tilde{b}_{2, n-m+1}, \ldots, \tilde{b}_{n, n-m+1})^T$}
must be an integer solution to the problem defined in Equation~\eqref{eq:1_prob_main}, but may out of the bounded range
$[0, u_i]$, $i=1,2,\ldots,n$ of unknowns. Then branch-and-bound methods are used to enumerate the feasible integer solution.
\end{itemize}

\section{Building Blocks of Our Solution Framework}\label{sec:main_technique_part}

\subsection{The lattice formulation}\label{subsec:improvements_of_lattice}

As we know, subset-sum problems are special cases of systems of linear Diophantine equations (LDEs),
thus any lattice formulation for systems of LDEs can be readily applied
to subset-sum problems.
In this section, we propose our lattice formulation for systems of LDEs, $Ax=b$,
in which the lattice is generated by
column vectors of matrix $B$ defined as follows,
\begin{align}\label{eq:4_modification_1}
B=\left(\begin{array}{c}
I^{n\times n}\\
A^{m\times n}N
\end{array}\right),
\end{align}
where $I^{n\times n}$ denotes identity matrix of dimension ${n\times n}$,
$A^{m\times n}$ denotes the coefficient matrix with dimension ${m\times n}$
in the LDEs problem $A\emph{\textbf{x}} = \emph{\textbf{b}}$ (defined in Equation~\eqref{eq:1_prob_main}),
and $N$ denotes a large enough and polynomially finite positive integer.
In the proof of Theorem~\ref{thm:matrix B_reduced}, the usage of $N$ would be clear.

\qquad
As a note, in order to make the notations in this section clearer, sometimes, the dimension of
a matrix is added as a superscript. For example, $D^{n\times (n-m)}$
denotes a matrix of dimension $n\times (n-m)$. Whether the matrix is a real number matrix
or integer number matrix will be clarified or can be identified based on the context.

\qquad
Let $\tilde{B}$ denote the column-wise LLL-reduced matrix of $B$,
and let $\tilde{b}_{i,j}$ denote the $(i,j)$th entry of $\tilde{B}$.
We next explore and discuss properties of
the column-wise LLL-reduced matrix
$\tilde{B}$.
The properties are summarized in the following Theorem~\ref{thm:matrix B_reduced}.

\begin{thm}\label{thm:matrix B_reduced}
There exists polynomially finite $N_0\in \Z_+$
to guarantee that if $N > N_0$ then
the column-wise LLL-reduced
matrix $\tilde{B}$ of $B$ defined in \eqref{eq:4_modification_1} has the following form,
\begin{align}\label{eq:5_modification_1_reduced}
\tilde{B} =
\left(
\begin{array}{cc}
D^{n\times (n-m)}   &  C^{n\times m }\\
\textbf{0}^{ m\times (n-m) }   & E^{ m\times m }N
\end{array}\right),
\end{align}
i.e.,
$\tilde{b}_{i,j}=0$, $\forall i,~j$ with $n+1\leq i \leq n+m$ and $1\leq j\leq n-m$.

Moreover, when $\tilde{b}_{i,j}=0$, $\forall i, ~j$ with $n+1\leq i \leq n+m$ and $1\leq j\leq n-m$,
then the following properties regarding matrices $D$, $C$, and $E$ can be derived.
\vspace{-3mm}
\begin{enumerate}
\item [(a)]
{Matrix $D$ in \eqref{eq:5_modification_1_reduced} satisfies that,
$$
{\rm ker}_{\Z}(A) =  \mathcal{L}(D),
$$
where $\mathcal{L}(D) := \{D\textbf{z}~|~\textbf{z}\in\Z^{n-m}\}$
denotes the lattice generated by columns of $D$, and
${\rm ker}_{\Z}(A):=\{\textbf{x}\in \mathbb{Z}^n~|~A\textbf{x}=\textbf{0}\}$ denotes
the kernel lattice of $A\textbf{x}=\textbf{b}$. 
}
\item [(b)]
{$E^{-1}$ exists, if and only if $A$ is of full row rank.
}
\item [(c)]
{There exists an integer solution to $A\textbf{x}=\textbf{b}$ defined in \eqref{eq:1_prob_main} if and only if $E^{-1}\textbf{b}\in\mathbb{Z}^m$ that is $E^{-1}\textbf{b}$ is an integer column vector.
}
\item [(d)]
{If $E^{-1}\textbf{b}\in\mathbb{Z}^m$, then $\textbf{x}_\textbf{b} := CE^{-1}\textbf{b}$ is a special integer solution to
$A\textbf{x}=\textbf{b}$. Note that, the notation $\textbf{x}_\textbf{b}$ means that the special solution
depends on $\textbf{b}$.
}
\end{enumerate}
\end{thm}
\proof
We first prove that $\tilde{b}_{i,j}=0$, $\forall i, ~j$ with $n+1\leq i \leq n+m$ and $1\leq j\leq n-m$.
Suppose that $AU=(H~|~0)$, where $U$ is a unimodular matrix and $H$ is the Hermite normal form of $A$, (see \cite{Murray:2011} for unimodular transformation and the concept of Hermite normal form).
Then the last $n-m$ columns of $U$ form a basis of ${\rm ker}_{\Z}(A)$.
Let us denote the last $n-m$
columns of $U$ as $\emph{\textbf{x}}_0^1, \emph{\textbf{x}}_0^2, \ldots, \emph{\textbf{x}}_0^{n-m}$. Then,
\begin{align*}
\emph{\textbf{v}}_j:=\left(\begin{array}{c}
\emph{\textbf{x}}_0^j\\
\textbf{0}^{m\times 1}
\end{array}\right)
 = B\emph{\textbf{x}}_0^j\in \mathcal{L}(B), \quad \forall j\text{~~with~~} 1\leq j \leq n-m,
\end{align*}
and $\emph{\textbf{v}}_1$, $\emph{\textbf{v}}_2$, $\ldots$, $\emph{\textbf{v}}_{n-m}$ are linearly independent.
According to Lemma~\ref{lem:6_LLL_3},
we have that
\begin{align*}
||\tilde{B}_j||^2 \leq 2^{(n-1)}\max \{||\emph{\textbf{v}}_1||^2, ||\emph{\textbf{v}}_2||^2, \ldots, ||\emph{\textbf{v}}_{n-m}||^2\}, \quad \forall j\text{~~with~~} 1\leq j \leq n-m,
\end{align*}
where $\tilde{B}_j$ denotes the $j$th column of $\tilde{B}$.

We choose $N_0$ which satisfies that $N_0^2 > 2^{(n-1)}\max \{||\emph{\textbf{v}}_1||^2, ||\emph{\textbf{v}}_2||^2, \ldots, ||\emph{\textbf{v}}_{n-m}||^2\}$,
then when $N > N_0$,
we must have that $\tilde{b}_{i,j}=0$
$\forall i,~j$ with $n+1\leq i \leq n+m$ and $1\leq j\leq n-m$.
Otherwise, there exist indices $i$ and $j$ with $n+1\leq i \leq n+m$ and $1\leq j\leq n-m$
such that $\tilde{b}_{i,j}\neq 0$ and it must be a non-zero multiple of $N$,
then $||\tilde{B}_j||^2\geq |\tilde{b}_{i,j}|^2\geq N^2 > N_0^2$, which is a contradiction.

Next, given that
$\tilde{b}_{i,j}=0$ $\forall i, ~j$ with $n+1\leq i \leq n+m$ and $1\leq j\leq n-m$,
we prove items $(a)$, $(b)$, $(c)$, and $(d)$.
\vspace{-2mm}
\begin{itemize}
  \item [(a)]
For unimodular matrix $U = [D~|~C] \in \Z^{n\times n}$, we have that
$$
\tilde{B}=BU,
$$
since the LLL algorithm consists of a sequence of unimodular vector operations.
Next we prove item (a) in two directions.\\
(i) We prove that $\mathcal{L}(D)\subseteq {\rm ker}_{\Z}(A)$. Since $AD=0$, the conclusion is readily obtained.\\
(ii) We prove that $ {\rm ker}_{\Z}(A)\subseteq \mathcal{L}(D)$. For any $\emph{\textbf{x}}\in {\rm ker}_{\Z}(A)$, let $\emph{\textbf{y}} = U^{-1}\emph{\textbf{x}}$.
Then
\begin{align*}
\emph{\textbf{0}}^{m\times 1} = A\emph{\textbf{x}} = AUU^{-1}\emph{\textbf{x}} = AU\emph{\textbf{y}} = (\emph{\textbf{0}}^{m\times (n-m)}~|~E^{m\times m})\emph{\textbf{y}}.
\end{align*}
Hence we have that $E(y_{n-m+1}, y_{n-m+2},\ldots, y_{n})^T = \emph{\textbf{0}}^{m\times 1}$.
Since $A$ is of full row rank, and
$\tilde{b}_{i,j}=0$ for $n+1\leq i \leq n+m$ and $1\leq j\leq n-m$,
$E$ must be a nonsingular matrix.
Hence the following holds true,
$$
(y_{n-m+1},y_{n-m+2},\ldots,y_n)^T = \emph{\textbf{0}}^{m\times 1}.
$$
Therefore $\emph{\textbf{x}} = U\emph{\textbf{y}} = D(y_1,\ldots,y_{n-m})^T$, which implies that $\emph{\textbf{x}}\in \mathcal{L}(D)$.
\item [(b)]
{
$A$ is of full row rank if and only if the sub-matrix consisting of
the last $m$ rows of $B$ is of full row rank.
The sub-matrix consisting of
the last $m$ rows of $B$ is of full row rank, is equivalent to that the sub-matrix consisting of
the last $m$ rows of $\tilde{B}$ is of full row rank.
While the sub-matrix consisting of
the last $m$ rows of $\tilde{B}$ is of full row rank, is equivalent to that $E$ is of full row rank.
As $E$ is a square matrix of dimension $n\times {n}$.
This is equivalent to that $E^{-1}$ exists.
}
\item [(c)]
{
We prove this item in two directions.\\
i) If $E^{-1}\emph{\textbf{b}}\in\mathbb{Z}^m$ then $CE^{-1}\emph{\textbf{b}}\in\mathbb{Z}^m$ is an integral solution
to
problem~\eqref{eq:1_prob_main}, since $ACE^{-1}\emph{\textbf{b}} = EE^{-1}\emph{\textbf{b}} = \emph{\textbf{b}}$.\\
ii) We prove that $E^{-1}\emph{\textbf{b}}$ must be an integer vector, if there exists $\emph{\textbf{x}}^*\in\Z^n$ such that $A\emph{\textbf{x}}^* = \emph{\textbf{b}}$.
Since
$A\left(D^{n\times (n-m)}~|~C^{n\times m}\right)=
\left(\emph{\textbf{0}}^{m\times (n-m)}~|~E^{m\times m}\right)$,
and
$\left(D^{n\times (n-m)}~|~C^{n\times m}\right)$ is a unimodular matrix,
then based on Theorem~\ref{thm:lattice_equivalence},
$\mathcal{L}(A)=\mathcal{L}(E)$ is deduced.
Meanwhile, since $\emph{\textbf{b}}\in \mathcal{L}(A)$, there must exist
$\emph{\textbf{y}}\in\Z^m$ such that $E\emph{\textbf{y}} = \emph{\textbf{b}}$.
Therefore $E^{-1}\emph{\textbf{b}} = \emph{\textbf{y}}\in\Z^m$.
}
\item [(d)]
{
This has already been proved in item $(c)$.
}
\endproof
\end{itemize}

\qquad
Theorem~\ref{thm:matrix B_reduced} is selected from Theorem 3.1.1 of the Ph.D. thesis \cite{Bojun:2014}.
As a note, in Theorem~\ref{thm:matrix B_reduced}, when $m=1$, $A$ becomes a row vector with $A=(a_1,a_2,\ldots,a_n)$,
and matrix $E$ becomes a scaler. Moreover, $E$ must be the extended
greatest common divisor (GCD) of $a_1$, $a_2$, $\ldots$, $a_n$.
The proof can be found in \cite{HMM:1998}.

\qquad
In the next subsection, we will propose two algorithms, $\textbf{Reduce}(\emph{\textbf{x}}_\emph{\textbf{b}}, D)$
and $\textbf{Reduce}_{1/2}(\emph{\textbf{x}}_\emph{\textbf{b}}, D)$, which are based on our lattice formulation
in this part and the results proved in
Theorem~\ref{thm:matrix B_reduced}.
As a fact, our algorithm $\textbf{Reduce}(\emph{\textbf{x}}_\emph{\textbf{b}}, D)$ can achieve the same result
compared with the lattice formulation and column reduction procedure in \textbf{AHL-Alg}.
But in $\textbf{Reduce}(\emph{\textbf{x}}_\emph{\textbf{b}}, D)$ only one big integer number is involved;
while two big integers $N_1$ and $N_2$,
with $N_1>N_{01}$
and $N_2>2^{n+m}N^2_1 + N_{02}$,
must be involved in \textbf{AHL-Alg}.
Thus $\textbf{Reduce}(\emph{\textbf{x}}_\emph{\textbf{b}}, D)$ is more concise
compared with \textbf{AHL-Alg}.

\qquad
The algorithm $\textbf{Reduce}_{1/2}(\emph{\textbf{x}}_\emph{\textbf{b}}, D)$ is an improved version based on
$\textbf{Reduce}(\emph{\textbf{x}}_\emph{\textbf{b}}, D)$, which will be explained in detail in the next subsection.
$\textbf{Reduce}_{1/2}(\emph{\textbf{x}}_\emph{\textbf{b}}, D)$ can return binary solutions with significantly higher success ratio.

\subsubsection{Algorithms $\textbf{Reduce}(\emph{\textbf{x}}_\emph{\textbf{b}}, D)$ and $\textbf{Reduce}_{1/2}(\emph{\textbf{x}}_\emph{\textbf{b}}, D)$}
In this section we pursue beyond Theorem~\ref{thm:matrix B_reduced},
and propose two algorithms which are denoted as
$\textbf{Reduce}(\emph{\textbf{x}}_\emph{\textbf{b}}, D)$ and $\textbf{Reduce}_{1/2}(\emph{\textbf{x}}_\emph{\textbf{b}}, D)$,
respectively.
Both of these two algorithms enable us to get short, size-reduced integer solutions to
systems of LDEs, $A\emph{\textbf{x}} = \emph{\textbf{b}}$, defined in~\eqref{eq:1_prob_main}.
Recall that in this paper we use Euclidean norm of
vector to measure the length of a solution vector, which is consistent with the norm used
in Lemma~\ref{lem:6_LLL_3}.

\qquad
The most ideal output of these two algorithms would be a feasible binary solution to the hard subset-sum problems
proposed in Equation~\eqref{eq:prob_subset_sum}, when $m=1$.


\begin{algorithm}[!h]
\caption{${\rm sol}(\emph{\textbf{x}}_\emph{\textbf{b}}, D)$ = \textbf{Reduce}($\emph{\textbf{x}}_\emph{\textbf{b}}$, $D$)}
\label{alg:1_reduce(x,D)}
{\small
\textbf{input} A special solution $\emph{\textbf{x}}_\emph{\textbf{b}}\in \Z^n$ to $A\emph{\textbf{x}} = \emph{\textbf{b}}$, and an ordered basis $D\in\Z^{n\times(n-m)}$ of ${\rm ker}_{\Z}(A)$.\\
\textbf{output} A reduced short integer solution, denoted as ${\rm sol}(\emph{\textbf{x}}_\emph{\textbf{b}}, D)$,
to $A\emph{\textbf{x}} = \emph{\textbf{b}}$. \\
\vspace{-4mm}
\begin{algorithmic}[1]
\STATE $G$ $\leftarrow$ $(D~|~\emph{\textbf{x}}_\emph{\textbf{b}})^T$, conduct GSO process on rows of $G$
to decompose $G$ as $G = MG^*$, where $M = (\mu_{i,j})\in\Q^{(n-m+1)\times(n-m+1)}$ is a lower
triangular matrix and the rows of $G^*\in\Q^{(n-m+1)\times n}$ are orthogonal to each other.
\FOR {$j=n-m$ to $1$}
    \STATE $\lambda_j$ $\leftarrow$ $\lceil \mu_{n-m+1,j}\rfloor$
    \STATE $G_{n-m+1}$ $\leftarrow$ $G_{n-m+1} - \lambda_j G_{j}$
    \STATE $M_{n-m+1}$ $\leftarrow$ $M_{n-m+1} - \lambda_j M_{j}$
\ENDFOR
\STATE ${\rm sol}(\emph{\textbf{x}}_\emph{\textbf{b}}, D)$ $\leftarrow$ $G_{n-m+1}^T$
\RETURN ${\rm sol}(\emph{\textbf{x}}_\emph{\textbf{b}}, D)$
\end{algorithmic}
}
{\footnotesize Note: $M_{j}$ denotes the $j$th row of $M$, and $G_{j}$ denote the $j$th row of $G$.
$\mu_{i,j}$ denotes the $(i,j)$th entry of $M$.}
\end{algorithm}

\qquad
In Algorithm~\ref{alg:1_reduce(x,D)}, $\textbf{Reduce}(\emph{\textbf{x}}_\emph{\textbf{b}}, D)$, the notation ${\rm sol}(\emph{\textbf{x}}_\emph{\textbf{b}}, D)$ means that
the reduced integer solution depends on the input
$\emph{\textbf{x}}_\emph{\textbf{b}}$ and $D$.
For example, based on Theorem~\ref{thm:matrix B_reduced}, if $CE^{-1}\emph{\textbf{b}}\in\Z^n$, we can let
$\emph{\textbf{x}}_\emph{\textbf{b}} = CE^{-1}\emph{\textbf{b}}$.
In our latter numerical implementation,
we choose the LLL-reduced basis $D$ obtained
in Equation~\eqref{eq:5_modification_1_reduced} as part of the input
for $\textbf{Reduce}(\emph{\textbf{x}}_\emph{\textbf{b}}, D)$,
since the LLL-reduced basis has
desirable properties as we claimed and reviewed in Section 2.

\qquad
In fact,
the key purpose of $\textbf{Reduce}(\emph{\textbf{x}}_\emph{\textbf{b}}, D)$ is to reduce
any special integer solution $\emph{\textbf{x}}_\emph{\textbf{b}}$ to $A\emph{\textbf{x}} = \emph{\textbf{b}}$ by a basis $D$ of
its integer kernel space ${\rm ker}_{\Z}(A)$. Mathematically, there exist integer scalars
$\lambda_1$, $\lambda_2$, $\cdots$, $\lambda_{n-m}$, such that,
$$
{\rm sol}(\emph{\textbf{x}}_\emph{\textbf{b}}, D) = \emph{\textbf{x}}_\emph{\textbf{b}} - D(\lambda_1, \lambda_2,\ldots,\lambda_{n-m})^T.
$$

\qquad
As a note, we could also input other improved reduced basis of a lattice compared with the LLL reduced basis,
for instances the BKZ reduction algorithm (see \cite{aono2016improved}, \cite{schnorr2003lattice}, \cite{chen2011bkz}, and
\cite{schneider2010extended}),
as the input of $\textbf{Reduce}(\emph{\textbf{x}}_\emph{\textbf{b}}, D)$.
The numerical performances of $\textbf{Reduce}(\emph{\textbf{x}}_\emph{\textbf{b}}, D)$ should can be improved accordingly, in the sense that returning binary solution to hard subset-sum problems with higher success ratio.
Next we propose an improved variation, denoted as $\textbf{Reduce}_{1/2}(\emph{\textbf{x}}_\emph{\textbf{b}}, D)$,
based on $\textbf{Reduce}(\emph{\textbf{x}}_\emph{\textbf{b}}, D)$.
The variation algorithm is presented in
the following Algorithm~\ref{alg:2_reduce2(x,D)}.


\begin{algorithm}[h!]
\caption{${\rm sol}_{1/2}(\emph{\textbf{x}}_\emph{\textbf{b}}, D)$ = $\textbf{Reduce}_{1/2}(\emph{\textbf{x}}_\emph{\textbf{b}}, D)$}
\label{alg:2_reduce2(x,D)}
{\small
\textbf{input} A special solution $\emph{\textbf{x}}_\emph{\textbf{b}}\in \Z^n$ to $A\emph{\textbf{x}} = \emph{\textbf{b}}$, and an ordered basis $D\in\Z^{n\times(n-m)}$ of ${\rm ker}_{\Z}(A)$.\\
\textbf{output} A reduced short integer solution, denoted as ${\rm sol}_{1/2}(\emph{\textbf{x}}_\emph{\textbf{b}}, D)$,
to $A\emph{\textbf{x}} = \emph{\textbf{b}}$. \\
\vspace{-4mm}
\begin{algorithmic}[1]
\STATE $G$ $\leftarrow$ $(2D~|~2\emph{\textbf{x}}_\emph{\textbf{b}} - \textbf{1}^{n\times 1})^T$, conduct GSO process on rows of $G$
to decompose $G$ as $G = MG^*$, where $M = (\mu_{i,j})\in\Q^{(n-m+1)\times(n-m+1)}$ is a lower
triangular matrix and the rows of $G^*\in\Q^{(n-m+1)\times n}$ are orthogonal to each other.
\FOR {$j=n-m$ to $1$}
    \STATE $\lambda_j$ $\leftarrow$ $\lceil \mu_{n-m+1,j}\rfloor$
    \STATE $G_{n-m+1}$ $\leftarrow$ $G_{n-m+1} - \lambda_j G_{j}$
    \STATE $M_{n-m+1}$ $\leftarrow$ $M_{n-m+1} - \lambda_j M_{j}$
\ENDFOR
\STATE ${\rm sol}_{1/2}(\emph{\textbf{x}}_\emph{\textbf{b}}, D)$ $\leftarrow$ $\frac{G_{n-m+1}^T + \textbf{1}^{n\times 1} }{2}$
\RETURN ${\rm sol}_{1/2}(\emph{\textbf{x}}_\emph{\textbf{b}}, D)$
\end{algorithmic}
}
{\footnotesize Note: $M_{j}$ and $G_{j}$ are the $j$'th row of $M$ and $G$, respectively.
$\mu_{i,j}$ is the $(i,j)$th entry of $M$.}
\end{algorithm}

\qquad
In fact,
the key purpose of $\textbf{Reduce}_{1/2}(\emph{\textbf{x}}_\emph{\textbf{b}}, D)$ is also to reduce
any special integer solution $\emph{\textbf{x}}_\emph{\textbf{b}}$ to $A\emph{\textbf{x}} = \emph{\textbf{b}}$ by a basis $D$ of
its integer kernel space ${\rm ker}_{\Z}(A)$. Another series of
integer scalars $\lambda_1$, $\lambda_2$, $\cdots$, $\lambda_{n-m}$ are generated
by $\textbf{Reduce}_{1/2}(\emph{\textbf{x}}_\emph{\textbf{b}}, D)$ to yield that,
\begin{align*}
{\rm sol}_{1/2}(\emph{\textbf{x}}_\emph{\textbf{b}}, D) = \emph{\textbf{x}}_\emph{\textbf{b}} - D(\lambda_1, \lambda_2,\ldots,\lambda_{n-m})^T.
\end{align*}

\qquad
Next we propose Theorem~\ref{thm:1_algorithm_reduce}
and
Theorem~\ref{thm:2_algorithm_reduce}, to
explore how $\text{sol}(\emph{\textbf{x}}_\emph{\textbf{b}}, D)$
and $\text{sol}_{1/2}(\emph{\textbf{x}}_\emph{\textbf{b}}, D)$
respond to
the changes in the input vector $\emph{\textbf{x}}_\emph{\textbf{b}}$ and matrix $D$.
Specifically,
Theorem~\ref{thm:1_algorithm_reduce} shows that
$\text{sol}(\emph{\textbf{x}}_\emph{\textbf{b}}, D)$ and $\text{sol}_{1/2}(\emph{\textbf{x}}_\emph{\textbf{b}}, D)$
will not change
accordingly when another special integer solution $\emph{\textbf{x}}_\emph{\textbf{b}}$ is used as the input vector.
Meanwhile,
Theorem~\ref{thm:2_algorithm_reduce} shows that
$\text{sol}(\emph{\textbf{x}}_\emph{\textbf{b}}, D)$ and
$\text{sol}_{1/2}(\emph{\textbf{x}}_\emph{\textbf{b}}, D)$
will also not change
accordingly when only directions of the basis vectors in $D$ change.

\vspace{3mm}

\begin{thm}[Theorem 3.1.2 of \cite{Bojun:2014}]\label{thm:1_algorithm_reduce}
Given a basis $D\in\Z^{n\times(n-m)}$ of ${\rm ker}_{\Z}(A)$, then
$$
{\rm sol}(\tilde{\textbf{y}},D) = {\rm sol}(\textbf{y},D),~\text{and}~{\rm sol}_{1/2}(\tilde{\textbf{y}},D) = {\rm sol}_{1/2}(\textbf{y},D),
$$
for any $\tilde{\textbf{y}}$, $\textbf{y}$ $\in\{\textbf{x}\in\Z^n~|~A\textbf{x}=\textbf{b}\}$.
\end{thm}

\proof Based on Algorithm~\ref{alg:1_reduce(x,D)},
${\rm sol}(\emph{\textbf{y}}, D)$ can be expressed as,
$${\rm sol}(\emph{\textbf{y}}, D) = \emph{\textbf{y}} - \lambda_{n-m}D_{n-m} - \cdots - \lambda_1 D_1,$$
and ${\rm sol}(\tilde{\emph{\textbf{y}}}, D)$ can be expressed as,
$${\rm sol}(\tilde{\emph{\textbf{y}}}, D) = \tilde{\emph{\textbf{y}}} - \tilde{\lambda}_{n-m}D_{n-m} - \cdots - \tilde{\lambda}_1 D_1,$$
where $D_{i}$ denotes the $i$th column of $D$ and $\lambda_{i}$, $\tilde{\lambda}_{i}\in\Z$, $1\leq i\leq n-m$.
Meanwhile there must exist integer numbers $z_i$, $i=1,2,\ldots,n-m$, such that $\tilde{\emph{\textbf{y}}}
= \emph{\textbf{y}} + \sum_{i=1}^{n-m}z_iD_i$, since $\emph{\textbf{y}}$ and $\tilde{\emph{\textbf{y}}}$ are in the same solution space.
Later, by showing that $\tilde{\lambda}_i =
\lambda_i + z_i$, $i=1,2,\ldots,n-m$, we prove ${\rm sol}(\tilde{y},D) = {\rm sol}(y,D)$.

\qquad
In the process of computing ${\rm sol}(\emph{\textbf{y}},D)$ and ${\rm sol}(\tilde{\emph{\textbf{y}}},D)$, we use notation
$\mu_{i,j}$ and $\tilde{\mu}_{i,j}$ respectively in the GSO process in Step~1
of Algorithm~\ref{alg:1_reduce(x,D)}.
It is easy to find that $\tilde{\mu}_{i,j} = \mu_{i,j}$, $1\leq j<i\leq n-m$. Next we analyze the relation between $\tilde{\mu}_{n-m+1,j}$ and $\mu_{n-m+1,j}$, $j=1,2,\ldots,n-m$.
Initially, we have,
\begin{align*}
\tilde{\mu}_{n-m+1,j}& = \frac{\tilde{\emph{\textbf{y}}}\cdot D_j^*}{D_j^*\cdot D_j^*}
=\frac{ (\emph{\textbf{y}} + \sum_{i=1}^{n-m}z_iD_i) \cdot D_j^*}{D_j^*\cdot D_j^*}
=\frac{\emph{\textbf{y}}\cdot D_j^*}{D_j^*\cdot D_j^*}+\sum_{i=j}^{n-m}z_i\frac{D_i\cdot D_j^*}{D_j^*\cdot D_j^*}\\
&= \mu_{n-m+1,j} + z_j + \sum_{i=j+1}^{n-m}z_i\mu_{i,j}, \quad j=1,2,\ldots,n-m.
\end{align*}
Thus, we have,
$$
\tilde{\lambda}_{n-m} = \lceil \tilde{\mu}_{n-m+1,n-m} \rfloor
= \lceil \mu_{n-m+1,n-m} + z_{n-m} \rfloor =
\lceil \mu_{n-m+1,n-m}\rfloor + z_{n-m}  = \lambda_{n-m} + z_{n-m}.
$$

\qquad
After subtracting $\lambda_{n-m}D_{n-m}$ from $\emph{\textbf{y}}$
and subtracting $\tilde{\lambda}_{n-m}D_{n-m}$ from $\tilde{\emph{\textbf{y}}}$,
we could update the values of
$\mu_{n-m+1,j}$ and $\tilde{\mu}_{n-m+1,j}$, $j=n-m, n-m-1,\ldots,1$ as follows,
\begin{align*}
&\mu_{n-m+1,n-m} \leftarrow \mu_{n-m+1,n-m} - \lambda_{n-m},\\
&\mu_{n-m+1,j} \leftarrow \mu_{n-m+1,j} - \lambda_{n-m}\mu_{n-m,j},\quad
j=n-m-1,\ldots,1,
\end{align*}
and
\begin{align*}
&\tilde{\mu}_{n-m+1,n-m} \leftarrow \tilde{\mu}_{n-m+1,n-m} - \tilde{\lambda}_{n-m} = \mu_{n-m+1,n-m} - \lambda_{n-m},\\
&\tilde{\mu}_{n-m+1,j} \leftarrow
\tilde{\mu}_{n-m+1,j} - \tilde{\lambda}_{n-m}\tilde{\mu}_{n-m,j}\\
=& \mu_{n-m+1,j} + z_j + \sum_{i=j+1}^{n-m-1}z_i\mu_{i,j} - \lambda_{n-m}\mu_{n-m,j},\quad
j=n-m-1,\ldots,1.
\end{align*}
Note that, $\mu_{n-m+1,n-m+1}$ and $\tilde{\mu}_{n-m+1,n-m+1}$ always equal 1.

\qquad
Therefore after subtracting $\lambda_{n-m}D_{n-m}$ from $\emph{\textbf{y}}$
and subtracting $\tilde{\lambda}_{n-m}D_{n-m}$ from $\tilde{\emph{\textbf{y}}}$, we have,
\begin{align*}
&\tilde{\mu}_{n-m+1,n-m} = \mu_{n-m+1,n-m},\\
&\tilde{\mu}_{n-m+1,j} = \mu_{n-m+1,j} + z_j + \sum_{i=j+1}^{n-m-1}z_i\mu_{i,j},
\quad j=n-m-1,\ldots,1.
\end{align*}
Thus, we have,
\begin{align*}
\tilde{\lambda}_{n-m-1}
&= \lceil \tilde{\mu}_{n-m+1,n-m-1} \rfloor
= \lceil \mu_{n-m+1,n-m-1} + z_{n-m-1} \rfloor\\
&=\lceil \mu_{n-m+1,n-m-1}\rfloor + z_{n-m-1}  = \lambda_{n-m-1} + z_{n-m-1}.
\end{align*}

\qquad
Recursively, we subtract $\lambda_{n-m-1}D_{n-m-1}$ from $\emph{\textbf{y}}$ and subtract $\tilde{\lambda}_{n-m-1}D_{n-m-1}$ from $\tilde{\emph{\textbf{y}}}$ and so on,
until the last step within which we
subtract $\lambda_{1}D_{1}$ from $\emph{\textbf{y}}$ and subtract $\tilde{\lambda}_{1}D_{1}$ from $\tilde{\emph{\textbf{y}}}$.
Conduct recursively similar analyses as
that above for deriving
$\tilde{\lambda}_{n-m-1} = \lambda_{n-m-1} + z_{n-m-1}$,
we get
$\tilde{\lambda}_{i} = \lambda_i + z_i$, $i=n-m-2,\ldots,1$.
Thus, it is proved that
${\rm sol}(\tilde{\emph{\textbf{y}}}, D) = {\rm sol}(\emph{\textbf{y}},D)$.
\endproof

 \begin{thm}[Theorem 3.1.3 of \cite{Bojun:2014}]\label{thm:2_algorithm_reduce}
Let $D$ and $\tilde{D}$ be two bases of ${\rm ker}_{\Z}(A)$, with $\tilde{D}_{i} = D_i$ or $-D_i$, and ${\rm sign}(i) = 1$ if $\tilde{D}_{i} = D_i$,
${\rm sign}(i) = -1$ if $\tilde{D}_{i} = -D_i$, $i=1,2,\ldots,n-m$.
Assume that in the reduction process of Algorithm~\ref{alg:1_reduce(x,D)},
we get parameters
$\tilde{\mu}_{i,j}$, $\tilde{D}_i^*$ and $\tilde{\textbf{x}}^*$ for input $\tilde{D}$ and $\textbf{x}$,
and parameters $\mu_{i,j}$, $D_i^*$
and $\textbf{x}^*$ for input $D$ and $\textbf{x}$.
Then we have the followings,
 \vspace{-4mm}
 \begin{enumerate}
 \item [(a)]In every recursive step, $\tilde{\mu}_{i,j} = {\rm sign}(i)\times {\rm sign}(j) \times \mu_{i,j}$, $1\leq j<i\leq n-m+1$.
 \item [(b)]$\tilde{D}_{i}^* = {\rm sign}(i) \times D_i^*$, $i=1,2,\ldots,n-m$, and $\tilde{\textbf{x}}^* = \textbf{x}^*$.
 \item [(c)]${\rm sol}(\textbf{x}, \tilde{D}) = {\rm sol}(\textbf{x}, D)$ and
            ${\rm sol}_{1/2}(\textbf{x}, \tilde{D}) = {\rm sol}_{1/2}(\textbf{x}, D)$.
 \end{enumerate}
 \end{thm}

\proof
 $a),~b)$ Without loss of generality and for simplicity,
 we can assume that there is only one vector in $\tilde{D}$
 with different sign from vectors in $D$. For example, $\tilde{D}_h=-D_h$.
 If this case can be proved, then the general case can be readily proved as well.

 For $1\leq j<i\leq h-1$, we have $\tilde{\mu}_{i,j}=\mu_{i,j}$ and $\tilde{D}_i^*=D_i^*$,
 since $\tilde{D}_i = D_i$, $i=1,2,\ldots,h-1$.

 For $i=h$, we have,
 \begin{align*}
 \tilde{\mu}_{h,j}= \frac{\tilde{D}_h\cdot \tilde{D}_j^*}{\tilde{D}_j^*\cdot \tilde{D}_j^*}
 =\frac{-D_h\cdot D_j^*}{D_j^*\cdot D_j^*} = -\mu_{h,j}, \quad j=1,2,\ldots, h-1,
 \end{align*}
 and
 \begin{align*}
 \tilde{D}_h^* &= \tilde{D_h} - \tilde{\mu}_{h,1}\tilde{D}_1^* - \tilde{\mu}_{h,2}\tilde{D}_2^* - \cdots -\tilde{\mu}_{h,h-1}\tilde{D}_{h-1}^*\\
 &= -D_h - \tilde{\mu}_{h,1}D_1^* - \tilde{\mu}_{h,2}D_2^* - \cdots -\tilde{\mu}_{h,h-1}D_{h-1}^*\\
 &= -D_h + \mu_{h,1}D_1^* + \mu_{h,2}D_2^* + \cdots +\mu_{h,h-1}D_{h-1}^*\\
 &= -D_h^*.
 \end{align*}

 For $i = h+1$, we have,
 \begin{align*}
 &\tilde{\mu}_{h+1,j} = \frac{\tilde{D}_{h+1}\cdot \tilde{D}_j^*}{\tilde{D}_j^*\cdot \tilde{D}_j^*}
 =\frac{D_{h+1}\cdot D_j^*}{D_j^*\cdot D_j^*}
 =\mu_{h+1,j},
 \quad j=1,2,\ldots,h-1,\\
 &\tilde{\mu}_{h+1,h} = \frac{\tilde{D}_{h+1}\cdot \tilde{D}_h^*}{\tilde{D}_h^*\cdot \tilde{D}_h^*}
 = \frac{D_{h+1}\cdot (-D_h^*)}{(-D_h^*)\cdot (-D_h^*)} = -\mu_{h+1,h},
 \end{align*}
 and
 \begin{align*}
 \tilde{D}_{h+1}^* &= \tilde{D}_{h+1} - \tilde{\mu}_{h+1,1}\tilde{D}_1^* - \tilde{\mu}_{h+1,2}\tilde{D}_2^* - \cdots -\tilde{\mu}_{h+1,h-1}\tilde{D}_{h-1}^*
 -\tilde{\mu}_{h+1,h}\tilde{D}_{h}^*\\
 &=D_{h+1} - \mu_{h+1,1}D_1^* - \mu_{h+1,2}D_2^* - \cdots -\mu_{h+1,h-1}D_{h-1}^*
 - (-\mu_{h+1,h})(-D_{h}^*) \\
 &=D_{h+1}^*.
 \end{align*}

 For $i=h+2,h+3,\ldots,n-m$, we have,
 \begin{align*}
 &\tilde{\mu}_{i,j} = \frac{\tilde{D}_{i}\cdot \tilde{D}_j^*}{\tilde{D}_j^*\cdot \tilde{D}_j^*}
 =\frac{D_{i}\cdot D_j^*}{D_j^*\cdot D_j^*}
 =\mu_{i,j},
 \quad j=1,2,\ldots,h-1,\\
 &\tilde{\mu}_{i,h} = \frac{\tilde{D}_{i}\cdot \tilde{D}_h^*}{\tilde{D}_h^*\cdot \tilde{D}_h^*}
 = \frac{D_{i}\cdot (-D_h^*)}{(-D_h^*)\cdot (-D_h^*)} = -\mu_{i,h},\\
 &\tilde{\mu}_{i,j} = \frac{\tilde{D}_{i}\cdot \tilde{D}_j^*}{\tilde{D}_j^*\cdot \tilde{D}_j^*}
 =\frac{D_{i}\cdot D_j^*}{D_j^*\cdot D_j^*}
 =\mu_{i,j},
 \quad j=h+1,\ldots,i-1,\\
 \end{align*}
 and
 \begin{align*}
 \tilde{D}_{i}^* &= \tilde{D}_{i} - \tilde{\mu}_{i,1}\tilde{D}_1^* - \tilde{\mu}_{i,2}\tilde{D}_2^* - \cdots -\tilde{\mu}_{i,h-1}\tilde{D}_{h-1}^*
 -\tilde{\mu}_{i,h}\tilde{D}_{h}^* - \cdots - \tilde{\mu}_{i,i-1}\tilde{D}_{i-1}^*\\
 &=D_{i} - \mu_{i,1}D_1^* - \mu_{i,2}D_2^* - \cdots -\mu_{i,h-1}D_{h-1}^*
 - (-\mu_{i,h})(-D_{h}^*) - \cdots -  \mu_{i,i-1}D_{i-1}^*\\
 &=D_{i}^*.
 \end{align*}

 For $i=n-m+1$, we have,
 \begin{align*}
 &\tilde{\mu}_{n-m+1,j} = \frac{\emph{\textbf{x}} \cdot \tilde{D}_j^*}{\tilde{D}_j^*\cdot \tilde{D}_j^*}
 =\frac{\emph{\textbf{x}} \cdot D_j^*}{D_j^*\cdot D_j^*}
 =\mu_{n-m+1,j},
 \quad j=1,2,\ldots,h-1,\\
 &\tilde{\mu}_{n-m+1,h} = \frac{\emph{\textbf{x}} \cdot \tilde{D}_h^*}{\tilde{D}_h^*\cdot \tilde{D}_h^*}
 = \frac{\emph{\textbf{x}} \cdot (-D_h^*)}{(-D_h^*)\cdot (-D_h^*)} = -\mu_{n-m+1,h},\\
 &\tilde{\mu}_{n-m+1,j} = \frac{\emph{\textbf{x}} \cdot \tilde{D}_j^*}{\tilde{D}_j^*\cdot \tilde{D}_j^*}
 =\frac{\emph{\textbf{x}} \cdot D_j^*}{D_j^*\cdot D_j^*}
 =\mu_{n-m+1,j},
 \quad j=h+1,\ldots,n-m.
 \end{align*}
Thus the special case is proved.
Each time, we just change the sign of one vector,
then the general case
in items 1) and 2) can be derived readily.

 $c)$ ${\rm sol}(\emph{\textbf{x}}, D)$ can be expressed as
 $${\rm sol}(\emph{\textbf{x}}, D) = \emph{\textbf{x}} - \lambda_{n-m}D_{n-m} - \cdots - \lambda_1 D_1,$$
 and ${\rm sol}(\emph{\textbf{x}}, \tilde{D})$ can be expressed as
 $${\rm sol}(\emph{\textbf{x}}, \tilde{D}) = \emph{\textbf{x}} - \tilde{\lambda}_{n-m}\tilde{D}_{n-m} - \cdots - \tilde{\lambda}_1 \tilde{D}_1,$$
 where $\lambda_{i}$ and $\tilde{\lambda}_{i}$ for $i=1,2,\ldots,n-m$ are integer numbers.
 We will prove that $\tilde{\lambda}_i = {\rm sign}(i)\times\lambda_i$, $i=n-m, \ldots,1$,
 thus to achieve ${\rm sol}(\emph{\textbf{x}}, \tilde{D}) = {\rm sol}(\emph{\textbf{x}}, D)$.

\qquad
First, $\lambda_{n-m} = \lceil \mu_{n-m+1,n-m}\rfloor$, and $\tilde{\lambda}_{n-m} =
\lceil \tilde{\mu}_{n-m+1,n-m}\rfloor =
\lceil {\rm sign}(n-m) \times \mu_{n-m+1,n-m}\rfloor =
{\rm sign}(n-m) \times\lceil \mu_{n-m+1,n-m}\rfloor =
{\rm sign}(n-m)\times \lambda_{n-m}$.

\qquad
Next we will show that if $\tilde{\lambda}_{i} = {\rm sign}(i)\times\lambda_{i}$,
for some $i\leq n-m$, then it holds that
$\tilde{\lambda}_{i-1} = {\rm sign}(i-1)\times\lambda_{i-1}$.
If this is the case, then based on mathematical induction, it follows
that $\tilde{\lambda}_i = {\rm sign}(i)\times \lambda_i$, $i=n-m,\ldots, 1$.
Thus ${\rm sol}(\emph{\textbf{x}}, \tilde{D}) = {\rm sol}(\emph{\textbf{x}}, D)$ can be proved.

\qquad
Given that $\tilde{\lambda}_{i} = {\rm sign}(i)\times\lambda_{i}$
for some $i\leq n-m$, in Step~4 and Step~5 of Algorithm~\ref{alg:1_reduce(x,D)},
subtracting $\lambda_i$ times $D_i$ from $\emph{\textbf{x}}$ and
$\tilde{\lambda}_i$ times $\tilde{D}_i$ from $\emph{\textbf{x}}$, respectively,
gives us the updated values for $\mu_{n-m+1,j}$ and
 $\tilde{\mu}_{n-m+1,j}$, $j=i, i-1, \ldots,1$,
 \begin{align*}
 &\mu_{n-m+1,i} \leftarrow \mu_{n-m+1,i} - \lambda_i, \\
 &\mu_{n-m+1,j} \leftarrow \mu_{n-m+1,j} - \lambda_i \mu_{i,j}, \quad j=i-1,\ldots,1,
 \end{align*}
 and
 \begin{align*}
 &\tilde{\mu}_{n-m+1,i} \leftarrow \tilde{\mu}_{n-m+1,i} - \tilde{\lambda}_i
 ={\rm sign}(i)\times (\mu_{n-m+1,i} - \lambda_i), \\
 &\tilde{\mu}_{n-m+1,j} \leftarrow \tilde{\mu}_{n-m+1,j} - \tilde{\lambda}_i\tilde{\mu}_{i,j}
 ={\rm sign}(j)\times (\mu_{n-m+1,j} - \lambda_i\mu_{i,j}), \quad j=i-1,\ldots,1.
 \end{align*}
Thus, in the next round of the calculation within the loop from
Step~2 to Step~6 of Algorithm~\ref{alg:1_reduce(x,D)},
we get that $\tilde{\lambda}_{i-1} = \lceil \tilde{\mu}_{n-m+1,i-1}\rfloor$
$=\lceil{\rm sign}(i-1)\times\mu_{n-m+1,i-1}\rfloor ={\rm sign}(i-1)\times\lceil\mu_{n-m+1,i-1}\rfloor$ $={\rm sign}(i-1)\times \lambda_{i-1}$.
\endproof

\qquad
As a note, we mention here that the tie-breaking issue may arise in the calculation of nearest integer,  $\lceil{\bullet}\rfloor$. For example, should we round $\lceil{4.5}\rfloor$ to $4$ or to $5$; and should we round
$\lceil{-4.5}\rfloor$ to $-4$ or to $-5$?
There are different tie-breaking rules in the literature (see \cite{rounding} and
Appendix A of \cite{Bojun:2014}).
Later in our numerical simulation part, we will adopt the same tie-breaking rule as that in \textbf{AHL-Alg} (see \cite{Aardal:2000}).
That is, $\lceil{\mu_{jk}}\rfloor = \lceil{\mu_{jk}-\frac{1}{2}}\rceil$, then
$\lceil{4.5}\rfloor$ would be rounded to $4$, and $\lceil{-4.5}\rfloor$ would be rounded to $-5$.

\qquad
Also as a note, in the proof for item (a) and item (b) in Theorem~\ref{thm:2_algorithm_reduce}, tie-breaking rule is adopted as that, if $\lceil{4.5}\rfloor$ is rounded to $4$
then $\lceil{-4.5}\rfloor$ is rounded to $-4$ but not $-5$;
or if $\lceil{4.5}\rfloor$ is rounded to $5$
then $\lceil{-4.5}\rfloor$ should be rounded to $-5$ but not $-4$.
More detailed discussions about the tie-breaking issues can be found in Appendix A of \cite{Bojun:2014}.

\subsection{Modular disaggregation technique}\label{subsec:disaggregation_tech}
In this section we propose modular disaggregation techniques (abbreviated as ``DAG" in this paper) for subset-sum problems defined in Equation~\eqref{eq:prob_subset_sum}.
In this text we always assume that
$b\leq \frac{1}{2}\sum_{i=1}^{n}a_i$.
Otherwise, setting $y_i=1-x_i$, $i=1,2,\ldots,n$, yields
the complementary subset-sum problem,
\begin{align*}
  \emph{\textbf{a}}\emph{\textbf{y}} := a_1y_1+a_2y_2+\cdots+a_ny_n = \sum_{i=1}^{n}a_i - b = \tilde{b},
\end{align*}
with $\tilde{b} < \frac{1}{2}\sum_{i=1}^{n}a_i$ and
$\emph{\textbf{y}}=(y_1, y_2, \ldots,y_n)\in \{0,1\}^n$.
As a note, since $\tilde{b}<b$, after transformation some $a_i$ may be larger than $\tilde{b}$
and thus the corresponding $y_i$ can be fixed at zero value.

\qquad
In our modular disaggregation techniques, firstly two positive integer parameters, $t$ and $M$,
will be introduced. Then after detailed analysis and deduction, we would see that only one
rational parameter $r:=\frac{t}{M}$ is sufficient.
Equipped with disaggregation techniques, more equations and more information can be revealed
for a given equation system.

\qquad
Let $t$ and $M$ be two positive integers with $t<M$,
for a given subset-sum problem~\eqref{eq:prob_subset_sum},
modular transformations are conducted as follows,
\begin{align}
\left\{
  \begin{array}{l}
    c_i:=ta_i\pmod{M},\qquad i=1,2,\ldots,n\vspace{2mm}\\\vspace{2mm}
    d:=tb\pmod{M}\\\vspace{2mm}
    v_i := \left\lfloor\frac{t}{M}a_i\right\rfloor,\qquad i=1,2,\ldots,n\\\vspace{2mm}
    w := \left\lfloor\frac{t}{M}b\right\rfloor
  \end{array}
  \right.
\end{align}
with the notation $\lfloor{\bullet}\rfloor$ denoting the floor truncate integer of a rational number,
for instance, $\lfloor 2\frac{2}{3}\rfloor = 2$ and
$\lfloor -2\frac{2}{3}\rfloor = -3$.
It is not hard to observe that
$0\leq d<M$, $0\leq w < b$, and
$0\leq c_i<M$, $0\leq v_i < a_i$ for $i=1,2,\ldots,n$.
The relationship between $c$ and $v$, and between $d$ and $w$,
can be derived as follows,
\begin{align}\label{eq:c_v}
  c_i = ta_i - Mv_i,\qquad i=1,2,\ldots,n,
\end{align}
and
\begin{align}\label{eq:d_w}
  d = tb - Mw.
\end{align}

\qquad
For a given subset-sum problem~\eqref{eq:prob_subset_sum},
any $\emph{\textbf{x}}\in \{0,1\}^n$ satisfying $\emph{\textbf{a}}\emph{\textbf{x}}=b$ must also satisfy
the following modular equation,
\begin{equation}\label{eq:6_mod_pro}
\emph{\textbf{c}}\emph{\textbf{x}} := \sum_{i=1}^{n}c_ix_i\equiv d\pmod{M},
\end{equation}
and the following algebraic equation,
\begin{align}\label{eq:cd_k}
\emph{\textbf{c}}\emph{\textbf{x}} := \sum_{i=1}^{n}c_ix_i = d + Mk,~\text{with }k\in\Z_+,
\end{align}
with $\emph{\textbf{c}}=(c_1,c_2,\ldots,c_n)$.
Substituting \eqref{eq:c_v} and \eqref{eq:d_w} into \eqref{eq:cd_k} yields that,
\begin{align}\label{eq:vw_k}
  \emph{\textbf{v}}\emph{\textbf{x}}:=\sum_{i=1}^{n}v_ix_i  = w - k,
\end{align}
with $\emph{\textbf{v}}=(v_1,v_2,\ldots,v_n)$.

\qquad
In the following theorem, we derive an upper bound
for
the non-negative integer $k$ introduced in the algebraic equations
\eqref{eq:cd_k} and $\eqref{eq:vw_k}$.

\begin{thm}\label{lem:3_bound for k_xBinary}
If $\textbf{x}\in \mathcal{X} = \{0,1\}^n$, then for the integer $k$ introduced in
\eqref{eq:cd_k} and \eqref{eq:vw_k}, an upper bound $u_k$ which depends on
$\frac{t}{M}$, can be derived as follows,
\begin{align}\label{eq:10_expression for uk}
u_k(\frac{t}{M})
= \left\lfloor
\tilde{b}\frac{t}{M}
\right\rfloor
+ \left\lfloor b\frac{t}{M} \right\rfloor
- \sum_{i=1}^{n}\left\lfloor a_i\frac{t}{M}\right\rfloor,
\end{align}
where $\tilde{b} := \sum_{i=1}^{n}a_i - b$ is the right-hand-side of
the complementary problem~\eqref{eq:complem_subset_sum}.
\end{thm}
\proof
Firstly, $k$ must be greater than or equal to zero which is implied
from \eqref{eq:6_mod_pro} and \eqref{eq:cd_k}.
Secondly, based on \eqref{eq:cd_k}, the following inequality can be derived,
\begin{align*}
k = (\emph{\textbf{c}}\emph{\textbf{x}} -d)/M \leq \sum_{i=1}^{n}\frac{c_i}{M} - \frac{d}{M},
\end{align*}
as $\emph{\textbf{x}}\in\{0,1\}^n$.
The fact that $k$ is an integer number implies that,
\begin{align}\label{eq:def_g_uk}
k\leq \left\lfloor \sum_{i=1}^{n}\frac{c_i}{M} - \frac{d}{M}\right\rfloor.
\end{align}

\qquad
We denote the term within the floor function in \eqref{eq:def_g_uk} as $g(\frac{t}{M})$, that is,
$$
 g(\frac{t}{M}) := \sum_{i=1}^{n}\frac{c_i}{M} - \frac{d}{M}.
$$
Substituting \eqref{eq:c_v} and \eqref{eq:d_w} into the above expression
gives rise to the expression of $g(\frac{t}{M})$ 
as follows,
\begin{align}\label{eq:9_expression for g}
g(\frac{t}{M}) =
\tilde{b}\frac{t}{M} + \left\lfloor b\frac{t}{M} \right\rfloor
- \sum_{i=1}^{n}\left\lfloor a_i\frac{t}{M}\right\rfloor,
\end{align}
and thus an upper bound of $k$ as follows,
\begin{align*}
  u_k(\frac{t}{M}) = \left\lfloor {g(\frac{t}{M})} \right\rfloor =
  \left\lfloor\tilde{b}\frac{t}{M}\right\rfloor
  + \left\lfloor b\frac{t}{M} \right\rfloor
  - \sum_{i=1}^{n}\left\lfloor a_i\frac{t}{M}\right\rfloor.
\end{align*}
\endproof

\qquad
Based on the expression of $u_k(\frac{t}{M})$ derived in Theorem~\ref{lem:3_bound for k_xBinary}
and the fact that $k$ is non-negative,
Corollary~\ref{cor:1} can be derived. The inequality derived in Corollary~\ref{cor:1} is neat and elegant,
which exhibits an unadorned, plain, and important relation among coefficients, $a_i$s, $b$ and $\tilde{b}$, of the given subset-sum problem and its complementary problem.
\begin{cor}\label{cor:1}
  Given a subset-sum problem~\eqref{eq:prob_subset_sum} and its complementary problem~\eqref{eq:complem_subset_sum},
  the following inequality can be derived,
\begin{align}
  \left\lfloor{r}\tilde{b}\right\rfloor
  + \left\lfloor {r}b \right\rfloor
  - \sum_{i=1}^{n}\left\lfloor {r}a_i\right\rfloor \geq 0,
\end{align}
with $r\in[0,1]\cap\mathbb{R}$.
\end{cor}

\qquad
Next in Theorem~\ref{thm:k=0}, three equivalent conditions are derived, under which $k\equiv 0$ holds.
In fact, $k$ depends on $r:=\frac{t}{M}$ as well. When we choose proper values of $t$ and $M$, thus proper value of $r:=\frac{t}{M}$, one more equation is revealed for the given subset-sum problem, but meanwhile no more unknown variables have been introduced. This is the most ideal scenario.
\begin{thm}\label{thm:k=0}
The following three inequalities are equivalent to each other,
\begin{itemize}
  \item [(a)] $\sum_{i=1}^{n}c_i< M+d$,
  \item [(b)] $g(\frac{t}{M})<1$,
  \item [(c)] $u_k(\frac{t}{M}) = 0$.
\end{itemize}
\end{thm}
\proof
$(a)~\Leftrightarrow~(b)$. Because $g(\frac{t}{M}) = (\sum_{i=1}^{n}c_i - d)/M$ and $M>0$.
\\
$(b)~\Leftrightarrow~(c)$. Because $u_k(\frac{t}{M}) = \left\lfloor{g(\frac{t}{M})}\right\rfloor$
 and $u_k(\frac{t}{M})\geq 0$.
\endproof

\qquad
The following example illustrates the most ideal situation that modular disaggregation techniques can achieve, that is when $k\equiv{0}$
under some specific chosen values of $t/M$.

\begin{exam} We consider the example in Part II of Merkle and Hellman's work in 1978
(see \cite{Merkle:1978})
with
$$
\textbf{a} = (171,196,457,1191,2410)\qquad\text{and}\qquad b = 3797.
$$
\begin{itemize}
\vspace{-3mm}
  \item [(i)]
    Let $M_1=4426$ and $t_1=79$. Set then
    $\textbf{c}^{(1)}$ $=$ $t_1\textbf{a}\pmod{M_1}$ = (231,2206,695,1143,72) and
    $d_1$ $=$ $t_1b\pmod{M_1}$ = 3421.
  \item [(ii)]
    Let $M_2=4348$ and $t_2=69$. Set then
    $\textbf{c}^{(2)}$ $=$ $t_2\textbf{c}^{(1)}\pmod{M_2}$ = (2895,34,127,603,620) and
    $d_2$ $=$ $t_2d_1\pmod{M_2}$ = 1257.
  \item [(iii)]
    Let $M_3=4280$ and $t_3=3$. Set then
    $\textbf{c}^{(3)}$ $=$ $t_3\textbf{c}^{(2)}\pmod{M_3}$ = (125,102,381,1809,1860) and
    $d_3$ $=$ $t_3d_2\pmod{M_3}$ = 3771.
  \item [(iv)]
    Let $M_4=4278$ and $t_4=5$. Set then
    $\textbf{c}^{(4)}$ $=$ $t_4\textbf{c}^{(3)}\pmod{M_4}$ = (625,510,1905,489,744) and
    $d_4$ $=$ $t_4d_3\pmod{M_4}$ = 1743.
\end{itemize}
Here, $\textbf{c}^{(j)} = (c_1^{(j)}, c_2^{(j)}, \ldots, c_5^{(j)})$ with $j = 1, 2, 3, 4$.
It is easy to verify that $\sum_{i=1}^5c^{(j)}_i < M_j + d_j$, for all $j = 1, 2, 3, 4$.
Therefore, based on Theorem~\ref{thm:k=0}, for any binary $\textbf{x}$ $\in$ $\{0,1\}^5$, the following subset-sum problem,
$$
\textbf{a} \textbf{x} = b
$$
is equivalent to the following system of linear equations,
$$
E\textbf{x} = F,
$$
where matrix $E$ and column vector $F$ can be represented as,
\begin{align*}
  E = \left(\begin{array}{c}
    \textbf{a}\\
    \textbf{c}^{(1)}\\
    \textbf{c}^{(2)}\\
    \textbf{c}^{(3)}\\
    \textbf{c}^{(4)}
  \end{array}\right),\qquad
  F = \left(\begin{array}{c}
    b\\
    d_1\\
    d_2\\
    d_3\\
    d_4
  \end{array}\right).
\end{align*}

Since the rank of $E$ equals 5,
we can solve and obtain the binary solution as
$$
\textbf{x} = E^{-1}F = (0,1,0,1,1)^{T}.
$$
\endsolution
\end{exam}

\qquad
In the general scenario, for arbitrary value of $r=t/M$,
based on the upper bound $u_k$ derived in Theorem~\ref{lem:3_bound for k_xBinary},
we further decompose the corresponding $k$ into its binary representation form,

\begin{align}\label{eq:10_k_decompose}
k = k_1 + 2k_2 + 4k_3 + \cdots + 2^{n^{(k)}-1}k_{n^{(k)}},
\end{align}
where $k_1,\ldots,k_{n^{(k)}}$ $\in\{0,1\}$ and
\begin{align}\label{eq:expression_for_nk}
n^{(k)} = \lceil \log_{2}(u_k+1)\rceil
\end{align}
with $u_k$ being a compact notation for $u_k(\frac{t}{M})$.
Substituting \eqref{eq:10_k_decompose} into \eqref{eq:vw_k} yields
the following disaggregated system with two Diophantine equations,
\begin{align}\label{eq:10_decompose_k_combine with ax=b}
\left\{\begin{array}{l}
\emph{\textbf{a}}\emph{\textbf{x}} = b\\
\emph{\textbf{v}}\emph{\textbf{x}} + k_1 + 2k_2 + 4k_3 + \cdots + 2^{n^{(k)}-1}k_{n^{(k)}} = w
\end{array}\right.
\end{align}
with $x_1,\ldots,x_n $, $k_1,\ldots,k_{n^{(k)}}$ $\in\{0,1\}$.
As a note, later in this paper, the bold symbol
$\emph{\textbf{k}}$ will be used to denote the decomposed vector $(k_1,k_2,\ldots,k_{n^{(k)}})^T$.

\qquad
Next in Lemma~\ref{lem:4_x_k BinarySolution}, we establish the relation between the original subset-sum problem \eqref{eq:prob_subset_sum}
and the new system \eqref{eq:10_decompose_k_combine with ax=b} obtained via our disaggregation techniques.

\begin{lem}\label{lem:4_x_k BinarySolution}
$\textbf{x}^*$ is a binary solution to \eqref{eq:prob_subset_sum}
if and only if there exists $\textbf{k}^*$
such that $(\textbf{x}^*,\textbf{k}^*)$ is a binary solution to \eqref{eq:10_decompose_k_combine with ax=b}.
\end{lem}
\proof
The logic is clear and the proof is readily obtained.
\endproof

\subsubsection{Jump points}

We re-write down the expressions of functions $g(\frac{t}{M})$ and $u_k(\frac{t}{M})$ as follows,
$$
g(\frac{t}{M}) = \tilde{b}\frac{t}{M} + \left\lfloor{b\frac{t}{M}}\right\rfloor
-\sum_{i=1}^{n}\left\lfloor{a_i\frac{t}{M}}\right\rfloor,
$$
$$
u_k(\frac{t}{M}) = \left\lfloor{\tilde{b}\frac{t}{M}}\right\rfloor + \left\lfloor{b\frac{t}{M}}\right\rfloor
-\sum_{i=1}^{n}\left\lfloor{a_i\frac{t}{M}}\right\rfloor,
$$
and also the disaggregation equation \eqref{eq:vw_k}, that is $\emph{\textbf{v}}\emph{\textbf{x}}+k=w$, as follows,
$$
\sum_{i=1}^{n}{\left\lfloor{a_i\frac{t}{M}}\right\rfloor}x_i + k = \left\lfloor{b\frac{t}{M}}\right\rfloor.
$$
We could see that, $g$, $u_k$, $\emph{\textbf{v}}$, and $w$ are all functions which depend on the parameter $\frac{t}{M}$.
As a note, $\emph{\textbf{v}}=(v_1,v_2,\ldots,v_n)$ is a vector function of
$\frac{t}{M}$
with
$v_i = {\left\lfloor{a_i\frac{t}{M}}\right\rfloor}$.
Basic observations are presented in the following.

\begin{obs}\label{obs:1}
\begin{itemize}
\item [(a)] $\textbf{v}$ is discontinuous if and only if the parameter $r=\frac{t}{M}$ takes value at any of the points, $\frac{j}{a_i}$, with $j=1,2,\ldots,a_i-1$ and $i=1,2,\ldots,n$.
\item [(b)] $w$ is discontinuous if and only if the parameter $r=\frac{t}{M}$ takes value at any of the points,
            $\frac{j}{b}$, with $j=1,2,\ldots,b-1$.
\item [(c)] $u_k$ is discontinuous if and only if the parameter $r=\frac{t}{M}$ takes value at any of the points, $\frac{j}{a_i}$ with $j=1,2,\ldots,a_i-1$ and $i=1,2,\ldots,n$, or
$\frac{j}{b}$ with $j=1,2,\ldots,b-1$, or $\frac{j}{\tilde{b}}$ with $j=1,2,\ldots,\tilde{b}-1$,
\end{itemize}
\end{obs}

\qquad
Based on Observation~\ref{obs:1}, the concept \emph{jump points of subset-sum problems}
is introduced and defined in Definition~\ref{def:1}.
Later some basic properties and benefits of these \emph{jump points} are derived in
Section~\ref{subsec:cutting-off_short_solution}
and Section~\ref{subsec:NeighbouringJP}.

\begin{defn}\label{def:1}[Jump points of subset-sum problems]
Consider coefficients in Problem~\eqref{eq:prob_subset_sum}, the following points,
\vspace{-3mm}
  \begin{itemize}
    \item [(a)] rational numbers $\frac{j}{a_i}$, $j=1,2,\ldots,a_i-1$, $i=1,2,\ldots,n$; and
    \item [(b)] rational numbers $\frac{j}{b}$, $j=1,2,\ldots,b-1$; and
    \item [(c)] rational numbers $\frac{j}{\tilde{b}}$, $j=1,2,\ldots,\tilde{b}-1$, with
    $\tilde{b}:=\sum_{i=1}^{n}a_i - b$,
  \end{itemize}
\vspace{-3mm}
are called \emph{jump points of subset-sum problems}.
\end{defn}

In fact, $\emph{\textbf{v}}$, $w$, upper bound $u_k$ of $k$,
and number $n^{(k)}$ of the newly introduced unknowns in the disaggregated equation,
are all piecewise linear functions.
Only at jump points of subset-sum problems, coefficients of the disaggregated equation,
$$
\emph{\textbf{v}}\emph{\textbf{x}} + \emph{\textbf{u}}\emph{\textbf{k}} = w
$$
jump and change, with $\emph{\textbf{u}} = (1,2,\ldots,2^{n^{(k)}-1})$
and $\emph{\textbf{k}} = (k_1, k_2, \ldots, k_{n^{(k)}})^T\in\{0,1\}^{n^{(k)}}$.

\subsubsection{Cutting-off short integer solutions}\label{subsec:cutting-off_short_solution}

In this section, we introduce our idea of trying to cut-off
some non-binary integer solutions to Problem~\eqref{eq:prob_subset_sum} with small Euclidean lengths,
thus to achieve the ultimate goal that to increase the probability of returning binary solutions to the given subset-sum problems.

\qquad
The disaggregation techniques introduced previously can actually divide the feasible solution set of
Problem~\eqref{eq:prob_subset_sum} into several subsets,
where $k$ introduced in the disaggregated equation plays an important role.
Next we introduce and define five sets, denoted as \circled{i}, $i=1,2,\ldots,5$,
which are generated during the disaggregation process, as follows,
\begin{align*}
&\circled{1} :=
\{\emph{\textbf{x}}\in\Z^n~|~\emph{\textbf{a}}\emph{\textbf{x}} = b\},\\
&\circled{2} :=
\{\emph{\textbf{x}}\in\Z^n~|~\emph{\textbf{a}}\emph{\textbf{x}} = b,~\emph{\textbf{v}}\emph{\textbf{x}} = w-k,~k\in\Z\},\\
&\circled{3} :=
\{\emph{\textbf{x}}\in\Z^n~|~\emph{\textbf{a}}\emph{\textbf{x}} = b,~\emph{\textbf{v}}\emph{\textbf{x}} = w-k,~k\in\Z~\text{and}~0\leq k\leq u_k\},\\
&\circled{4} :=
\{\emph{\textbf{x}}\in\{0,1\}^n~|~\emph{\textbf{a}}\emph{\textbf{x}} = b\},\\
&\circled{5} :=
\{\emph{\textbf{x}}\in\{0,1\}^n~|~\emph{\textbf{a}}\emph{\textbf{x}} = b,~\emph{\textbf{v}}\emph{\textbf{x}} = w-k,~k\in\Z~\text{and}~0\leq k\leq u_k\}.
\end{align*}
As a note,
\begin{itemize}
\item [$(a)$] Set $\circled{3}$ depends on the value of $r=\frac{t}{M}$ and thus is also denoted
as $\circled{3}|r$ for clarity purpose;
\item [$(b)$] There may exist $k^*\in\{0,1,\ldots,u_k\}$,
such that, in terms of $\emph{\textbf{x}}$, the integer solution set of the following system of two equations,
$$
\left\{\begin{array}{l}
  \emph{\textbf{a}}\emph{\textbf{x}} = b\\
  \emph{\textbf{v}}\emph{\textbf{x}} = w - k^*
\end{array}\right.
$$
is empty.
\end{itemize}

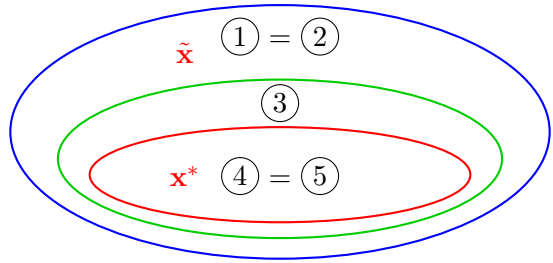
\begin{figure}[H]
    \begin{center}
    \begin{tikzpicture}[scale=0.6]
    \draw[blue, thick] (0,0)  ellipse (170pt and 80pt);
    \draw (0,60pt) node{\circled{1} = \circled{2}};
    \draw[DarkGreen, thick] (0,-17pt)  ellipse (140pt and 50pt);
    \draw (0,18pt) node{\circled{3}};
    \draw[red, thick] (0,-27pt)  ellipse (120pt and 30pt);
    \draw (0,-28pt) node{\circled{4} = \circled{5}};
    \draw (-60pt,50pt) node{{\color{Red}$\tilde{\textbf{x}}$}};
    \draw (-60pt,-28pt) node{{\color{Red}$\textbf{x}^*$}};
    \end{tikzpicture}
    \caption{Inclusion relation.}
    \label{fig:1_Disaggregation_sets}
    \end{center}
\end{figure}

\qquad
The set inclusion relation among these 5 sets can be described in Figure~\ref{fig:1_Disaggregation_sets},
which neatly illustrate that,
$$
\circled{5} = \circled{4}\subset \circled{3}\subset\circled{2}=\circled{1}.
$$

If lattice attack algorithms are applied to Problem~\eqref{eq:prob_subset_sum},
sometimes the returned solution is not binary although the Euclidean length of it is small.
Actually the returned solution generally belongs
to set  $\circled{1}$. However, the desired binary solution should always belong
to set  $\circled{4}$.
Our target now is to generate the desired set $\circled{3}|r$ that could increase the probability
of returning binary solutions.

\qquad
Next we use a concrete example to illustrate how disaggregation techniques can help generate
desired set $\circled{3}|r$, thus can cut-off the initially returned non-binary integer solutions with small Euclidean lengths. Thereafter, binary solutions are successfully found.

\begin{exam}\label{exam:1_excluding_non-binary_solution}
We use the following toy problem
$$
\textbf{a} = (3,  15,   6),~\text{and}~b = 9
$$
to illustrate how some values of $r=\frac{t}{M}$ can help generate
desired set $\circled{3}|r$, thus to
cut-off the initially returned non-binary solutions with small Euclidean lengths.
Thereafter successfully search and return the binary solution.
\end{exam}
\solution
\begin{enumerate}
\item [$(a)$]
A general representation of all integer solutions to $\emph{\textbf{a}}\emph{\textbf{x}}=b$ is
presented as below,
\begin{align*}
 \emph{\textbf{x}} = (1, 0, 1)^T +
\lambda_1 (-2, 0, 1)^T +
\lambda_2(-1, 1, -2)^T~\text{with}~\lambda_1,\lambda_2\in\Z,
\end{align*}
and thus
$\circled{1} = \circled{2} = \{\emph{\textbf{x}}~|~
\emph{\textbf{x}} = (1,0,1)^T + \lambda_1(-2, 0, 1)^T + \lambda_2(-1, 1, -2)^T,~\lambda_1,\lambda_2\in\Z\}$.
\item [$(b)$] $\tilde{\emph{\textbf{x}}} = (0, 1, -1)^T$ is a non-binary solution
to the given problem, with small Euclidean length, which is initially returned
via the lattice attack algorithm, for instance, Algorithm~\ref{alg:1_reduce(x,D)}.
\item [$(c)$] The given problem has only one binary solution $(1,0,1)^T$, and thus
$
\circled{4} = \circled{5} =
\{(1, 0, 1)^T\}.
$
\item [$(d)$] In our modular disaggregation process,
different values of $r := \frac{t}{M}$ result in different sets~$\circled{3}|r$,
for example,
\begin{itemize}
    \item [(i)] Let $(t,M) = (3,  6)$, $r = 3/6 = 1/2$, then $u_k = 0$,
    $\emph{\textbf{v}} = (1,  7,  3)$, and $w = 4$.
    Thus
    $\circled{3}|\frac{1}{2} = \{\emph{\textbf{x}}~|~\emph{\textbf{x}} = (1,0,1)^T + \lambda_1(-1, 1, -2 )^T, \lambda_1\in\Z\}$,
    and,
    $$\tilde{\emph{\textbf{x}}} = (0, 1, -1)^T = (1,0,1)^T + (-1, 1, -2 )^T\in\circled{3}|\frac{1}{2}.$$
    \item [(ii)] Let $(t,M) = (6, 15)$, $r=6/15=2/5$, then $u_k = 0$,
    $\emph{\textbf{v}} = (1,  6,  2)$, and $w = 3$.
    Thus
    $\circled{3}|\frac{2}{5} = \{\emph{\textbf{x}}~|~\emph{\textbf{x}} = (1,0,1)^T + \lambda_1(-2, 0, 1)^T, \lambda_1\in\Z\}$,
    and,
    $$
    \tilde{\emph{\textbf{x}}} = (0, 1, -1)^T \not\in\circled{3}|\frac{2}{5}.
    $$
\end{itemize}
\item [$(e)$] After disaggregation, under $r=\frac{2}{5}$, now Algorithm 3 can successfully return the binary solution
$\emph{\textbf{x}} = (1,0,1)^T$.\endsolution
\end{enumerate}


\begin{prop}\label{lem:5}
  Given a non-binary solution $\tilde{\textbf{x}}$ of subset-sum problem~\eqref{eq:prob_subset_sum},
  if value of the parameter $r=\frac{t}{M}$ is chosen such that,
  \begin{align}\label{eq:condition_cutoff_x}
    w - \textbf{v}\tilde{\textbf{x}} > u_k \quad \text{or} \quad w - \textbf{v}\tilde{\textbf{x}} <0,
\end{align}
then there must have,
$$
\tilde{\textbf{x}}\not\in\circled{3}|r.
$$
\end{prop}

\qquad
Proposition~\ref{lem:5} summarizes the conditions, under which after disaggregation process,
a given non-binary solution with small Euclidean length must can be excluded from the new feasible solution set $\circled{3}|r$.

\subsubsection{Neighbouring jump points}\label{subsec:NeighbouringJP}
In the last section, the idea of cutting-off a given non-binary integer solution
$\tilde{\textbf{x}}$ with
small Euclidean length has been introduced.
In this section, we investigate further the properties of jump points of subset-sum problems,
and study
the relations between two neighbouring
jump points (NJPs).

\qquad
Consider all the jump points of subset-sum problems defined in
Definition~\ref{def:1}, which are ordered in sequence based on their magnitudes.
Then NJPs are the two jump points that just next to one another.
Basic theorems will be derived,
and some notations are introduced as follows to facilitate our analysis,
\begin{align}\label{eq:definition_for_Delta_v_w}
    \Delta \emph{\textbf{v}}:=\emph{\textbf{v}}^{(2)}-\emph{\textbf{v}}^{(1)},~\Delta w:=w^{(2)} - w^{(1)},
    ~\text{and}~\Delta \tilde{w}:=\tilde{w}^{(2)} - \tilde{w}^{(1)},
\end{align}
where
$$
\emph{\textbf{v}}^{(i)} := \lfloor{\emph{\textbf{a}}r_i}\rfloor
=(\lfloor{a_1r_i}\rfloor, \lfloor{a_2r_i}\rfloor,
\ldots, \lfloor{a_nr_i}\rfloor)\in\Z^{n},
\quad i=1,2,
$$
$$
\Delta \emph{\textbf{v}} = (\Delta v_1, \Delta v_2, \ldots, \Delta v_n),
$$
and
$$
w^{(i)} := \lfloor{br_i}\rfloor,
\quad \tilde{w}^{(i)} := \lfloor{\tilde{b}r_i}\rfloor,\quad i=1,2,
$$
with $\tilde{b} = \sum_{i=1}^{n}a_i-b$, and
$r_2 > r_1\in(0,1)$ being two NJPs.

\qquad
Next we conduct analysis on the difference between upper bounds $u_k^{(1)}$ and $u_k^{(2)}$
of two neighbouring jump points $r_1$ and $r_2$.
The difference between $u_k^{(2)}$ and $u_k^{(1)}$ is defined as follows,
\begin{align}\label{eq:NJP_difference_u_k_withoutAssumption}
\Delta u_k := u_k^{(2)}-u_k^{(1)} = {\Delta \tilde{w}} + {\Delta w} - \sum_{i=1}^{n}{\Delta v}_i.
\end{align}
For a given non-binary integer vector $\tilde{\emph{\textbf{x}}}$ and jump point $r_i$,
the corresponding value of $\tilde{k}^{(i)}$ can be calculated as
follows,
\begin{align*}
   \tilde{k}^{(i)} := w^{(i)} - \emph{\textbf{v}}^{(i)}\tilde{\emph{\textbf{x}}},\quad i=1,2,
\end{align*}
from which the difference between $\tilde{k}^{(2)}$ and $\tilde{k}^{(1)}$
can be derived as follows,
\begin{align}\label{eq:NJP_difference_k_withoutAssumption}
   \Delta \tilde{k}:= \tilde{k}^{(2)} -  \tilde{k}^{(1)} = {\Delta w} - {\Delta \emph{\textbf{v}}}\tilde{\emph{\textbf{x}}}.
\end{align}

\qquad
Now it is ready to present our observations and basic theorems regarding NJPs.
Consider coefficients of the disaggregated equations $\emph{\textbf{v}}^{(1)}\emph{\textbf{x}}+k^{(1)}=w^{(1)}$
and $\emph{\textbf{v}}^{(2)}\emph{\textbf{x}}+k^{(2)}=w^{(2)}$,
our observations are summarized in Observation~\ref{obs:a_h and LHS NJP}
and Observation~\ref{obs:b and LHS NJP}.

\begin{obs}\label{obs:a_h and LHS NJP}
  If $r_2 = \frac{j}{a_h}$ and $r_1$ is the left-hand-side (\emph{LHS}) \emph{NJP}
  of $r_2$, then we have the following observations.
  \vspace{-2mm}
  \begin{itemize}
    \item [(a)] $\Delta \textbf{v} \in\{0,1\}^n$.
    \item [(b)] ${\Delta v}_h=1$.
    \item [(c)] For $1\leq i\neq h\leq n$,
    ${\Delta v}_i = 1$ if and only if $\frac{j}{a_h} = \frac{\hat{j}}{a_i}$ for some $\hat{j}\in\{1,2,\ldots,a_i-1\}$.
    \item [(d)] $\Delta w \in\{0,1\}$.
    \item [(e)]$\Delta w =1$ if and only if $\frac{j}{a_h} = \frac{\hat{j}}{b}$
    for some $\hat{j}\in\{1,2,\ldots,b-1\}$.
  \end{itemize}
\end{obs}

\begin{obs}\label{obs:b and LHS NJP}
  If $r_2 = \frac{j}{b}$ and $r_1$ is the \emph{LHS} \emph{NJP}
  of $r_2$, then we have the following observations.
  \vspace{-2mm}
  \begin{itemize}
    \item [(a)] $\Delta \textbf{v} \in\{0,1\}^n$.
    \item [(b)] For $1\leq i\leq n$, ${\Delta v}_i = 1$ if and only if
        $\frac{j}{b} = \frac{\hat{j}}{a_i}$ for some $\hat{j}\in\{1,2,\ldots,a_i-1\}$.
    \item [(c)] $\Delta w = 1$.
  \end{itemize}
\end{obs}

\qquad
Next, we derive Theorem~\ref{lem:1_NJP_withoutAssumption} and
Theorem~\ref{lem:2_NJP_withoutAssumption}
which are about the properties of NJPs,
in terms of cutting-off $\tilde{\emph{\textbf{x}}}$,
where $\tilde{\emph{\textbf{x}}}$ denotes a given non-binary integer solution
to Problem~\eqref{eq:prob_subset_sum}
with
small Euclidean length.

\begin{thm}\label{lem:1_NJP_withoutAssumption}
  Given two NJPs $r_1,r_2\in(0,1)$ with $r_1<r_2$,
  and given a non-binary integer solution $\tilde{\textbf{x}}$ to subset-sum problem~\eqref{eq:prob_subset_sum},
  then we have the following three equivalent statements.
  \vspace{-2mm}
  \begin{itemize}
    \item [(a)]$
    {\Delta w}\leq {\Delta \textbf{v}}\tilde{\textbf{x}} \leq \sum_{i=1}^{n}{\Delta v}_i - {\Delta \tilde{w}}.
    $
    \item [(b)] $\tilde{\textbf{x}} \in \circled{3}|r_2~\Rightarrow~\tilde{\textbf{x}} \in \circled{3}|r_1$.
    \item [(c)]$\tilde{\textbf{x}} \not\in\circled{3}|r_1~\Rightarrow~\tilde{\textbf{x}} \not\in\circled{3}|r_2$.
    \end{itemize}
\end{thm}
\proof
$(a)~\Leftrightarrow~(b)$.
We have that,
\begin{align*}
&\tilde{\emph{\textbf{x}}}\in \circled{3}|r_2
~\Leftrightarrow~
\tilde{k}^{(2)} \in\{0,1,\ldots,u_k^{(2)}\}\\
\Leftrightarrow \quad
&
\tilde{k}^{(2)}-{\Delta w} + {\Delta \emph{\textbf{v}}}\tilde{\emph{\textbf{x}}}
\in\{-{\Delta w} + {\Delta \emph{\textbf{v}}}\tilde{\emph{\textbf{x}}},
1-{\Delta w} + {\Delta \emph{\textbf{v}}}\tilde{\emph{\textbf{x}}},
\ldots,
u_k^{(2)}-{\Delta w} + {\Delta \emph{\textbf{v}}}\tilde{\emph{\textbf{x}}}\}.
\end{align*}
From \eqref{eq:NJP_difference_k_withoutAssumption} and
\eqref{eq:NJP_difference_u_k_withoutAssumption}, we have that,
$$
\tilde{k}^{(1)} = \tilde{k}^{(2)} -{\Delta w} + {\Delta \emph{\textbf{v}}}\tilde{\emph{\textbf{x}}},
\text{ and }
u_k^{(1)} = u_k^{(2)} - {\Delta \tilde{w}} + \sum_{i=1}^{n}{\Delta v}_i - {\Delta w},
$$
which implies that,
$$
\tilde{\emph{\textbf{x}}}\in \circled{3}|r_2
~\Leftrightarrow~
\tilde{k}^{(1)} \in
\{-{\Delta w} + {\Delta \emph{\textbf{v}}}\tilde{\emph{\textbf{x}}},
1-{\Delta w} + {\Delta \emph{\textbf{v}}}\tilde{\emph{\textbf{x}}},
\ldots,
u_k^{(1)}+{\Delta \tilde{w}}-\sum_{i=1}^{n}{\Delta v}_i + {\Delta \emph{\textbf{v}}}\tilde{\emph{\textbf{x}}}\}.
 $$
Meanwhile, we have that,
$$
\tilde{\emph{\textbf{x}}}\in \circled{3}|r_1
~\Leftrightarrow~
\tilde{k}^{(1)} \in
\{0,1,\ldots,u_k^{(1)}\}.
$$

\qquad
Based on the derivations shown above, it is now ready to obtain that,
\begin{align*}
  & \tilde{\emph{\textbf{x}}}\in \circled{3}|r_2 ~\Rightarrow~ \tilde{\emph{\textbf{x}}}\in \circled{3}|r_1 \\
  \Leftrightarrow \quad
  & \tilde{k}^{(1)} \in
        \{-{\Delta w} + {\Delta \emph{\textbf{v}}}\tilde{\emph{\textbf{x}}}, 1-{\Delta w} + {\Delta \emph{\textbf{v}}}\tilde{\emph{\textbf{x}}},
        \ldots,
        u_k^{(1)}+{\Delta \tilde{w}}-\sum_{i=1}^{n}{\Delta v}_i + {\Delta \emph{\textbf{v}}}\tilde{\emph{\textbf{x}}}\}
        ~\Rightarrow~
        \tilde{k}^{(1)} \in\{0,1,\ldots,u_k^{(1)}\}\\
  \Leftrightarrow \quad
  & \{-{\Delta w} + {\Delta \emph{\textbf{v}}}\tilde{\emph{\textbf{x}}},
  1-{\Delta w} + {\Delta \emph{\textbf{v}}}\tilde{\emph{\textbf{x}}},
        \ldots,
        u_k^{(1)}+{\Delta \tilde{w}}-\sum_{i=1}^{n}{\Delta v}_i + {\Delta \emph{\textbf{v}}}\tilde{\emph{\textbf{x}}}\}
        \subseteq \{0,1,\ldots,u_k^{(1)}\}\\
  \Leftrightarrow \quad
  & -{\Delta w} + {\Delta \emph{\textbf{v}}}\tilde{\emph{\textbf{x}}}\geq 0\text{ and }
        {\Delta \tilde{w}}-\sum_{i=1}^{n}{\Delta v}_i + {\Delta \emph{\textbf{v}}}\tilde{\emph{\textbf{x}}} \leq 0\\
  \Leftrightarrow \quad
  & {\Delta w}\leq {\Delta \emph{\textbf{v}}}\tilde{\emph{\textbf{x}}} \leq \sum_{i=1}^{n}{\Delta v}_i - {\Delta \tilde{w}}.
\end{align*}
$(b)~\Leftrightarrow~(c)$.
By contradiction, it is straightforward.
\endproof

\begin{thm}\label{lem:2_NJP_withoutAssumption}
  Given two NJPs $r_1,r_2\in(0,1)$ with $r_1<r_2$, and
  given a non-binary integer solution $\tilde{\textbf{x}}$ to subset-sum problem~\eqref{eq:prob_subset_sum},
  then we have the following three equivalent statements.
  \vspace{-2mm}
  \begin{itemize}
    \item [(a)]
    $
    \sum_{i=1}^{n}{\Delta v}_i - {\Delta \tilde{w}} \leq
    {\Delta \textbf{v}}\tilde{\textbf{x}}
    \leq {\Delta w}.
    $
    \item [(b)] $\tilde{\textbf{x}}\in\circled{3}|r_1~\Rightarrow~\tilde{\textbf{x}}\in\circled{3}|r_2$.
    \item [(c)] $\tilde{\textbf{x}}\not\in\circled{3}|r_2~\Rightarrow~\tilde{\textbf{x}}\not\in\circled{3}|r_1$.
  \end{itemize}
\end{thm}
\proof
$(a)~\Leftrightarrow~(b)$.
We have that,
\begin{align*}
&\tilde{\emph{\textbf{x}}}\in\circled{3}|r_1
~\Leftrightarrow~  \tilde{k}^{(1)} \in\{0,1,\ldots,u_k^{(1)}\}\\
\Leftrightarrow \quad
&
\tilde{k}^{(1)} + {\Delta w} - {\Delta \emph{\textbf{v}}}\tilde{\emph{\textbf{x}}}
\in\{{\Delta w} - {\Delta \emph{\textbf{v}}}\tilde{\emph{\textbf{x}}},
1 + {\Delta w} - {\Delta \emph{\textbf{v}}}\tilde{\emph{\textbf{x}}},
\ldots, u_k^{(1)} + {\Delta w} - {\Delta \emph{\textbf{v}}}\tilde{\emph{\textbf{x}}}\}\\
\Leftrightarrow \quad
&
\tilde{k}^{(2)}\in
\{{\Delta w} - {\Delta \emph{\textbf{v}}}\tilde{\emph{\textbf{x}}},
1 + {\Delta w} - {\Delta \emph{\textbf{v}}}\tilde{\emph{\textbf{x}}},
\ldots,
u_k^{(2)}-{\Delta \tilde{w}} + \sum_{i=1}^{n}{\Delta v_i} - {\Delta \emph{\textbf{v}}}\tilde{\emph{\textbf{x}}}.
\end{align*}
Meanwhile, we have that,
$$
\tilde{\emph{\textbf{x}}}\in\circled{3}|r_2 \Leftrightarrow
\tilde{k}^{(2)} \in
\{0,1,\ldots,u_k^{(2)}\}.
$$
Based on the derivations shown above, it is now ready to obtain that,
\begin{align*}
  & \tilde{\emph{\textbf{x}}}\in\circled{3}|r_1 ~\Rightarrow~
  \tilde{\emph{\textbf{x}}}\in\circled{3}|r_2\\
  \Leftrightarrow \quad
  & \tilde{k}^{(2)}\in
        \{{\Delta w} - {\Delta \emph{\textbf{v}}}\tilde{\emph{\textbf{x}}},
        1 + {\Delta w} - {\Delta \emph{\textbf{v}}}\tilde{\emph{\textbf{x}}},
        \ldots,
        u_k^{(2)}-{\Delta \tilde{w}} + \sum_{i=1}^{n}{\Delta v_i} -
        {\Delta \emph{\textbf{v}}}\tilde{\emph{\textbf{x}}}\}
        \Rightarrow
        \tilde{k}^{(2)} \in\{0,1,\ldots,u_k^{(2)}\}\\
  \Leftrightarrow \quad
  & \{{\Delta w} - {\Delta \emph{\textbf{v}}}\tilde{\emph{\textbf{x}}} ,
  1 + {\Delta w} - {\Delta \emph{\textbf{v}}}\tilde{\emph{\textbf{x}}},
        \ldots,
        u_k^{(2)}-{\Delta \tilde{w}} + \sum_{i=1}^{n}{\Delta v_i} -
        {\Delta \emph{\textbf{v}}}\tilde{\emph{\textbf{x}}} \}
        \subseteq \{0,1,\ldots,u_k^{(2)}\}\\
  \Leftrightarrow \quad
  & {\Delta w} - {\Delta \emph{\textbf{v}}}\tilde{\emph{\textbf{x}}}
  \geq 0,
  \text{ and }
        -{\Delta \tilde{w}} + \sum_{i=1}^{n}{\Delta v}_i - {\Delta \emph{\textbf{v}}}\tilde{\emph{\textbf{x}}}
         \leq 0\\
  \Leftrightarrow \quad
  & \sum_{i=1}^{n}{\Delta v}_i - {\Delta \tilde{w}} \leq {\Delta \emph{\textbf{v}}}\tilde{\emph{\textbf{x}}} \leq {\Delta w}.
\end{align*}
$(b)~\Leftrightarrow~(c)$.
By contradiction, it is straightforward.
\endproof


\qquad
Theorem~\ref{lem:1_NJP_withoutAssumption} proofs the equivalent conditions,
under which if the jump point $r_1$ can help cut-off $\tilde{\emph{\textbf{x}}}$,
then its RHS NJP $r_2$ must can as well. Therefore, the condition derived in Theorem~\ref{lem:1_NJP_withoutAssumption}
actually can identify that when the incumbent jump point
is stronger than its RHS NJP.
Meanwhile, Theorem~\ref{lem:2_NJP_withoutAssumption} proofs the equivalent conditions,
under which if the jump point $r_2$ can help cut-off $\tilde{\emph{\textbf{x}}}$,
then its LHS NJP $r_1$ must can as well. Therefore, the condition derived in Theorem~\ref{lem:2_NJP_withoutAssumption}
actually can identify that when the incumbent jump point
is stronger than its LHS NJP.
Here, $\tilde{\textbf{x}}$ normally represents a vector with short Euclidean length, which
easily leads the possible failure of lattice attack algorithms
in searching a valid binary solution to
subset-sum problems~\eqref{eq:prob_subset_sum}.

\section{Numerical Tests: Integration of Modular Disaggregation Technique with
Lattice Attacks}\label{sec:lattice+disaggregation}

\subsection{Systems with Single Subset-Sum Equation}
\begin{figure}[H]
\begin{center}
    \begin{tikzpicture}[scale=0.75]
    \draw (40pt,70pt) node{{\color{Blue}\footnotesize{Input a subset-sum problem,
    $\emph{\textbf{a}}\emph{\textbf{x}} = b$}}};
    \draw [->, rounded corners, thick, black] (40pt, 60pt)--(40pt, 22pt);
    \draw[blue, thick] (0pt,0pt) rectangle (80pt, 20pt);
    \draw (40pt,10pt) node{\footnotesize{Lattice Attack}};
    \draw[blue, thick] (200pt,0pt) rectangle (280pt, 20pt);
    \draw (240pt,10pt) node{\footnotesize{Disaggregation}};
    \draw[blue, thick] (200pt,-80pt) rectangle (280pt, -60pt);
    \draw (240pt,-70pt) node{\footnotesize{Search {\color{Red}$r:=\frac{t}{M}$}}};
    \draw[blue, thick] (100pt,-160pt) rectangle (180pt, -140pt);
    \draw (140pt,-150pt) node{\footnotesize{Return}};
    \draw [->, rounded corners, thick, black] (140pt,-162pt)--(140pt, -200pt);
    \draw (140pt, -210pt) node{{\color{Red}\footnotesize{Output a binary solution}}};
    \draw [->, rounded corners, thick, DarkRed] (40pt, 0pt)--(40pt, -150pt)--(100pt, -150pt);
    \draw [->, rounded corners, thick, DarkRed] (230pt, -80pt)--(230pt, -150pt)--(180pt, -150pt);
    \draw [->, rounded corners, thick, DarkGreen] (80pt, 10pt)--(200pt, 10pt);
    \draw [->, rounded corners, thick, DarkGreen] (240pt, 0pt)--(240pt, -60pt);
    \draw [DarkGreen, thick] (250pt, -80pt) to [out=-90,in=180] (290pt, -120pt);
    \draw [DarkGreen, thick] (290pt, -120pt) to [out=0,in=-90] (310pt, -80pt);
    \draw [->, DarkGreen, thick] (310pt, -80pt) to [out=90,in=0] (280pt, -70pt);
    \draw (17pt,-70pt) node{{\color{Red}\footnotesize{Succeed}}};
    \draw (207pt,-110pt) node{{\color{Red}\footnotesize{Succeed}}};
    \draw (140pt,15pt) node{{\color{Blue}\footnotesize{Fail}}};
    \draw (297pt,-100pt) node{{\color{Blue}\footnotesize{Fail}}};
    \end{tikzpicture}
\caption{Illustration for the process of ``DAG + lattice attack".}
\label{fig:Illustrate_DAG+LatticeMethod}
\end{center}
\end{figure}
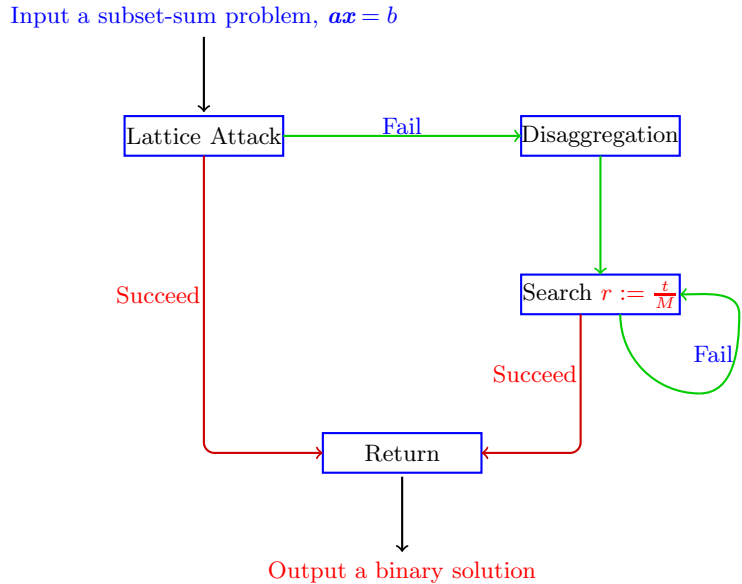

In this section, plenty of subset-sum problems
with density one are randomly generated to test our algorithms proposed in this paper.
Our progress is achieved by invoking the integration of modular disaggregation technique
(DAG)
with lattice attack algorithms \textbf{CJLOSS-Alg} and $\textbf{Reduce}_{1/2}$,
respectively. Figure~\ref{fig:Illustrate_DAG+LatticeMethod} is used to illustrate the procedure.
Later, numerical results are summarized in
Table~\ref{table:1_comparison_lattice_attacks},
Table~\ref{table:alg:modification_2+DIS},
and Table~\ref{table:alg:CJ+DIS}.
Our numerical experiment confirms that the success ratio
of finding valid binary solutions to the hard subset-sum problems with density one
can increase dramatically, by applying this integration procedure.

\qquad
The designation method of how to randomly generate systems with single subset-sum equation,
which are in the form of Problem~\eqref{eq:prob_subset_sum}, is described as follows,
\vspace{-2mm}
\begin{enumerate}
   \item [(1).]  100 systems are randomly generated with dimension $n=16$, $20$,
$26$, $30$, $36$, $40$, $50$, $60$, and $70$, respectively;
    \item [2).] $a_i$, $i=1,2,\ldots,n$, follows discrete uniform distribution
        on the interval $[1, 2^n]$, where $a_i$ is the $i$th entry of $\emph{\textbf{a}}$;
    \item [3).] $\emph{\textbf{x}}$ is randomly generated with cardinality $n/2$, which is fixed
    for the systems with the same dimension $n$;
    \item [4).] $density := \frac{n}{\max_{1\leq i\leq n}(\log_{2}a_i)} \in (0.99,1.00)$;
    \item [5).] $b:=\emph{\textbf{a}}  \emph{\textbf{x}}$, which satisfies that $b>\max(\emph{\textbf{a}})$
    and $b\leq \text{sum}(\emph{\textbf{a}})/2$.
\end{enumerate}

The reason for generating $\emph{\textbf{x}}$ with cardinality $n/2$ is that these systems are
even more difficult than the other systems with the same dimension $n$.
Otherwise
the information contained in the binary solution is sparse, either for the
original problem~\eqref{eq:prob_subset_sum} or
for the complementary problem~\eqref{eq:complem_subset_sum}.

\begin{table}[!h]
\caption{Computational performance of different lattice attack algorithms.}
\label{table:1_comparison_lattice_attacks}
\vspace{2mm}
\centering
\begin{tabular}{c||c|c|c|c|c}
  \hline
  $n$ & $\textbf{Reduce}$  & $\textbf{Reduce}_{1/2}$
  & \textbf{CJLOSS-Alg} & \textbf{LO-Alg} &\textbf{AHL-Alg} \\\hline\hline
  \hspace{3mm} 16 \hspace{3mm} & \hspace{3mm} 26\% \hspace{3mm} & \hspace{3mm} 67\% \hspace{3mm}
  &\hspace{3mm} 100\% \hspace{3mm} &\hspace{3mm} 56\% \hspace{3mm} &\hspace{3mm} 26\% \hspace{3mm}\\
  \hspace{3mm} 20 \hspace{3mm} & \hspace{3mm} 15\% \hspace{3mm} & \hspace{3mm} 44\% \hspace{3mm}
  &\hspace{3mm} 99\% \hspace{3mm} &\hspace{3mm} 37\% \hspace{3mm} &\hspace{3mm} 15\% \hspace{3mm}\\
  \hspace{3mm} 26 \hspace{3mm} & \hspace{3mm} 6\% \hspace{3mm} & \hspace{3mm} 15\% \hspace{3mm}
  &\hspace{3mm} 84\% \hspace{3mm} &\hspace{3mm} 11\% \hspace{3mm} &\hspace{3mm} 6\% \hspace{3mm}\\
  \hspace{3mm} 30 \hspace{3mm} & \hspace{3mm} 2\% \hspace{3mm} & \hspace{3mm} 10\% \hspace{3mm}
  &\hspace{3mm} 58\% \hspace{3mm} &\hspace{3mm} 7\% \hspace{3mm} &\hspace{3mm} 2\% \hspace{3mm}\\
  \hspace{3mm} 36 \hspace{3mm} & \hspace{3mm} 0\% \hspace{3mm} & \hspace{3mm} 4\% \hspace{3mm}
  &\hspace{3mm} 20\% \hspace{3mm} &\hspace{3mm} 1\% \hspace{3mm} &\hspace{3mm} 0\% \hspace{3mm}\\
  \hspace{3mm} 40 \hspace{3mm} & \hspace{3mm} 0\% \hspace{3mm} & \hspace{3mm} 1\% \hspace{3mm}
  &\hspace{3mm} 8\% \hspace{3mm} &\hspace{3mm} 0\% \hspace{3mm} &\hspace{3mm} 0\% \hspace{3mm}\\
  \hspace{3mm} 50 \hspace{3mm} & \hspace{3mm} 0\% \hspace{3mm} & \hspace{3mm} 0\% \hspace{3mm}
  &\hspace{3mm} 0\% \hspace{3mm} &\hspace{3mm}  0\% \hspace{3mm} &\hspace{3mm} 0\% \hspace{3mm}\\
  \hspace{3mm} 60 \hspace{3mm} & \hspace{3mm} 0\% \hspace{3mm} & \hspace{3mm} 0\% \hspace{3mm}
  &\hspace{3mm} 0\% \hspace{3mm} &\hspace{3mm}  0\% \hspace{3mm} &\hspace{3mm} 0\% \hspace{3mm}\\
  \hspace{3mm} 70 \hspace{3mm} & \hspace{3mm} 0\% \hspace{3mm} & \hspace{3mm} 0\% \hspace{3mm}
  &\hspace{3mm} 0\% \hspace{3mm} &\hspace{3mm}  0\% \hspace{3mm} &\hspace{3mm} 0\% \hspace{3mm}\\
  \hline
\end{tabular}
\end{table}

\begin{table}[!h]
\caption{Computational performance of ``DAG + $\textbf{Reduce}_{1/2}$".}
\label{table:alg:modification_2+DIS}
\vspace{2mm}
\centering
\begin{tabular}{c||c|c|c}
  \hline
  $n$ & DAG + $\textbf{Reduce}_{1/2}$ & Average value of valid $t$ searched & $M$  \\\hline\hline
  \hspace{3mm} 16 \hspace{3mm}  & \hspace{3mm} 100\% \hspace{3mm} & \hspace{3mm} 4.939 \hspace{3mm} & \hspace{3mm} $10^3$ \hspace{3mm}\\
  \hspace{3mm} 20 \hspace{3mm} & \hspace{3mm} 100\%\hspace{3mm} & \hspace{3mm} 4.875 \hspace{3mm} & \hspace{3mm} $10^4$ \hspace{3mm}\\
  \hspace{3mm} 26 \hspace{3mm} & \hspace{3mm} 100\% \hspace{3mm} & \hspace{3mm} 6.682 \hspace{3mm} & \hspace{3mm} $10^4$ \hspace{3mm}\\
  \hspace{3mm} 30 \hspace{3mm} & \hspace{3mm} 100\% \hspace{3mm} & \hspace{3mm} 22.222  \hspace{3mm} & \hspace{3mm} $10^4$ \hspace{3mm}\\
  \hspace{3mm} 36 \hspace{3mm} & \hspace{3mm} 100\%\hspace{3mm} &
  \hspace{3mm} 76.354 \hspace{3mm} & \hspace{3mm} $10^5$ \hspace{3mm}\\
  \hspace{3mm} 40 \hspace{3mm} & \hspace{3mm} 100\%\hspace{3mm} &
  \hspace{3mm} 216.899 \hspace{3mm} & \hspace{3mm} $10^5$ \hspace{3mm}\\
 \hline
\end{tabular}
\end{table}

\begin{table}[!h]
\caption{Computational performance of ``DAG + \textbf{CJLOSS-Alg}".}
\label{table:alg:CJ+DIS}
\vspace{2mm}
\centering
\begin{tabular}{c||c|c|c}
  \hline
  $n$ & DAG + \textbf{CJLOSS-Alg} & Average value of valid $t$ searched & $M$  \\\hline\hline
  \hspace{3mm} 20 \hspace{3mm} & \hspace{3mm} 100\% \hspace{3mm} &
  \hspace{3mm} 1.000 \hspace{3mm} & \hspace{3mm} $10^4$ \hspace{3mm}\\
  \hspace{3mm} 26 \hspace{3mm} & \hspace{3mm} 100\% \hspace{3mm} &
  \hspace{3mm} 2.125 \hspace{3mm} & \hspace{3mm} $10^4$ \hspace{3mm}\\
  \hspace{3mm} 30 \hspace{3mm} & \hspace{3mm} 100\% \hspace{3mm} &
  \hspace{3mm}4.310  \hspace{3mm} & \hspace{3mm} $10^4$ \hspace{3mm}\\
  \hspace{3mm} 36 \hspace{3mm} & \hspace{3mm} 100\% \hspace{3mm} &
  \hspace{3mm} 22.200 \hspace{3mm} & \hspace{3mm} $10^5$ \hspace{3mm}\\
  \hspace{3mm} 40 \hspace{3mm} & \hspace{3mm} 100\% \hspace{3mm} &
  \hspace{3mm} 114.141 \hspace{3mm} & \hspace{3mm} $10^5$ \hspace{3mm}\\
 \hline
\end{tabular}
\end{table}

\newpage
\qquad
Table~\ref{table:1_comparison_lattice_attacks} reports the computational performance of variant
lattice attacks including the revisited methods proposed in the literature
and the two algorithms proposed in this paper, where
column 2 to column 6 record the success ratio when calling different lattice attack
algorithms.

\qquad
Table~\ref{table:alg:modification_2+DIS} and
Table~\ref{table:alg:CJ+DIS} report the computational performance of
integrating modular disaggregation technique with algorithms
$\textbf{Reduce}_{1/2}$
and $\textbf{CJLOSS-Alg}$,
respectively.
The reason of choosing these two lattice attack algorithms
to integrate with modular disaggregation technique
is that based on data reported in
Table~\ref{table:1_comparison_lattice_attacks},
they are with the better performance among variant lattice attack
algorithms. We would like to further enhance their performance via
integrating with module disaggregation technique.

\qquad
Specifically,
in both Table~\ref{table:alg:modification_2+DIS} and
Table~\ref{table:alg:CJ+DIS}:
1).~Column 1 records the number of unknown variables;
2).~Column 2 with algorithm's name records the success ratio
which is the number of successful problems divided by the total number of tested problems;
3).~Column 3 records the average value of valid $t$ that have been searched, only concerning initially failed problems;
4).~Column 4 records the fixed value of $M$.
Note that, the value of $\frac{t}{M}$ is just the value of parameter $r$ introduced in Section 3.2.

\qquad
The codes used in this section are implemented in C++ computer language,
utilizing packages in the C++ library named
NTL (see \cite{NTL}) which is the most cutting-edge library for doing number theory
and for dealing with arbitrarily large integer numbers.
For parameter setting, we set
$N = 10^8$ in the matrices $B_{CJLOSS}$ and $B$
defined in Equation~\eqref{eq:matrix_B_CJ}
and Equation~\eqref{eq:4_modification_1}, respectively,
and $\alpha = 99/100$ in the LLL algorithm (refer to Algorithm~\ref{alg:LLL} in this paper).

\subsection{Systems with Multiple Subset-Sum Equations}

\begin{figure}[H]
\begin{center}
    \begin{tikzpicture}[scale=0.8]
    \draw (40pt,70pt) node{{\color{Blue}\footnotesize{Input a system, $A\emph{\textbf{x}}=\emph{\textbf{b}}$}}};
    \draw [->, rounded corners, thick, black] (40pt, 60pt)--(40pt, 22pt);
    \draw[blue, thick] (0pt,0pt) rectangle (80pt, 20pt);
    \draw (40pt,10pt) node{\footnotesize{Lattice Attack}};
    \draw[blue, thick] (200pt,0pt) rectangle (280pt, 20pt);
    \draw (240pt,10pt) node{\footnotesize{Disaggregation}};
    \draw[blue, thick] (200pt,-80pt) rectangle (280pt, -60pt);
    \draw (240pt,-70pt) node{\footnotesize{Search {\color{Red}$r:=\frac{t}{M}$}}};
    \draw[blue, thick] (100pt,-160pt) rectangle (180pt, -140pt);
    \draw (140pt,-150pt) node{\footnotesize{Return}};
    \draw [->, rounded corners, thick, black] (140pt,-162pt)--(140pt, -200pt);
    \draw (140pt, -210pt) node{{\color{Red}\footnotesize{Output a binary solution}}};
    \draw [->, rounded corners, thick, DarkRed] (40pt, 0pt)--(40pt, -150pt)--(100pt, -150pt);
    \draw [->, rounded corners, thick, DarkRed] (230pt, -80pt)--(230pt, -150pt)--(180pt, -150pt);
    \draw [->, rounded corners, thick, DarkGreen] (80pt, 10pt)--(200pt, 10pt);
    \draw [->, rounded corners, thick, DarkGreen] (240pt, 0pt)--(240pt, -60pt);
    \draw [DarkGreen, thick] (250pt, -80pt) to [out=-90,in=180] (290pt, -120pt);
    \draw [DarkGreen, thick] (290pt, -120pt) to [out=0,in=-90] (310pt, -80pt);
    \draw [->, DarkGreen, thick] (310pt, -80pt) to [out=90,in=0] (280pt, -70pt);
    \draw (17pt,-70pt) node{{\color{Red}\footnotesize{Succeed}}};
    \draw (207pt,-110pt) node{{\color{Red}\footnotesize{Succeed}}};
    \draw (140pt,15pt) node{{\color{Blue}\footnotesize{Fail}}};
    \draw (297pt,-100pt) node{{\color{Blue}\footnotesize{Fail}}};
    \end{tikzpicture}
\caption{Illustration for the process of ``DAG + lattice attack".}
\label{fig2:Illustrate_DAG+LatticeMethod}
\end{center}
\end{figure}
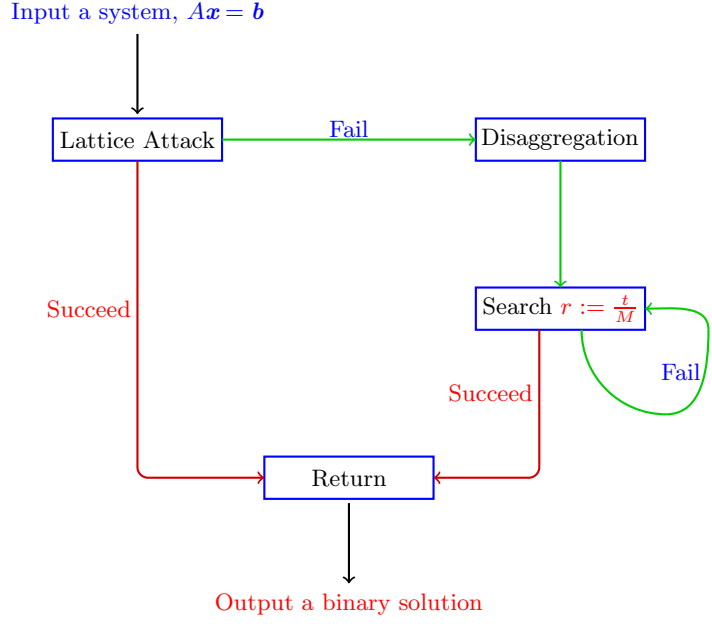

In this section, systems with multiple hard subset-sum equations are further tested,
which are formulated in Problem~\eqref{eq:1_prob_main}.
These systems are with $m$ equations and $n$ unknown variables.
We invoke again the procedure of integrating our modular disaggregation technique
(DAG) with lattice attack algorithms \textbf{CJLOSS-Alg} and $\textbf{Reduce}_{1/2}$,
respectively. Figure~\ref{fig2:Illustrate_DAG+LatticeMethod} is used to illustrate the procedure.
Later, numerical results are summarized in
Table~\ref{table:2_m_n_lattice_attacks},
Table~\ref{table:3_alg:modification_2+DAG},
and Table~\ref{table:4alg:CJ+DIS}.
Our numerical experiment confirms that the success ratio
of finding valid binary solutions to the systems with multiple hard subset-sum equations
can increase dramatically, by applying this integration procedure.

\qquad
The designation method of how to randomly generate systems with multiple subset-sum equations,
which are in the form of Problem~\eqref{eq:1_prob_main}, is described as follows,
\begin{enumerate}
\vspace{-2mm}
    \item [1).] 100 systems are randomly generated for each fixed dimension $(m,n)$;
    \item [2).] $A_{i,j}$, $i=1,2,\ldots,m$, $j=1,2,\ldots,n$, follows discrete uniform distribution on the interval $[1, 2^n]$, where $A_{i,j}$ is the $i$th row
        and $j$th column entry of $A$;
    \item [3).] $\emph{\textbf{x}}$ is randomly generated with cardinality $n/2$, which is fixed
    for the systems with the same dimension $n$;
    \item [4).] $density_{i} := \frac{n}{\max_{1\leq j\leq n}(\log_{2}A_{i,j})} \in (0.99,1.00)$, $i=1,2,\ldots,m$;
    \item [5).] $\emph{\textbf{b}} := A\emph{\textbf{x}}$, which satisfies that $b_i>\max(A_i)$
    and $b_i\leq \text{sum}(A_i)/2$, $i=1,2,\ldots,m$,
    where $b_i$ is the $i$th entry of $\emph{\textbf{b}}$
    and $A_i$ is the $i$th row of $A$.
\end{enumerate}


\newpage

\begin{table}[!h]
\caption{Systems with multiple subset-sum equations.}
\label{table:2_m_n_lattice_attacks}
\vspace{2mm}
\centering
\begin{threeparttable}
\begin{tabular}{c|c||c|c}
  \hline
  $m$ & $n$ & $\textbf{Reduce}_{1/2}$ & \textbf{CJLOSS-Alg} \\\hline\hline
  \hspace{3mm} 2 \hspace{3mm} & \hspace{3mm} 30 \hspace{3mm} & \hspace{3mm} 100\% \hspace{3mm} & \hspace{3mm} 100\% \hspace{3mm}\\
    \hspace{3mm} 2 \hspace{3mm} & \hspace{3mm} 40 \hspace{3mm} & \hspace{3mm} 94\% \hspace{3mm} & \hspace{3mm} 100\% \hspace{3mm}\\
    \hspace{3mm} 2 \hspace{3mm} & \hspace{3mm} 50 \hspace{3mm} & \hspace{3mm} 59\% \hspace{3mm} & \hspace{3mm} 100\% \hspace{3mm}\\
    \hspace{3mm} 2 \hspace{3mm} & \hspace{3mm} 60 \hspace{3mm} & \hspace{3mm} 24\% \hspace{3mm} & \hspace{3mm} 84\% \hspace{3mm}\\
    \hspace{3mm} 2 \hspace{3mm} & \hspace{3mm} 70 \hspace{3mm} & \hspace{3mm} 5\% \hspace{3mm} & \hspace{3mm} 30\% \hspace{3mm}\\
    \hspace{3mm} 2 \hspace{3mm} & \hspace{3mm} 80 \hspace{3mm} & \hspace{3mm} 0\% \hspace{3mm} & \hspace{3mm} 3\% \hspace{3mm}\\
    \hspace{3mm} 2 \hspace{3mm} & \hspace{3mm} 90 \hspace{3mm} & \hspace{3mm} 0\% \hspace{3mm} & \hspace{3mm} 0\% \hspace{3mm}\\
    \hspace{3mm} 2 \hspace{3mm} & \hspace{3mm} 100 \hspace{3mm} & \hspace{3mm} 0\% \hspace{3mm} & \hspace{3mm} 0\% \hspace{3mm}\\
  \hline
  \hspace{3mm} 3 \hspace{3mm} & \hspace{3mm} 40 \hspace{3mm} & \hspace{3mm} 100\% \hspace{3mm} & \hspace{3mm} 100\% \hspace{3mm}\\
  \hspace{3mm} 3 \hspace{3mm} & \hspace{3mm} 50 \hspace{3mm} & \hspace{3mm} 100\% \hspace{3mm} & \hspace{3mm} 100\% \hspace{3mm}\\
  \hspace{3mm} 3 \hspace{3mm} & \hspace{3mm} 60 \hspace{3mm} & \hspace{3mm} 99\% \hspace{3mm} & \hspace{3mm} 100\% \hspace{3mm}\\
  \hspace{3mm} 3 \hspace{3mm} & \hspace{3mm} 70 \hspace{3mm} & \hspace{3mm} 97\% \hspace{3mm} & \hspace{3mm} 100\% \hspace{3mm}\\
  \hspace{3mm} 3 \hspace{3mm} & \hspace{3mm} 80 \hspace{3mm} & \hspace{3mm} 82\% \hspace{3mm} & \hspace{3mm} 100\% \hspace{3mm}\\
  \hspace{3mm} 3 \hspace{3mm} & \hspace{3mm} 90 \hspace{3mm} & \hspace{3mm} 47\% \hspace{3mm} & \hspace{3mm} 94\% \hspace{3mm}\\
  \hspace{3mm} 3 \hspace{3mm} & \hspace{3mm} 100 \hspace{3mm} & \hspace{3mm} 10\% \hspace{3mm} & \hspace{3mm} 55\% \hspace{3mm}\\
  \hline
  \hspace{3mm} 4 \hspace{3mm} & \hspace{3mm} 60 \hspace{3mm} & \hspace{3mm} 100\% \hspace{3mm} & \hspace{3mm} 100\% \hspace{3mm}\\
  \hspace{3mm} 4 \hspace{3mm} & \hspace{3mm} 70 \hspace{3mm} & \hspace{3mm} 100\% \hspace{3mm} & \hspace{3mm} 100\% \hspace{3mm}\\
  \hspace{3mm} 4 \hspace{3mm} & \hspace{3mm} 80 \hspace{3mm} & \hspace{3mm} 100\% \hspace{3mm} & \hspace{3mm} 100\% \hspace{3mm}\\
  \hspace{3mm} 4 \hspace{3mm} & \hspace{3mm} 90 \hspace{3mm} & \hspace{3mm} 100\% \hspace{3mm} & \hspace{3mm} 100\% \hspace{3mm}\\
  \hspace{3mm} 4 \hspace{3mm} & \hspace{3mm} 100 \hspace{3mm} & \hspace{3mm} 100\% \hspace{3mm} & \hspace{3mm} 100\% \hspace{3mm}\\
  \hline
  \hspace{3mm} 5 \hspace{3mm} & \hspace{3mm} 100 \hspace{3mm} & \hspace{3mm} 100\% \hspace{3mm} & \hspace{3mm} 100\% \hspace{3mm}\\
  \hspace{3mm} 6 \hspace{3mm} & \hspace{3mm} 100 \hspace{3mm} & \hspace{3mm} 100\% \hspace{3mm} & \hspace{3mm} 100\% \hspace{3mm}\\
  \hspace{3mm} 7 \hspace{3mm} & \hspace{3mm} 100 \hspace{3mm} & \hspace{3mm} 100\% \hspace{3mm} & \hspace{3mm} 100\% \hspace{3mm}\\
  \hspace{3mm} 8 \hspace{3mm} & \hspace{3mm} 100 \hspace{3mm} & \hspace{3mm} 100\% \hspace{3mm} & \hspace{3mm} ${100\%}^*$ \hspace{3mm}\\
  \hspace{3mm} 9 \hspace{3mm} & \hspace{3mm} 100 \hspace{3mm} & \hspace{3mm} ${100\%}^*$ \hspace{3mm} & \hspace{3mm} ${100\%}^*$  \hspace{3mm}\\
  \hspace{3mm} 10 \hspace{3mm} & \hspace{3mm} 100 \hspace{3mm} & \hspace{3mm} ${100\%}^*$ \hspace{3mm} & \hspace{3mm} ${100\%}^*$  \hspace{3mm}\\
  \hline
  \hline
\end{tabular}
\begin{tablenotes}
        \footnotesize
        \item Note: Numbers with superscript $^*$ mean that, these success ratios are obtained by logical inference, since systems with the same $n$, but with less equations, i.e., smaller $m$, can achieve success ratio 100\%.
      \end{tablenotes}
\end{threeparttable}
\end{table}

\newpage

\begin{table}[!h]
\caption{Computational performance of ``DAG + $\textbf{Reduce}_{1/2}$".}
\label{table:3_alg:modification_2+DAG}
\vspace{2mm}
\centering
\begin{threeparttable}
\begin{tabular}{c|c||c|c|c}
  \hline
  $m$ & $n$ & DAG + $\textbf{Reduce}_{1/2}$ & Average value of valid $t$ searched & M  \\\hline\hline
   \hspace{3mm} 2 \hspace{3mm}  & \hspace{3mm} 40 \hspace{3mm}  & \hspace{3mm} 100\%\tnote{*} \hspace{3mm} & \hspace{3mm} 1.000 \hspace{3mm} & \hspace{3mm} $10^5$ \hspace{3mm}\\
   \hspace{3mm} 2 \hspace{3mm}  & \hspace{3mm} 50 \hspace{3mm}  & \hspace{3mm} 100\%\tnote{*} \hspace{3mm} & \hspace{3mm} 1.756 \hspace{3mm} & \hspace{3mm} $10^5$ \hspace{3mm}\\
   \hspace{3mm} 2 \hspace{3mm}  & \hspace{3mm} 60 \hspace{3mm}  & \hspace{3mm} 100\%\tnote{*} \hspace{3mm} & \hspace{3mm} 7.184 \hspace{3mm} & \hspace{3mm} $10^5$ \hspace{3mm}\\
   \hspace{3mm} 2 \hspace{3mm}  & \hspace{3mm} 70 \hspace{3mm}  & \hspace{3mm} 100\%\tnote{*} \hspace{3mm} & \hspace{3mm} 66.189
 \hspace{3mm} & \hspace{3mm} $10^5$ \hspace{3mm}\\
   \hspace{3mm} 2 \hspace{3mm}  & \hspace{3mm} 80 \hspace{3mm}  & \hspace{3mm} --\%\tnote{**} \hspace{3mm} & \hspace{3mm}  --\tnote{**} \hspace{3mm} & \hspace{3mm} $10^5$ \hspace{3mm}\\
   \hspace{3mm} 2 \hspace{3mm}  & \hspace{3mm} 90 \hspace{3mm}  & \hspace{3mm} --\%\tnote{**} \hspace{3mm} & \hspace{3mm}  --\tnote{**} \hspace{3mm} & \hspace{3mm} $10^5$ \hspace{3mm}\\
   \hspace{3mm} 2 \hspace{3mm}  & \hspace{3mm} 100 \hspace{3mm}  & \hspace{3mm} --\%\tnote{**} \hspace{3mm} & \hspace{3mm}  --\tnote{**} \hspace{3mm} & \hspace{3mm} $10^5$ \hspace{3mm}\\
   \hline
   \hspace{3mm} 3 \hspace{3mm}  & \hspace{3mm} 60 \hspace{3mm}  & \hspace{3mm} 100\%\tnote{*} \hspace{3mm} & \hspace{3mm} 1.000 \hspace{3mm} & \hspace{3mm} $10^5$ \hspace{3mm}\\
   \hspace{3mm} 3 \hspace{3mm}  & \hspace{3mm} 70 \hspace{3mm}  & \hspace{3mm} 100\%\tnote{*} \hspace{3mm} & \hspace{3mm} 1.000 \hspace{3mm} & \hspace{3mm} $10^5$ \hspace{3mm}\\
   \hspace{3mm} 3 \hspace{3mm}  & \hspace{3mm} 80 \hspace{3mm}  & \hspace{3mm} 100\%\tnote{*} \hspace{3mm} & \hspace{3mm} 1.111 \hspace{3mm} & \hspace{3mm} $10^5$ \hspace{3mm}\\
   \hspace{3mm} 3 \hspace{3mm}  & \hspace{3mm} 90 \hspace{3mm}  & \hspace{3mm}  100\%\tnote{*} \hspace{3mm} & \hspace{3mm} 3.623 \hspace{3mm} & \hspace{3mm} $10^5$ \hspace{3mm}\\
 \hline
 \hline
\end{tabular}
\begin{tablenotes}
        \footnotesize
        \item
        Note: Numbers with superscript $^*$ tell that, one new equation is generated via disaggregation, and then is added to the original system. Normally, the new equation is generated based on the first equation of the original system.
        \item
        Note: Numbers with superscript $^{**}$ tell that, since lattice attack to all the original systems fail initially, this implies that these systems are more difficult to be solved.
      \end{tablenotes}
\end{threeparttable}
\end{table}

\begin{table}[!h]
\caption{Computational performance of ``DAG + \textbf{CJLOSS-Alg}".}
\label{table:4alg:CJ+DIS}
\vspace{2mm}
\centering
\begin{threeparttable}
\begin{tabular}{c|c||c|c|c}
  \hline
  $m$ & $n$ & DAG + \textbf{CJLOSS-Alg} & Average value of valid $t$ searched & M  \\\hline\hline
  \hspace{3mm} 2 \hspace{3mm} & \hspace{3mm} 60 \hspace{3mm} & \hspace{3mm}  100\%\tnote{*} \hspace{3mm} &
  \hspace{3mm}  2.813 \hspace{3mm} & \hspace{3mm} $10^5$ \hspace{3mm}\\
  \hspace{3mm} 2 \hspace{3mm} & \hspace{3mm} 70 \hspace{3mm} & \hspace{3mm} 100\%\tnote{*} \hspace{3mm} &
  \hspace{3mm} 45.714 \hspace{3mm} & \hspace{3mm} $10^5$ \hspace{3mm}\\
  \hspace{3mm} 2 \hspace{3mm} & \hspace{3mm} 80 \hspace{3mm} & \hspace{3mm} --\%\tnote{**} \hspace{3mm} &
  \hspace{3mm}  --\tnote{**} \hspace{3mm} & \hspace{3mm} $10^5$ \hspace{3mm}\\
  \hspace{3mm} 2 \hspace{3mm} & \hspace{3mm} 90 \hspace{3mm} & \hspace{3mm} --\%\tnote{**} \hspace{3mm} &
  \hspace{3mm}  --\tnote{**} \hspace{3mm} & \hspace{3mm} $10^5$ \hspace{3mm}\\
  \hspace{3mm} 2 \hspace{3mm} & \hspace{3mm} 100 \hspace{3mm} & \hspace{3mm} --\%\tnote{**} \hspace{3mm} &
  \hspace{3mm}  --\tnote{**} \hspace{3mm} & \hspace{3mm} $10^5$ \hspace{3mm}\\
 \hline
   \hspace{3mm} 3 \hspace{3mm} & \hspace{3mm} 90 \hspace{3mm} & \hspace{3mm} 100\%\tnote{*} \hspace{3mm} &
  \hspace{3mm} 1.500 \hspace{3mm} & \hspace{3mm} $10^5$ \hspace{3mm}\\
  \hspace{3mm} 3 \hspace{3mm} & \hspace{3mm} 100 \hspace{3mm} & \hspace{3mm} 100\%\tnote{*} \hspace{3mm} &
  \hspace{3mm} 16.968 \hspace{3mm} & \hspace{3mm} $10^5$ \hspace{3mm}\\
 \hline
 \hline
\end{tabular}
\begin{tablenotes}
        \footnotesize
        \item
        Note: Numbers with superscript $^*$ tell that, one new equation is generated via disaggregation, and then is added to the original system. Normally, the new equation is generated based on the first equation of the original system.
        \item
        Note: Numbers with superscript $^{**}$ tell that, since lattice attack to all the original systems fail initially, this implies that these systems are more difficult to be solved.
      \end{tablenotes}
\end{threeparttable}
\end{table}

\newpage
\qquad
Table~\ref{table:2_m_n_lattice_attacks}
records the success ratios
of Algorithm~$\textbf{Reduce}_{1/2}$ and
Algorithm~$\textbf{CJLOSS-Alg}$, respectively, i.e., the number of systems which successfully
return a binary solution divided by the total number of randomly generated systems with fixed dimension.
We use these two algorithms to test systems with multiple subset-sum equations,
as they are
with better performance among variant lattice attack algorithms reported in
Table~\ref{table:1_comparison_lattice_attacks}. Therefore, these two lattice attack
algorithms are used as benchmarks.
The number of unknown variables, $n$, has been tested up to 100,
where the coefficients of the tested systems are already as large as $2^n=2^{100}$.

\qquad
Table~\ref{table:3_alg:modification_2+DAG}
and Table~\ref{table:4alg:CJ+DIS}
report the computational performance of
integrating modular disaggregation technique with algorithms
$\textbf{Reduce}_{1/2}$
and $\textbf{CJLOSS-Alg}$,
respectively.
Specifically,
in both Table~\ref{table:3_alg:modification_2+DAG}
and Table~\ref{table:4alg:CJ+DIS}:
1).~Column 1 and Column 2 record the dimension of systems;
2).~Column 3 with algorithm's name records the success ratio
which is the number of successful systems divided by the total number of tested systems;
3).~Column 4 records the average value of valid $t$ that have been searched, only concerning initially failed systems;
4).~Column 5 records the fixed value of $M$.
Note that, the value of $\frac{t}{M}$ is just the value of parameter $r$ introduced in Section 3.2.


\qquad
Similarly to that of Section~4.1, the codes used in this section are implemented in C++ computer language,
utilizing packages in the C++ library named
NTL (see \cite{NTL}) which is the most cutting-edge library for doing number theory
and for dealing with arbitrarily large integer numbers.
For parameter setting, we set
$N = 10^8$ in the matrices $B_{CJLOSS}$ and $B$
defined in Equation~\eqref{eq:matrix_B_CJ}
and Equation~\eqref{eq:4_modification_1}, respectively,
and $\alpha = 99/100$ in the LLL algorithm (refer to Algorithm~\ref{alg:LLL} in this paper).

\section{Statistical Analysis of Numerical Tests}\label{sec:analysis}
Numerical tests in the previous section
exhibit the efficiency
of the algorithm that integrates modular disaggregation techniques with lattice attack algorithm.
In this section conjectures are proposed to explain this efficiency,
and we also try to partially confirm our conjectures via simulation.
Recall that, the vector $\emph{\textbf{v}}$ introduced in
Section~3.2 depends on the parameter $r$.

\begin{conjec}\label{conjecture 1}
\begin{itemize}
  \item [(a)]
  If after disaggregation, set $\circled{3}|r$, which are determined by parameter $r$ introduced in Section~3.2, can cut-off the first returned non-binary integer solution $\tilde{\textbf{x}}$
  with small Euclidean length, then
  the probability of returning a valid binary solution $\textbf{x}^*$ increases.
  \item [(b)]
  After disaggregation, whether set $\circled{3}|r$ can cut-off the
  non-binary integer solution
  $\tilde{\textbf{x}}$ with small Euclidean length, depends on the structure of
  the corresponding new kernel lattice ${\rm ker}_{\Z}(\textbf{a}^T, \textbf{v}^T)^T)$.
  Note that, the concept of kernel lattice has been defined in item (a) of Theorem~\ref{thm:matrix B_reduced}, which has also been studied in \cite{aardal2013structure}.
  \item [(c)]
   Regarding item (b), to be even more specific, we conjecture that if the kernel lattice ${\rm ker}_{\Z}((\textbf{a}^T, \textbf{v}^T)^T)$ is
  sparser and more rectangular, then
  the corresponding set $\circled{3}|r$ can cut-off the non-binary solution $\tilde{\textbf{x}}$ with small Euclidean length more easily.
\end{itemize}
\end{conjec}

\subsection{Volume and Minimum Volume Ellipsoid of a Lattice}\label{deduc_ellipsoid}

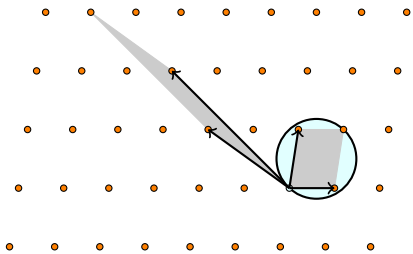
\begin{figure}[H]
\begin{center}
    \begin{tikzpicture}[scale=0.6]
    \draw [-, thick, fill=LightCyan](7.8,1.95) circle(0.88459);
    \draw [Gray, fill=Gray] (7.2,1.3)--(8.2,1.3)--(8.4,2.6)--(7.4,2.6);
    \draw [Gray, fill=Gray] (7.2,1.3)--(4.6,3.9)--(2.8,5.2)--(5.4,2.6);
    \draw [fill=orange](1,0) circle(2pt);
    \draw [fill=orange](2,0) circle(2pt);
    \draw [fill=orange](3,0) circle(2pt);
    \draw [fill=orange](4,0) circle(2pt);
    \draw [fill=orange](5,0) circle(2pt);
    \draw [fill=orange](6,0) circle(2pt);
    \draw [fill=orange](7,0) circle(2pt);
    \draw [fill=orange](8,0) circle(2pt);
    \draw [fill=orange](9,0) circle(2pt);
    \draw [fill=orange](1.2,1.3) circle(2pt);
    \draw [fill=orange](2.2,1.3) circle(2pt);
    \draw [fill=orange](3.2,1.3) circle(2pt);
    \draw [fill=orange](4.2,1.3) circle(2pt);
    \draw [fill=orange](5.2,1.3) circle(2pt);
    \draw [fill=orange](6.2,1.3) circle(2pt);
    \draw [fill=LightCyan](7.2,1.3) circle(2pt); 
    \draw [fill=orange](8.2,1.3) circle(2pt);
    \draw [fill=orange](9.2,1.3) circle(2pt);
    \draw [fill=orange](1.4,2.6) circle(2pt);
    \draw [fill=orange](2.4,2.6) circle(2pt);
    \draw [fill=orange](3.4,2.6) circle(2pt);
    \draw [fill=orange](4.4,2.6) circle(2pt);
    \draw [fill=orange](5.4,2.6) circle(2pt);
    \draw [fill=orange](6.4,2.6) circle(2pt);
    \draw [fill=orange](7.4,2.6) circle(2pt);
    \draw [fill=orange](8.4,2.6) circle(2pt);
    \draw [fill=orange](9.4,2.6) circle(2pt);
    \draw [fill=orange](1.6,3.9) circle(2pt);
    \draw [fill=orange](2.6,3.9) circle(2pt);
    \draw [fill=orange](3.6,3.9) circle(2pt);
    \draw [fill=orange](4.6,3.9) circle(2pt);
    \draw [fill=orange](5.6,3.9) circle(2pt);
    \draw [fill=orange](6.6,3.9) circle(2pt);
    \draw [fill=orange](7.6,3.9) circle(2pt);
    \draw [fill=orange](8.6,3.9) circle(2pt);
    \draw [fill=orange](9.6,3.9) circle(2pt);
    \draw [fill=orange](1.8,5.2) circle(2pt);
    \draw [fill=orange](2.8,5.2) circle(2pt);
    \draw [fill=orange](3.8,5.2) circle(2pt);
    \draw [fill=orange](4.8,5.2) circle(2pt);
    \draw [fill=orange](5.8,5.2) circle(2pt);
    \draw [fill=orange](6.8,5.2) circle(2pt);
    \draw [fill=orange](7.8,5.2) circle(2pt);
    \draw [fill=orange](8.8,5.2) circle(2pt);
    \draw [fill=orange](9.8,5.2) circle(2pt);
    \draw [->, thick] (7.2,1.3)--(8.2,1.3);
    \draw [->, thick] (7.2,1.3)--(7.4,2.6);
    \draw [->, thick] (7.2,1.3)--(4.6,3.9);
    \draw [->, thick] (7.2,1.3)--(5.4,2.6);
    \end{tikzpicture}
\caption{Minimum volume ellipsoid of a basis of kernel lattice.}
\label{fig1:lattice_MinVolumeEllipsoid}
\end{center}
\end{figure}

\begin{figure}[H]
\begin{center}
    \begin{tikzpicture}[scale=0.6]
    \draw[rotate = -38.7] [-, thick, fill=LightCyan](1.95, 5.70) ellipse(2.93981 and 0.390); 
    \draw [Gray, fill=Gray] (7.2,1.3)--(4.6,3.9)--(2.8,5.2)--(5.4,2.6);
    \draw [Gray, fill=Gray] (7.2,1.3)--(8.2,1.3)--(8.4,2.6)--(7.4,2.6);
    \draw[rotate = -38.7] [-, thick](1.95, 5.70) ellipse(2.93981 and 0.390); 
    \draw [fill=orange](1,0) circle(2pt);
    \draw [fill=orange](2,0) circle(2pt);
    \draw [fill=orange](3,0) circle(2pt);
    \draw [fill=orange](4,0) circle(2pt);
    \draw [fill=orange](5,0) circle(2pt);
    \draw [fill=orange](6,0) circle(2pt);
    \draw [fill=orange](7,0) circle(2pt);
    \draw [fill=orange](8,0) circle(2pt);
    \draw [fill=orange](9,0) circle(2pt);
    \draw [fill=orange](1.2,1.3) circle(2pt);
    \draw [fill=orange](2.2,1.3) circle(2pt);
    \draw [fill=orange](3.2,1.3) circle(2pt);
    \draw [fill=orange](4.2,1.3) circle(2pt);
    \draw [fill=orange](5.2,1.3) circle(2pt);
    \draw [fill=orange](6.2,1.3) circle(2pt);
    \draw [fill=LightCyan](7.2,1.3) circle(2pt); 
    \draw [fill=orange](8.2,1.3) circle(2pt);
    \draw [fill=orange](9.2,1.3) circle(2pt);
    \draw [fill=orange](1.4,2.6) circle(2pt);
    \draw [fill=orange](2.4,2.6) circle(2pt);
    \draw [fill=orange](3.4,2.6) circle(2pt);
    \draw [fill=orange](4.4,2.6) circle(2pt);
    \draw [fill=orange](5.4,2.6) circle(2pt);
    \draw [fill=orange](6.4,2.6) circle(2pt);
    \draw [fill=orange](7.4,2.6) circle(2pt);
    \draw [fill=orange](8.4,2.6) circle(2pt);
    \draw [fill=orange](9.4,2.6) circle(2pt);
    \draw [fill=orange](1.6,3.9) circle(2pt);
    \draw [fill=orange](2.6,3.9) circle(2pt);
    \draw [fill=orange](3.6,3.9) circle(2pt);
    \draw [fill=orange](4.6,3.9) circle(2pt);
    \draw [fill=orange](5.6,3.9) circle(2pt);
    \draw [fill=orange](6.6,3.9) circle(2pt);
    \draw [fill=orange](7.6,3.9) circle(2pt);
    \draw [fill=orange](8.6,3.9) circle(2pt);
    \draw [fill=orange](9.6,3.9) circle(2pt);
    \draw [fill=orange](1.8,5.2) circle(2pt);
    \draw [fill=orange](2.8,5.2) circle(2pt);
    \draw [fill=orange](3.8,5.2) circle(2pt);
    \draw [fill=orange](4.8,5.2) circle(2pt);
    \draw [fill=orange](5.8,5.2) circle(2pt);
    \draw [fill=orange](6.8,5.2) circle(2pt);
    \draw [fill=orange](7.8,5.2) circle(2pt);
    \draw [fill=orange](8.8,5.2) circle(2pt);
    \draw [fill=orange](9.8,5.2) circle(2pt);
    \draw [->, thick] (7.2,1.3)--(8.2,1.3);
    \draw [->, thick] (7.2,1.3)--(7.4,2.6);
    \draw [->, thick] (7.2,1.3)--(4.6,3.9);
    \draw [->, thick] (7.2,1.3)--(5.4,2.6);
    \end{tikzpicture}
\caption{Minimum volume ellipsoid of another basis of kernel lattice.}
\label{fig3:lattice_MinVolumeEllipsoid}
\end{center}
\end{figure}
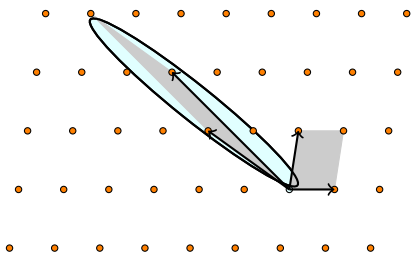

In order to further study and confirm the conjectures, we first
explain the concept of the volume of a lattice, and the concept of the minimum volume ellipsoid of a lattice. These concepts are
explained geometrically in Figure~\ref{fig1:lattice_MinVolumeEllipsoid}
and Figure~\ref{fig3:lattice_MinVolumeEllipsoid} as well.

\begin{defn}[see Definition~1.9 in \cite{Murray:2011}]
  Given that columns of matrix $D$ consist a basis of lattice $\mathcal{L}$, then the volume
  of lattice $\mathcal{L}$ is defined as follows,
  $$
  {\rm }vol(\mathcal{L}) = \sqrt{\det(D^T D)}.
  $$
\end{defn}

 \qquad
Next we study the minimum volume ellipsoid of a given lattice with basis $b_1, b_2, \ldots, b_m\in\R^{n}$ with $n>m$.
Our idea is that first transforming the lattice to another linear space with dimension $m$, meanwhile keeping the angles of each pair of the basis vector.
Let $D = (b_1~~b_2~~\cdots~~ b_m)\in\R^{n\times {m}}$, then find matrix $U= (u_1~~u_2~~\cdots~~ u_m)\in\R^{n\times {m}}$
and matrix $S = (s_1~~s_2~~\ldots~~ s_m) \in\R^{m\times {m}}$ with $U^TU=I$, i.e.,
the columns of $D$ form the basis of the new linear space that we want to transform the lattice to be within, moreover, we require the following,
\begin{align}\label{eq:1_sec4}
  D = US.
\end{align}

\qquad
We claim that for any such basis matrix $D$, such $U$ and $S$ can always be found to satisfy that $U^TU=I$. For example, the SVD decomposition of $D$
can achieve this goal.
Next we check the angles between each pair of $b_i$ and $b_j$, and each pair of $s_i$ and $s_j$.
Based on Eq.~\eqref{eq:1_sec4}, we have that,
\begin{align}
  b_i = Us_i,~~\forall~i\in\{1,2,\ldots,m\},
\end{align}
which yields that,
\begin{align}
  b_i^T b_j = (Us_i)^T (Us_j) = s_i^T(U^TU)s_j = s_i^T s_j,
\end{align}
and
\begin{align}
  ||b_i||^2 = b_i^Tb_i = s_i^Ts_i = ||s_i||^2,~~\forall~ i,j\in\{1,2,\ldots,m\}.
\end{align}
This directly shows that the angle between $b_i$ and $b_j$ is the same as the angle between $s_i$ and $s_j$.
We only transform the lattice generated by $D = (b_1~~b_2~~\cdots~~ b_m)$ to another linear space, but keep the shape structure of
the lattice all the same.

\qquad
Now instead of studying lattice $\mathcal{L}_D$ generated by $D$ in a dimension $n$ space, we study lattice $\mathcal{L}_S$ generated by $S$ in a dimension $m$ space.
Note that, $\mathcal{L}_D$ and $\mathcal{L}_S$ have the same lattice structure.
Meanwhile,
the minimum volume ellipsoid of lattice $\mathcal{L}_D$ should have the same structure as that of lattice $\mathcal{L}_S$.

\qquad
We adopt the algorithm proposed in Chapter 8.4.1 of \cite{Boyd:2004} 
to calculate the minimum volume ellipsoid of lattice $\mathcal{L}_S$, obtained by transforming $\mathcal{L}_D$ to a lower dimension space. Here $\mathcal{L}_D$ would be the kernel lattice of the new systems after disaggregation.

\qquad
In fact, under fixed dimension, the ratio between the volume of lattice and the volume of minimum volume ellipsoid is always a constant,
i.e.,
\begin{align}
  \gamma := \frac{\text{volume of minimum volume ellipsoid}}{\text{volume of covered lattice}}
  = \frac{m^{m/2}}{2^{m-1}}\frac{1}{m}\frac{\pi^{m/2}}{\Gamma(m/2)}
\end{align}
is always a constant, under the same dimension $m$ of the lattice.
The ratio values are listed in Table~\ref{table:ratio_lattice_ellipsoid} with dimensions from 2 up to 7.

\begin{table}[!ht]
\caption{Relation between volume of minimum volume ellipsoid and its relative covered lattice.}
\label{table:ratio_lattice_ellipsoid}
\vspace{2mm}
\small
\centering
\begin{tabular}{c||c}\hline
    Dimension of Lattice          &
    {\Large$\frac{\text{Volume of minimum volume ellipsoid}}{\text{volume of the covered lattice}}$}\\
    ($m$) & ($\gamma$) \\\hline
  2         &   1.5708  \\\hline 
  3          &  2.7207  \\\hline 
  4          &  4.9348  \\\hline 
  5          &  9.1955  \\\hline 
  6          &  17.4410  \\\hline 
  7          &  33.4976  \\       
  \hline\hline
\end{tabular}
\end{table}

\newpage
\subsection{Analysis on Conjecture (a)}

\subsubsection{Systems with single subset-sum equation}

In order to test conjecture item (a), logistic regressions are conducted on the preliminary small examples
recorded in Table~\ref{table:n6_10problems}, which initially fail to return binary solution
under $\textbf{Reduce}$.
All the possible values of ratio $\frac{t}{M}$ are enumerated, and
the statistical results are recorded in Table~\ref{table:logistic analysis3-1}.
The reason of choosing $\textbf{Reduce}$ in this section is that we need to enumerate all the jump points
of subset-sum problems, thus we want to control the magnitude of dimension $n$, while
for small $n$, it is very difficult to find initially failed problems under
$\textbf{Reduce}_{1/2}$ and \textbf{CJLOSS-Alg}.

\begin{table}[!ht]
\caption{Problems with given $\textbf{a}$ and $b$, dimension $n=6$.}
\label{table:n6_10problems}
\vspace{2mm}
\centering
\begin{tabular}{c||c|c}\hline
  Problem No. &  $\textbf{a}$                            &  $b$  \\\hline\hline
  1           &   (7,  26,  18,  43,  32,  10)  & 57 \\\hline
  2           &   (24,  31,   3,  29,  17,  18) & 44\\\hline
  3           &   (36,  21,   8,  63,  53,  52)  & 97 \\\hline
  4           &   (58,  56,   5,  50,  30,  62)  & 93 \\\hline
  5           &   (15,  55,  37,  11,  13,  43)  & 65 \\\hline
  6           &   (8,  51,  26,  32,  21,  25)  & 55 \\\hline
  7           &   (17,  52,  43,  45,  63,  40)  & 123 \\\hline
  8           &   (8,  39,  47,  35,  48,  63)  & 103 \\\hline
  9           &   (30,   2,   4,  47,  33,  36)  & 67 \\\hline
  10          &   (14,  57,  29,  38,  60,  11) & 103 \\
  \hline\hline
\end{tabular}
\end{table}

\begin{table}[!ht]
\caption{Success or Failure vs. Cut or Non-cut.}
\label{table:logistic analysis3-1}
\vspace{2mm}
\footnotesize
\centering
\begin{threeparttable}
\begin{tabular}{c||c|c}\hline
  \multirow{2}{*}{Problem No.}
  &  \multicolumn{2}{c}{Coefficients ($t$-Value, $p$-Value)}\\\cline{2-3}
  &  Constant          &  Cut or Non-cut     \\\hline
  1           &  0.4249 (1.3622, 0.1731)   &  -2.4200 (-5.9737, 2.3187$\times{10}^{-9}$) \\\hline
  2           &  0.2364 (0.6844, 0.4937) &  -0.4959 (-1.2715, 0.2035) \\\hline
  3           &  -0.0000 (-0.0000, 1.0000) &  -2.2773 (-6.9154, 4.6646$\times{10}^{-12}$) \\\hline
  4           &  0.1542 (0.2771, 0.7817) &  -0.1602 (-0.2824, 0.7776) \\\hline
  5           &  0.7577 (2.4214, 0.0155) &  -2.0070 (-5.5813, 2.3866$\times{10}^{-8}$) \\\hline
  6           &  -1.7918(-4.3889, 1.1393$\times{10}^{-5}$)  &  0.0000 (0.0000, 1.0000) \\\hline
  7           &  2.3979 (3.2468, 0.0012)   &  -1.7430 (-2.3331, 0.0196) \\\hline
  8           &  -0.4169(-1.7430, 0.0813)  &  -2.3844 (-6.6667, 2.6167$\times{10}^{-11}$) \\\hline
  9           &  0.5108 (1.7134, 0.0866)   &  -2.2192 (-6.0147, 1.8025$\times{10}^{-9}$) \\\hline
  10          &  0.1082 (0.4648, 0.6421)   &  -2.2580 (-7.1145, 1.1233$\times{10}^{-12}$) \\
  \hline\hline
\end{tabular}
    \begin{tablenotes}
        \footnotesize
        \item Note: Table~\ref{table:logistic analysis3-1} records
    the logistic regression test result \emph{with} constant,
    which can be compared with Table~\ref{table:logistic analysis3-2}.
    We can observe that \emph{without} constant, the relation
    between `Success', 'Failure' (1 or 0) and `Non-cut', 'Cut' (1 or 0) becomes even more significant.
    \end{tablenotes}
\end{threeparttable}
\end{table}

\begin{table}[!ht]
\caption{Success or Failure vs. Cut or Non-cut.}
\label{table:logistic analysis3-2}
\vspace{2mm}
\footnotesize
\centering
\begin{threeparttable}
\begin{tabular}{c||c|c}\hline
  \multirow{2}{*}{Problem No.}
  &  \multicolumn{2}{c}{Coefficients ($t$-Value, $p$-Value)}\\\cline{2-3}
  &  Constant          &  Cut or Non-cut     \\\hline
  1           &   - &   -1.9951 ( -7.7179,  1.1825$\times{10}^{-14}$) \\\hline
  2           &   - &   -0.2595 (-1.4328, 0.1519) \\\hline
  3           &   - &   -2.2773 ( -10.6247,  2.2868$\times{10}^{-26}$) \\\hline
  4           &   - &   -0.0060 ( -0.0548, 0.9563) \\\hline
  5           &   - &   -1.2493 ( -7.0519, 1.7645$\times{10}^{-12}$) \\\hline
  6           &   - &  -1.7918 ( -7.9556, 1.7834$\times{10}^{-15}$) \\\hline
  7           &   - &   0.6549 ( 5.8202, 5.8790$\times{10}^{-9}$) \\\hline
  8           &   - &   -2.8013 (-10.5344, 5.9977$\times{10}^{-26}$) \\\hline
  9           &   - &   -1.7084 (-7.8596, 3.8540$\times{10}^{-15}$) \\\hline
  10          &   - &   -2.1498 (-9.9673, 2.1189$\times{10}^{-23}$) \\
  \hline\hline
\end{tabular}
    \begin{tablenotes}
        \footnotesize
        \item Note: Table~\ref{table:logistic analysis3-2} records
    the logistic regression test result \emph{without} constant,
    which can be compared with Table~\ref{table:logistic analysis3-1}.
    We can observe that \emph{without} constant, the relation
    between `Success', 'Failure' (1 or 0) and `Non-cut', 'Cut' (1 or 0)' becomes even more significant,
    i.e., cutting-off $\tilde{\textbf{x}}$ strongly implies a success.
    \end{tablenotes}
\end{threeparttable}
\end{table}


\qquad
Note that for every possible scenario of $(t,M)$ pair, `Success'~$:=1$, `Failure'~$:=0$, `Non-cut'~$:=1$,
and `Cut'~$:=0$, where `Cut' means cutting off the initially returned
non-binary solution $\tilde{\textbf{x}}$, and `Noncut' means not cutting off $\tilde{\textbf{x}}$.
Based on data recorded in Table~\ref{table:logistic analysis3-1} and
Table~\ref{table:logistic analysis3-2}, our conjecture (a) proposed at the beginning of this section can be verified, i.e., cutting-off the initially returned non-binary short solution $\tilde{\textbf{x}}$ strongly correlates with a success in returning a valid binary solution.


\subsubsection{Systems with multiple subset-sum equations}

In this part,
we further test and verify conjecture (a) for systems with dimension $(m,n)$, to check
whether conjecture (a) also holds for systems with multiple subset-sum equations.
The small systems used are
recorded in Table~\ref{table:m2_n6_10problems}, which initially fail to return valid binary solutions
under $\textbf{Reduce}$.
Similarly to the reasons in Section 5.2.1, $\textbf{Reduce}$ is chosen other than
$\textbf{Reduce}_{1/2}$ and \textbf{CJLOSS-Alg}.
All the jump points of subset-sum problems are enumerated, and
the statistical results are recorded in Table~\ref{table:n2_logistic analysis}
and Table~\ref{table:n2_logistic analysis2}.

{\small
\begin{table}[!ht]
\caption{Systems with $A$ and $\emph{\textbf{b}}$, dimension $(m,n)=(2,6)$.}
\label{table:m2_n6_10problems}
\vspace{2mm}
\centering
\begin{tabular}{c||c|c}\hline
  Problem No. &  $A$                            &  $\emph{\textbf{b}}$  \\\hline\hline
  1           &   $\left(\begin{array}{c} 63,~ 9,~ 34,~ 46,~ 2,~ 55\\
                                          51,~ 19,~ 12,~ 44,~ 3,~ 25 \end{array}\right)$ &
                                          $\left(\begin{array}{c}
                                            99 \\ 66
                                          \end{array}\right)$ \\\hline 
  2           &   $\left(\begin{array}{c} 52,~ 43,~ 1,~ 10,~ 9,~ 11\\
                                          9,~ 21,~ 1,~ 37,~ 41,~ 43 \end{array}\right)$ &
                                          $\left(\begin{array}{c}
                                          62 \\ 51
                                          \end{array}\right)$ \\\hline 
  3           &   $\left(\begin{array}{c} 8,~ 37,~ 62,~ 62,~ 17,~ 32\\
                                          12,~ 29,~ 38,~ 51,~ 10,~ 20 \end{array}\right)$ &
                                          $\left(\begin{array}{c}
                                            87 \\ 60
                                          \end{array}\right)$ \\\hline 
  4           &   $\left(\begin{array}{c} 13,~ 52,~ 53,~ 38,~ 2,~ 23\\
                                            29,~ 25,~ 8,~ 42,~ 8,~ 5 \end{array}\right)$ &
                                          $\left(\begin{array}{c}
                                            68 \\ 45
                                          \end{array}\right)$ \\\hline 
  5           &   $\left(\begin{array}{c} 37,~ 18,~ 1,~ 39,~ 37,~ 22\\
                                          54,~ 59,~ 8,~ 43,~ 27,~ 9 \end{array}\right)$ &
                                          $\left(\begin{array}{c}
                                            75 \\ 89
                                          \end{array}\right)$ \\\hline 
  6           &   $\left(\begin{array}{c} 12,~ 13,~ 8,~ 32,~ 50,~ 49\\
                                          4,~ 51,~ 29,~ 37,~ 51,~ 27\end{array}\right)$ &
                                          $\left(\begin{array}{c}
                                            70  \\ 84
                                          \end{array}\right)$ \\\hline 
  7           &   $\left(\begin{array}{c} 4,~ 2,~ 49,~ 52,~ 7,~ 12\\
                                          13,~ 25,~ 4,~ 34,~ 53,~ 49\end{array}\right)$ &
                                          $\left(\begin{array}{c}
                                            60  \\ 70
                                          \end{array}\right)$ \\\hline 
  8           &   $\left(\begin{array}{c} 27,~ 1,~ 5,~ 2,~ 35,~ 64\\
                                          1,~ 21,~ 13,~ 8,~ 35,~ 44\end{array}\right)$ &
                                          $\left(\begin{array}{c}
                                            67  \\ 49
                                          \end{array}\right)$ \\\hline 
  9           &   $\left(\begin{array}{c} 31,~ 18,~ 3,~ 63,~ 61,~ 52\\
                                          19,~ 8,~ 36,~ 58,~ 50,~ 63\end{array}\right)$ &
                                          $\left(\begin{array}{c}
                                            95  \\ 105
                                          \end{array}\right)$ \\\hline 
  10           &   $\left(\begin{array}{c} 30,~ 59,~ 50,~ 7,~ 2,~ 34\\
                                          3,~ 58,~ 49,~ 42,~ 16,~ 35\end{array}\right)$ &
                                          $\left(\begin{array}{c}
                                            82 \\ 68
                                          \end{array}\right)$ \\ 
  \hline\hline
\end{tabular}
\end{table}
}

\newpage
\begin{table}[!ht]
\caption{Success or Failure vs. Cut or Non-cut.}
\label{table:n2_logistic analysis}
\vspace{2mm}
\footnotesize
\centering
\begin{threeparttable}
\begin{tabular}{c||c|c}\hline
  \multirow{2}{*}{Problem No.}
  &  \multicolumn{2}{c}{Coefficients ($t$-Value, $p$-Value)}\\\cline{2-3}
  &  Constant          &  Cut or Non-cut     \\\hline
  1           &  -1.9995 (-82.9485, 0) &  1.3583 (45.1497, 0) \\\hline
  2           &  -3.0770 (-61.0656, 0) &  3.0673 (55.7344, 0) \\\hline
  3           &  -2.6314 (-60.3586, 0) &  3.2163 (70.5206, 0) \\\hline
  4           &  -2.4893 (-64.8510, 0) &  3.1334 (71.4363, 0) \\\hline
  5           &  -3.6363 (-59.0880, 0) &  4.7840 (74.8191, 0) \\\hline
  6           &  -2.4159 (-77.1892, 0) &  2.7534 (78.7439, 0) \\\hline
  7           &  -2.1992 (-69.1417, 0) &  1.4579 (37.8164, 6.1188${\times}10^{-313}$) \\\hline
  8           &  -2.7930 (-55.0320, 0) &  3.2765 (58.6059, 0) \\\hline
  9           &  -2.3780 (-103.6493, 0)   & 2.9493 (112.1045, 0) \\\hline
  10          &  -3.9944 (-69.1250, 0)   & 4.1811 (70.0201, 0) \\
  \hline\hline
\end{tabular}
    \begin{tablenotes}
        \footnotesize
        \item Note: Table~\ref{table:n2_logistic analysis} records
    the logistic regression test result \emph{with} constant,
    which can be used together with Table~\ref{table:n2_logistic analysis2}.
    We can observe that the relation
    between `Success', 'Failure' (1 or 0) and 'Non-cut', `Cut' (1 or 0) is significant,
    i.e., cutting-off $\tilde{\textbf{x}}$ strongly correlates with a success.
    \end{tablenotes}
\end{threeparttable}
\end{table}

\begin{table}[!ht]
\caption{Success or Failure vs. Cut or Non-cut.}
\label{table:n2_logistic analysis2}
\vspace{2mm}
\footnotesize
\centering
\begin{threeparttable}
\begin{tabular}{c||c|c}\hline
  \multirow{2}{*}{Problem No.}
  &  \multicolumn{2}{c}{Coefficients ($t$-Value, $p$-Value)}\\\cline{2-3}
  &  Constant          &  Cut or Non-cut     \\\hline
  1           &   - &   -0.6412 ( -35.6229,  6.1922${\times}10^{-278}$) \\\hline
  2           &   - &   -0.0098 (-0.4425, 0.6581) \\\hline
  3           &   - &   0.5849 (43.6602, 0) \\\hline
  4           &   - &   0.6441 (30.3447, 2.9511${\times}10^{-202}$) \\\hline
  5           &   - &   1.1477 (66.1315, 0) \\\hline
  6           &   - &   0.3375 (21.6483, 6.3047${\times}10^{-104}$) \\\hline
  7           &   - &   -0.7413 (-34.0278, 8.6339${\times}10^{-254}$) \\\hline
  8           &   - &   0.4835 (20.6197, 1.8257${\times}10^{-94}$) \\\hline
  9           &   - &   0.5713 (44.3725, 0) \\\hline
  10          &   - &   0.1867 (12.4064, 2.4124${\times}10^{-35}$) \\
  \hline\hline
\end{tabular}
    \begin{tablenotes}
        \footnotesize
        \item Note: Table~\ref{table:n2_logistic analysis2} records
    the logistic regression test result \emph{without} constant,
    which can be used together with Table~\ref{table:n2_logistic analysis}.
    We can observe that the relation
    between `Success', 'Failure' (1 or 0) and 'Non-cut', `Cut' (1 or 0) is significant,
    i.e., cutting-off $\tilde{\textbf{x}}$ strongly correlates with a success.
    \end{tablenotes}
\end{threeparttable}
\end{table}

\qquad
Based on the data recorded in Table~\ref{table:n2_logistic analysis} and
Table~\ref{table:n2_logistic analysis2}, our conjecture (a) proposed at the beginning of this section can be verified,
i.e., cutting-off the initially returned non-binary short solution $\tilde{\textbf{x}}$ strongly correlates with a success in returning a valid binary solution.
Note that, for every possible scenario of $(t,M)$ pair, `Success'~$:=1$, `Failure'~$:=0$, `Non-cut'~$:=1$,
and `Cut'~$:=0$, where `Cut' means cutting off the initially returned
non-binary solution $\tilde{\textbf{x}}$, and `Non-cut' means not cutting off $\tilde{\textbf{x}}$.
Interesting phenomenon is that, when we compare
Table~\ref{table:logistic analysis3-1}
and
Table~\ref{table:n2_logistic analysis},
the signs of coefficients for variable 'Cut' or 'Non-cut' opposite to each other.

\qquad
Below, a concrete example is used to illustrate.

\begin{exam}\label{exam:m2_n6_prob1}
Consider the following problem,
\begin{align*}
  A = \left(\begin{array}{c}
  63,~ 9,~ 34,~ 46,~ 2,~ 55\\
  51,~ 19,~ 12,~ 44,~ 3,~ 25
 \end{array}\right),~~\text{and}~~
  \textbf{b} = \left(\begin{array}{c}
            99 \\ 66
            \end{array}\right),
\end{align*}
which helps to illustrate how some $r=\frac{t}{M}$ and the corresponding set
$\circled{3}|r$
enable the new system
to return a valid binary solution.
\end{exam}
\solution
Initially, the system returns a short non-binary solution $(0, 0, 0, 1, -1, 1)$
under Algorithm~$\textbf{Reduce}$.

Let $r_1 = \frac{t_1}{M_1} = \frac{1}{63}$, and apply $r_1$ to the first equation,
thus to obtain a new equation as follows,
\begin{align}
  v_{11}x_1 + v_{12}x_2 + v_{13}x_3 + v_{14}x_4 + v_{15}x_5 + v_{16}x_6 + \tilde{k}_1 = w_1,
\end{align}
with
\begin{align}
  v_{1i} := \left\lfloor\frac{t_1}{M_1}A_{1i}\right\rfloor,~\text{for } i=1,2,\ldots,n,~~\text{and}~~
   w_{1} := \left\lfloor\frac{t_1}{M_1}b_1\right\rfloor,
\end{align}
and the upper bound of $\tilde{k}_1$ is $u(\tilde{k}_1) = 1$.

Let $r_2 = \frac{t_2}{M_2} = \frac{22}{51}$, and apply $r_2$ to the second equation,
thus to obtain a new equation as follows,
\begin{align}
  v_{21}x_1 + v_{22}x_2 + v_{23}x_3 + v_{24}x_4 + v_{25}x_5 + v_{26}x_6 + \tilde{k}_2 = w_2,
\end{align}
with
\begin{align}
  v_{2i} := \left\lfloor\frac{t_2}{M_2}A_{2i}\right\rfloor,~\text{for } i=1,2,\ldots,n,~~\text{and}~~
   w_{2} := \left\lfloor\frac{t_2}{M_2}b_2\right\rfloor.
\end{align}
and the upper bound of $\tilde{k}_2$ is $u(\tilde{k}_2) = 1$.

The new system becomes,
\begin{align}\label{new_system}
\left\{
\begin{array}{l}
   63x_1+ 9x_2 +34x_3 + 46x_4 + 2x_5 + 55x_6 =  99\\
   51x_1 + 19x_2 + 12x_3 + 44x_4 + 3x_5 + 25x_6 = 66\\
   1x_1 + 0x_2 + 0x_3 + 0x_4 + 0x_5 + 0x_6 + \tilde{k}_1 = 1\\
   22x_1 + 8x_2 + 5x_3 + 18x_4 + 1x_5 + 10x_6 + \tilde{k}_2 = 28\\
\end{array}
\right.
\end{align}
with $(\textbf{x}, \tilde{k}_1,\tilde{k}_2) = (x_1,x_2,x_3,x_4,x_5,x_6,\tilde{k}_1,\tilde{k}_2)^T\in\{0,1\}^8$,
which returns the following solution,
$$
(x, \tilde{k}_1,\tilde{k}_2) = (1, 0, 1, 0, 1, 0, 0, 0)^T,
$$
under $\textbf{Reduce}$ (see Algorithm~\ref{alg:1_reduce(x,D)} in this paper).
Truncating the first $n=6$ elements yields,
$$
\textbf{x} = (1, 0, 1, 0, 1, 0)^T,
$$
which is a valid binary solution to the original system $A\textbf{x}=\textbf{b}$.

\quad
Specifically, next we enumerate all the jump points of subset-sum problems to generate the new system,
thus to test
and verify the reasons of why some $r$ can succeed in helping return a binary solution.
For Example~\ref{exam:m2_n6_prob1}, there are total $(a_{11}+a_{12}+\cdots+a_{1n} - n)\times (a_{21}+a_{22}+\cdots+a_{2n} - n) = 30,044$
possible scenarios for $(\frac{t_1}{M_1}, \frac{t_2}{M_2})$ in Example~\ref{exam:m2_n6_prob1}.
\endsolution

\subsection{Analysis on Conjecture (b) and Conjecture (c)}

In this part, we further verify our conjectures (b) and (c) proposed at the beginning of this section.
For each system in Section 5.2.2, we first generate the data of related variables, then conduct statistical regression to test the significance
of these variables. In the following, we use Example~\ref{exam:m2_n6_prob1} again to illustrate.

\begin{exam}
Consider the following problem,
\begin{align*}
  A = \left(\begin{array}{c}
  63,~ 9,~ 34,~ 46,~ 2,~ 55\\
  51,~ 19,~ 12,~ 44,~ 3,~ 25
 \end{array}\right),~~\text{and}~~
  \textbf{b} = \left(\begin{array}{c}
            99 \\ 66
            \end{array}\right),
\end{align*}
which helps to illustrate how some $r=\frac{t}{M}$ can enable the new systems
to return binary solution.
\end{exam}
\solution
a) Let $r_1 = \frac{t_1}{M_1} = \frac{1}{63}$ and $r_2 = \frac{t_2}{M_2} = \frac{22}{51}$, the new system in Eq.~\eqref{new_system} is as follows,
\begin{align*}
\left\{
\begin{array}{l}
   63x_1+ 9x_2 +34x_3 + 46x_4 + 2x_5 + 55x_6 =  99\\
   51x_1 + 19x_2 + 12x_3 + 44x_4 + 3x_5 + 25x_6 = 66\\
   1x_1 + 0x_2 + 0x_3 + 0x_4 + 0x_5 + 0x_6 + \tilde{k}_1 = 1\\
   22x_1 + 8x_2 + 5x_3 + 18x_4 + 1x_5 + 10x_6 + \tilde{k}_2 = 28\\
\end{array}
\right.
\end{align*}
which returns binary solution to the original system.
The reduced kernel basis of the new system consists of columns of $D$,
\begin{align*}
D = \left(
\begin{array}{cccc}
-1 & 1 & 0 & -5\\
0 & -1 & -9 & 5\\
-1 & 4 & -3 & 5\\
1 & 1 & 5 & 2\\
-2 & -8 & 4 & 4\\
1 & -4 & -1 & 0\\
1 & -1 & 0 & 5\\
1 &-4 &3 &5
\end{array}
\right)
\end{align*}
with volume of the kernel lattice $\sqrt{\text{det}(D^TD)} = 4112$.
The minimum volume ellipsoid of this kernel lattice is calculated with semi-axes,
\begin{align*}
  (13.4214, 12.6793, 7.8505, 3.0782),
\end{align*}
and with volume $=20,294$.


b) Here we consider another scenario to compare.
Let $r_1 = \frac{t_1}{M_1} = \frac{3}{63}$ and $r_2 = \frac{t_2}{M_2} = \frac{36}{51}$, the new system becomes,
\begin{align*}
\left\{
\begin{array}{l}
   63x_1+ 9x_2 +34x_3 + 46x_4 + 2x_5 + 55x_6 =  99\\
   51x_1 + 19x_2 + 12x_3 + 44x_4 + 3x_5 + 25x_6 = 66\\
   3x_1 + 0x_2 + x_3 + 2x_4 + 0x_5 + 2x_6 + \tilde{k}_1 = 4\\
   36x_1 + 13x_2 + 8x_3 + 31x_4 + 2x_5 + 17x_6 + \tilde{k}_2 = 46\\
\end{array}
\right.
\end{align*}
which returns a non-binary solution $(0, 0, 0, 1, -1, 1, 0, 0)$. The truncated solution $(0, 0, 0, 1, -1, 1)$ to the original system is non-binary as well.
The reduced kernel basis of the new system consists of columns of $D$,
\begin{align*}
D = \left(
\begin{array}{cccc}
-1 & -2 & -6 & -6\\
0 & 1 & -4 & 5\\
-1 & -5 & 1 & 4\\
1 & 0 & 8 & 3\\
-2 & 6 & 6 & 2\\
1 & 5 & 0 & 1\\
0 & 1 & 1 & 6\\
0 & 2 & 0 & 5
\end{array}
\right)
\end{align*}
with volume of the kernel lattice $\sqrt{\text{det}(D^TD)} = 3621$.
The minimum volume ellipsoid of this kernel lattice is calculated with semi-axes,
\begin{align*}
  (15.2642, 9.6705, 8.6903,  2.8230),
\end{align*}
and with volume $=17,870$.

c) Here we choose the other scenario to compare further.
Let $r_1 = \frac{t_1}{M_1} = \frac{3}{63}$ and $r_2 = \frac{t_2}{M_2} = \frac{49}{51}$, the new system becomes,
\begin{align*}
\left\{
\begin{array}{l}
   63x_1+ 9x_2 +34x_3 + 46x_4 + 2x_5 + 55x_6 =  99\\
   51x_1 + 19x_2 + 12x_3 + 44x_4 + 3x_5 + 25x_6 = 66\\
   3x_1 + 0x_2 + x_3 + 2x_4 + 0x_5 + 2x_6 + \tilde{k}_1 = 4\\
   49x_1 + 18x_2 + 11x_3 + 42x_4 + 2x_5 + 24x_6 + \tilde{k}_2 = 63\\
\end{array}
\right.
\end{align*}
which returns a binary solution $(1, 0, 1, 0, 1, 0, 0, 1)$.
The reduced kernel basis of the new system consists of columns of $D$,
\begin{align*}
D = \left(
\begin{array}{cccc}
1 & -3 & -5 & -7\\
0 & 1 & -5 & 5\\
1 & -6 & 5 & 3\\
-1 & 1 & 9 & 4\\
2 & 4 & -2 & 0\\
-1 & 6 & -4 & 2\\
0 & 1 & 0 & 6\\
2 & 1 & 2 & 4
\end{array}
\right)
\end{align*}
with volume of the kernel lattice $\sqrt{\text{det}(D^TD)} = 4493$.
The minimum volume ellipsoid of this kernel lattice is calculated with semi-axes,
\begin{align*}
  (15.1941, 12.4433, 7.1633, 3.3181),
\end{align*}
and with volume $=22,176$.
\endsolution

\qquad
Based on simulation data of the above example, we observe that normally, for scenarios
with the same dimension of kernel lattices of new systems:
1) the larger the volume of the kernel lattice, the easier the algorithm to succeed;
2) the smaller the ratio between the maximum axis and minimum axis which implies that
the ellipsoid is more like a spheroid, the easier the algorithm to succeed.

To conduct further analysis, for a single system, we enumerate all its scenarios of disaggregation
and list the following data variables thus to test the relations among them,
\begin{itemize}
  \item Dimension of kernel lattice of new system;
  \item Volume of kernel lattice of new system;
  \item New system cuts or non-cuts the original non-binary solution;
  \item New system succeeds or fails in returning binary solution;
  \item Minimum volume ellipsoid of kernel lattice of new system.
\end{itemize}

\qquad
Logistic regressions are performed in MATLAB, and simulation data are recorded in Tables in this Section.

\begin{table}[!ht]
\caption{Success/Failure vs. Volume of Kernel Lattice.}
\label{table2:logistic analysis3-1}
\vspace{2mm}
\small
\centering
\begin{threeparttable}
\begin{tabular}{c||c|c}\hline
  \multirow{2}{*}{Problem No. v.s. Dimension of Kernel}
  &  \multicolumn{2}{c}{Coefficients ($t$-Value, $p$-Value)}\\\cline{2-3}
  &  Constant          &  Volume of Kernel$\times10^{-3}$   \\\hline
  Prob. 1 with kernel dimension 2      &   -1.5748 (-2.1524, 0.0314) &  12.4943 (2.7779, 0.0055)  \\\hline
  Prob. 1 with kernel dimension 2      &   - &  6.0372 (4.0360, 5.4368$\times{10}^{-5}$)  \\\hline
  Prob. 1 with kernel dimension 3      &   -1.5698 (-8.6879, 3.6904$\times{10}^{-18}$) &  1.5518 (12.7746, 2.2727$\times{10}^{-37}$)  \\\hline
  Prob. 1 with kernel dimension 3      &   - &  0.5935 (14.3314, 1.3921$\times{10}^{-46}$)  \\\hline
  Prob. 1 with kernel dimension 4      &   -1.9937 (-15.1330, 9.8117$\times{10}^{-52}$) &  0.3988 (12.1653, 4.7583$\times{10}^{-34}$)  \\\hline
  Prob. 1 with kernel dimension 4      &   - &  -0.0909 (-14.3101, 1.8937$\times{10}^{-46}$)  \\\hline
  Prob. 1 with kernel dimension 5      &   -5.3278 (-22.2111, 2.6799$\times{10}^{-109}$) &  0.4945 (17.0606, 2.9148$\times{10}^{-65}$)  \\\hline
  Prob. 1 with kernel dimension 5      &   - &  -0.1541 (-59.7735, 0)  \\\hline
  Prob. 1 with kernel dimension 6      &   -9.0442 (-7.2874, 3.1605$\times{10}^{-13}$) &  0.3724 (4.7755, 1.7926$\times{10}^{-6}$)  \\\hline
  Prob. 1 with kernel dimension 6      &   - &  -0.1998 (-58.9966, 0)  \\
  \hline\hline
\end{tabular}
\end{threeparttable}
\end{table}

\qquad
Data in Table~\ref{table2:logistic analysis3-1} confirms our conjecture that, volume of kernel lattice after disaggregation has significant
relation with whether successfully returning a binary solution after disaggregation. Statistically, data in Table~\ref{table2:logistic analysis3-1}
also confirms our conjecture that
the \emph{larger} the volume of kernel lattice after disaggregation, the \emph{easier}
the algorithm to succeed, since the coefficient in front of volume feature is \emph{positive}.

\qquad
Next, we want to test our conjecture on the rectangularity feature of kernel lattices. Multiple ways are designed to capture the rectangularity feature of lattice. We define the ratio between maximum semi-axis and minimum semi-axis of the minimum volume ellipsoid of a given lattice,
\begin{align}\label{eq:lamda_2}
  \tilde{\lambda} := \frac{\text{Max semi-axis after normalization}}
  {\text{Min semi-axis after normalization}},
\end{align}
where the minimum volume ellipsoid is of the lattice after normalization, that is, the lengths of
basis vectors of a lattice are all first normalized to be 1, only the directions of basis vectors are preserved, respectively. Regression data are recorded in Table~\ref{table:logistic analysis3-2-2}.

\begin{table}[!ht]
\caption{Success/Failure vs. $\tilde{\lambda}$ in Equation~\eqref{eq:lamda_2}.}
\label{table:logistic analysis3-2-2}
\vspace{2mm}
\small
\centering
\begin{threeparttable}
\begin{tabular}{c||c|c}\hline
  \multirow{2}{*}{Problem No. v.s. Dimension of Kernel}
  &  \multicolumn{2}{c}{Coefficients ($t$-Value, $p$-Value)}\\\cline{2-3}
  &  Constant          &  $\tilde{\lambda} = \frac{\text{Max semi-axis after normalization}}{\text{Min semi-axis after normalization}}$     \\\hline
  Prob. 1 with kernel dimension 2      & -39.9368 (-2.6665, 0.0077)  &  39.1621 (2.6906, 0.0071) \\\hline
  Prob. 1 with kernel dimension 2      & -  & 1.2803 (4.8150, 1.4719$\times{10}^{-6}$)  \\\hline
  Prob. 1 with kernel dimension 3      &  -9.3810 (-15.4799, 4.7404$\times{10}^{-54}$) & 7.6354 (15.6488, 3.3857$\times{10}^{-55}$)  \\\hline
  Prob. 1 with kernel dimension 3      & -  & 0.6312 (14.2420, 5.0242$\times{10}^{-46}$)  \\\hline
  Prob. 1 with kernel dimension 4      & -3.8662 (-19.5329, 5.7704$\times{10}^{-85}$)  & 1.9946 (17.5799, 3.5132$\times{10}^{-69}$)  \\\hline
  Prob. 1 with kernel dimension 4      & -  & -0.2153 (-14.7667, 2.4031$\times{10}^{-49}$)  \\\hline
  Prob. 1 with kernel dimension 5      & -5.5635 (-27.8055, 3.7251$\times{10}^{-170}$)  & 2.4023 (21.7052, 1.8309$\times{10}^{-104}$)  \\\hline
  Prob. 1 with kernel dimension 5      & -  &  -0.6998 (-59.2329, 0) \\\hline
  Prob. 1 with kernel dimension 6      & -5.6999 (-13.3403, 1.3496$\times{10}^{-40}$)  & 1.4387 (6.1019, 1.0484$\times{10}^{-9}$)  \\\hline
  Prob. 1 with kernel dimension 6      & - & -1.7934 (-58.9660, 0)   \\
  \hline\hline
\end{tabular}
\end{threeparttable}
\end{table}

\qquad
Based on data in Table~\ref{table:logistic analysis3-2-2}, we could see that the
\emph{rectangularity} feature is significantly related with whether succeed or fail.
Since a rectangle is with $\tilde{\lambda}$ equal 1, and a rhomboid
is with $\tilde{\lambda}$ greater than 1. The \emph{positive} coefficient
in front of $\tilde{\lambda}$ tells that, the \emph{less} rectangular the kernel lattice,
the \emph{easier} for the algorithm to succeed and return a binary solution.

\quad
Next we define a distance quantity to capture the rectangularity structure of kernel lattices,
\begin{align}
  \text{minimize}\quad\quad & d:= || D^T D - \text{diag}(\lambda)||_2\label{eq:111}\\
  \text{subject to}\quad\quad & \lambda \geq 0,\nonumber
\end{align}
where columns of matrix $D$ form the basis of kernel lattice, $\lambda = (\lambda_1, \lambda_2, \ldots, \lambda_s)^T\in\R_+^s$ with $s\in\Z_+$ be the dimension of kernel lattice. In fact, if columns of $D$ are orthogonal to each other, then the optimized objective value in Eq.~\eqref{eq:111} is 0, otherwise,
the optimized objective value in Eq.~\eqref{eq:111}
is greater than 0. The results of statistical test are recorded in Table~\ref{table:logistic analysis3-3}.

\begin{table}[!ht]
\caption{Success/Failure vs. Rectangularity of Kernel Lattice.}
\label{table:logistic analysis3-3}
\vspace{2mm}
\small
\centering
\begin{threeparttable}
\begin{tabular}{c||c|c}\hline
  \multirow{2}{*}{Problem No. v.s. Dimension of Kernel}
  &  \multicolumn{2}{c}{Coefficients ($t$-Value, $p$-Value)}\\\cline{2-3}
  &  Constant          &  $d$: Rectangularity Distance    \\\hline
  Prob. 1 with kernel dimension 2      &   -1.6194 (-2.2612, 0.0237) & 0.3748 (1.8171, 1.8171)  \\\hline
  Prob. 1 with kernel dimension 2      &   - &  0.1434 (2.3768, 0.0175)  \\\hline
  Prob. 1 with kernel dimension 3      &   -1.0665 (-7.4854, 7.1341${\times}10^{-14}$) &  0.0305 (12.6218, 1.6008${\times}10^{-36}$)  \\\hline
  Prob. 1 with kernel dimension 3      &   - & 0.0150 (14.5127, 1.0074${\times}10^{-47}$)  \\\hline
  Prob. 1 with kernel dimension 4      &   -2.0257 (-23.2839, 6.4508${\times}10^{-120}$) & 0.0238 (19.4280, 4.4719${\times}10^{-84}$)  \\\hline
  Prob. 1 with kernel dimension 4      &   - &  -0.0037 (-10.2548, 1.1263${\times}10^{-24}$)  \\\hline
  Prob. 1 with kernel dimension 5      &  -2.5764 (-33.7724, 5.0216${\times}10^{-250}$) &  0.0205 (18.0961, 3.4221${\times}10^{-73}$) \\\hline
  Prob. 1 with kernel dimension 5      &   - &  -0.0182 (-54.7069, 0) \\\hline
  Prob. 1 with kernel dimension 6      &  -4.7381 (-25.7345, 4.8013${\times}10^{-146}$)  & 0.0280 (9.5833, 9.4002${\times}10^{-22}$)   \\\hline
  Prob. 1 with kernel dimension 6      &   - &  -0.0581 (-58.2613, 0)  \\
  \hline\hline
\end{tabular}
\end{threeparttable}
\end{table}


\vspace{2mm}
\quad
A variation of the rectangularity distance $d$ defined in Eq.~\eqref{eq:111} is defined as follows,
\begin{align}
  \text{minimize}\quad\quad & \tilde{d}:= || \tilde{D}^T \tilde{D} - \text{diag}(\lambda)||_2\label{eq:222}\\
  \text{subject to}\quad\quad & \lambda \geq 0,\nonumber
\end{align}
where columns of $\tilde{D}$ are normalized columns of $D$, all with length 1.
The results of statistical test are recorded in Table~\ref{table:logistic analysis3-4}.

\begin{table}[!ht]
\caption{Success/Failure vs. Rectangularity of Kernel Lattice.}
\label{table:logistic analysis3-4}
\vspace{2mm}
\small
\centering
\begin{threeparttable}
\begin{tabular}{c||c|c}\hline
  \multirow{2}{*}{Problem No. v.s. Dimension of Kernel}
  &  \multicolumn{2}{c}{Coefficients ($t$-Value, $p$-Value)}\\\cline{2-3}
  &  Constant          &  $\tilde{d}$: Normalized Rectangularity Distance     \\\hline
  Prob. 1 with kernel dimension 2      &  -0.7923 (-1.4310, 0.1524)  &  40.8038 (2.7501, 0.0060) \\\hline
  Prob. 1 with kernel dimension 2      &   - & 26.6884 (3.4789, 5.0345${\times}10^{-4}$)   \\\hline
  Prob. 1 with kernel dimension 3      &   -2.0401 (-13.4564, 2.8252${\times}10^{-41}$)  & 10.7147 (16.6736, 2.0399${\times}10^{-62}$)   \\\hline
  Prob. 1 with kernel dimension 3      &   - &  4.1293 (17.0258, 5.2867${\times}10^{-65}$) \\\hline
  Prob. 1 with kernel dimension 4      &  -2.2619 (-21.0078, 5.5669${\times}10^{-98}$)  &  3.8588 (17.7086, 3.5996${\times}10^{-70}$) \\\hline
  Prob. 1 with kernel dimension 4      &   - &  -0.6296 (-12.1393, 6.5419${\times}10^{-34}$)  \\\hline
  Prob. 1 with kernel dimension 5      &  -3.6053 (-31.3312, 1.7553${\times}10^{-215}$) &  4.6962 (20.9265, 3.0710${\times}10^{-97}$) \\\hline
  Prob. 1 with kernel dimension 5      &   - & -2.4374 (-57.2182, 0)  \\\hline
  Prob. 1 with kernel dimension 6      &  -5.5366 (-19.0951, 2.7707${\times}10^{-81}$)  & 4.9645 (8.6713, 4.2727${\times}10^{-18}$)   \\\hline
  Prob. 1 with kernel dimension 6      &   - &   -6.6999 (-58.7212, 0) \\
  \hline\hline
\end{tabular}
\end{threeparttable}
\end{table}

\qquad
Data in Table~\ref{table:logistic analysis3-3} and Table~\ref{table:logistic analysis3-4} reveal that $p$ values are small enough to guarantee that the rectangularity feature is significantly related with success (failure) output. As the coefficients for rectangularity feature are \emph{positive}, and success is of a higher value than failure (1 denotes success, 0 denotes failure), thus
statistically, the \emph{less rectangularity} of kernel lattice, the \emph{easier} the algorithm to succeed.

\qquad
In summary, we use Problem 1 recorded in Table~\ref{table:m2_n6_10problems} to further test the significance of some features that we conjecture can predict success or failure of returning binary solution statistically. We filter and test kernel lattices with same dimension after disaggregation, respectively. After disaggregation, there are total 72 kernel lattices with dimension 2,
total 1206 kernel lattices with dimension 3,
total 6574 kernel lattices with dimension 4,
total 13288 kernel lattices with dimension 5, and
total 8904 kernel lattices with dimension 6.
Generally, the \emph{larger} the volume of kernel lattice after disaggregation, the \emph{easier}
the algorithm to succeed.
Meanwhile, the \emph{less rectangularity} of kernel lattice after disaggregation,
the \emph{easier} the algorithm to succeed and return a binary solution.

\section{Conclusion and Further Study}
Generally speaking, the dimensions of solution space of subset-sum problems or
systems of equations depend on the magnitudes of
$m$ and $n$.
When restrict the general solution space to set of binary solutions, from the literature,
we know that subset-sum problems with density close to 1, and systems of equations with half-half split are the most difficult.
When increase the value of $m$ via disaggregation, more information will be revealed and the dimension of solution space will be lowered.
Moreover, by utilizing disaggregation techniques to add more equations, it is possible to cut off some invalid short solutions, thus to increase the probability of returning
valid binary solutions.
However, there do exist a balance between introducing new equation via disaggregation techniques, and including new variables after disaggregation. Here, the quantities of ``jump points" play a crucial role.

\qquad
Based on our numerical simulations, if the success ratio of original systems
with fixed dimension
is non-zero, then we always can utilize disaggregation technique to increase the success ratio
to 100\%. Otherwise, it is difficult to utilize disaggregation technique to increase the success ratio, for example,
for problems with dimension $m=1$ and $n=50$ and above, in Table~\ref{table:1_comparison_lattice_attacks};
and for problems with dimension $m=2$ and $n=80$ and above, in Table~\ref{table:2_m_n_lattice_attacks}.

\qquad
For further study, we observe that for some fixed dimensions, original systems all fail to return
valid binary solutions.
For such problem sets,
effective algorithms are in urgent need to be
further designed thus to return valid binary solutions.

\qquad
We also would like to examine the effects of adding more than one equation in disaggregation procedure.
Intuitively, adding more equations (if disaggregation is successful) will even further lower the dimension of solution space and
reveal even more information of the problem, and thus to increase the chance of returning binary solution after lattice transformation.

\bibliographystyle{acm}

\bibliography{literature}

\end{document}